\documentclass[11pt]{article}
\usepackage{amssymb}
\usepackage{graphicx}
\usepackage{amsmath}
\usepackage{makeidx}
\usepackage{indentfirst}
\usepackage[T1]{fontenc}
\usepackage[utf8]{inputenc}
\usepackage{authblk}
\usepackage{rotating}

\newcounter{resultnum}[section]
\newcounter{conclusionnum}[section]
\newcounter{conditionnum}[section]
\newcounter{conjecturenum}[section]
\newcounter{examplenum}[section]
\newcounter{exercisenum}[section]
\newcounter{lemmanum}[section]
\newcounter{notationnum}[section]
\newcounter{theoremnum}[section]
\newcounter{definitionnum}[section]
\newcounter{corollarynum}[section]
\newcounter{remarknum}[section]
\newcounter{propositionnum}[section]
\newcounter{acknowledgementnum}[section]
\newcounter{algorithmnum}[section]
\newcounter{axiomnum}[section]
\newcounter{casenum}[section]
\newcounter{claimnum}[section]
\newcounter{summarynum}[section]
\newcounter{problemnum}[section]
\begin{document}

\title{Off--Diagonal Deformations of Kerr Black Holes in Einstein\\
and Modified Massive Gravity and Higher Dimensions}
\date{September 26, 2014}
\author{ \textbf{Tamara Gheorghiu}\thanks{%
tamara.gheorghiu@yahoo.com}  \and  {\small {\textsl{\, University "Al. I.
Cuza" Ia\c si,\ Project IDEI;\ }} }  \and  {\small {\textsl{\ 14 Alexadnru
Lapu\c sneanu street, Corpus R, UAIC, office 323,} }}  \and  {\small {%
\textsl{I}a\c si, Romania 700057}}  \and  {\small and }  \and  {\small \ {%
\textsl{U}niversity of Medicine and Pharmacy "Gr. T. Popa" Ia\c si }} \\
{\small {\textsl{Faculty of Medicine;\ 16 University street, Ia\c si,\
Romania,\ 700115 }}} \\
{\qquad} \\
\textbf{Olivia Vacaru}\thanks{%
olivia.vacaru@yahoo.com} \\
{\small {\textsl{National College of Ia\c si;\ 4 Arcu street, Ia\c si,\
Romania,\ 700125 }}} \\
{\qquad} \\
\textbf{Sergiu I. Vacaru}\thanks{%
sergiu.vacaru@cern.ch; sergiu.vacaru@uaic.ro} \\
{\qquad} \\
{\small {\textsl{Theory Division, CERN, CH-1211, Geneva 23, Switzerland;}}}%
\thanks{%
visiting researcher}\\
{\small and } \\
{\small {\textsl{\, University "Al. I. Cuza" Ia\c si}, Rector's Office;\ } }%
\\
{\small {\textsl{\ 14 Alexadnru Lapu\c sneanu street, Corpus R, UAIC, office
323,} }}\\
{\small {\textsl{I}a\c si, Romania 700057}} }
\maketitle

\begin{abstract}
We find general parameterizations for generic off--diagonal spacetime
metrics and matter sources in general relativity (GR) and modified gravity
theories when the field equations decouple with respect to certain types of
nonholonomic frames of reference. This allows us to construct various
classes of exact solutions when the coefficients of the fundamental
geometric/physical objects depend on all spacetime coordinates via
corresponding classes of generating and integration functions and/or
constants. Such (modified) spacetimes display Killing and non--Killing
symmetries, describe nonlinear vacuum configurations and effective
polarizations of cosmological and interaction constants. Our method can be
extended to higher dimensions which simplifies some proofs for embedded and
nonholonomically constrained four-dimensional configurations. We reproduce
the Kerr solution and show how to deform it nonholonomically into new
classes of generic off--diagonal solutions depending on 3-8 spacetime
coordinates. Certain examples of exact solutions are analyzed and that are
determined by contributions of new type of interactions with sources in
massive gravity and/or modified f(R,T) gravity.  We conclude that by considering
generic off--diagonal nonlinear parametric interactions in GR it is possible
to mimic various effects in massive and/or modified gravity, or to
distinguish certain classes of "generic" modified gravity solutions which
cannot be encoded in GR.

\vskip0.1cm

\textbf{Keywords:}\

\vskip3pt MSC2010:\ 8C15, 8D99, 83E99

PACS2008:\ 04.20.Jb, 04.50.-h, 04.20.Cv
\end{abstract}

\tableofcontents

\section{Introduction}

The gravitational field equations in general relativity, GR, and modified
gravity theories, MGT, are very sophisticate systems of nonlinear partial
differential equations (PDEs). Advanced analytic and
numerical methods are necessary for constructing exact and approximate solutions of such
equations.  A number of examples of exact solutions are summarized in the
monographs \cite{kramer,griff} where the coefficients of the fundamental
geometric/physical objects depend on one and/or two coordinates in four
dimensional (4-d) spacetimes and when the diagonalization of the metrics is
possible via coordinate transformations. There are well known physically
important exact solutions for the Schwarzschild, Kerr, Friedman-Lema\^ itre
-Robertson-Worker (FLRW), wormhole spacetimes etc. These classes of
solutions are generated by certain ansatzes when the Einstein equations are
transformed into certain systems of nonlinear second order ordinary
equations (ODE), 2-d solitonic equations etc. Such systems of PDEs display
Killing vector symmetries which results in additional parametric symmetries
\cite{ger1,ger2,vpars}.

The problem of constructing generic off--diagonal exact solutions (which can
not be diagonalized via coordinate transformations) with metric coefficients
depending on three and/or four coordinates is much more difficult. There
are, in general, six independent components of a metric tensor from the ten components
in a 4-d (pseudo) Riemannian spacetime\footnote{%
four components of the ten can be fixed to zero  using coordinate
transformations, and which is related to the Bianchi identities}. Any such ansatz
transforms the Einstein equations into systems of nonlinear coupled PDEs
which cannot be integrated in a general analytic form if the constructions
are performed in local coordinate frames.

In a series of works \cite{vex1,vpars,vex2,vex3,veym}, we have shown that it
is possible to decouple the gravitational field equations and perform formal
analytic integrations in various theories of gravity with metric and
nonlinear, N-, and linear connections structures. To prove the decoupling
property in a simplest way we have to consider spacetime fibrations with
splitting of dimensions, 2 (or 3) + 2 + 2 + ..., introduce certain adapted
frames of reference, consider formal extensions/embeddings of 4-d spacetimes
into higher dimensional ones and work with necessary types of linear
connections. Such an (auxiliary, in Einstein gravity) adapted connection is
also completely defined in a compatible form by the metric structure and
contains a nonholonomically induced torsion field. This allows us to
decouple the gravitational field equations and generate various classes of
exact solutions in generalized/modified gravity theories. After a class of
generalized solutions has been constructed in explicit form, we can
constrain to zero the induced torsion fields and "extract" solutions in
Einstein gravity. We emphasize that it is important to impose the
zero--torsion conditions after we found a class of generalized solutions
(contrary, we cannot decouple the corresponding systems of PDEs).

It should be noted here that the off--diagonal solutions constructed
following the above described anholonomic frame deformation method, AFDM,
depend on various classes of generating and integration functions and
parameters. The Cauchy problem can be formulated with respect to necessary
types of N-adapted frames; it is possible to generate various stable or
un-stable solutions with singularities, nontrivial deformed horizons,
stochastic behavior, etc ...  which depends on the type of nonlinear couplings,
prescribed symmetries, asymptotic and boundary conditions, see a number of
examples in \cite{vp,vt,vsingl1,vgrg} and references therein. In general, it
is not clear what physical importance (if any ?) these classes of
such solutions may have. For some well defined conditions, we can speculate about
black hole/ellipsoid/wormhole configurations embedded, for instance, into
solitonic gravitational backgrounds or to consider small ellipsoidal
deformations of certain "primary" spherical/cylindrical configurations.

Our geometric techniques of constructing exact solutions can be applied to
four dimensional, 4-d, (pseudo) Riemannian spacetimes with one and two
Killing symmetries. For such configurations, the well known Kerr solution
can be generated as a particular case. Then these "primary" metrics can be
subjected to nonholonomic deformations to "target" off--diagonal exact
solutions depending on three, or four, spacetime coordinates.

The first goal of this paper is to show how certain primary physically
important solutions depending on two coordinates can be generalized to new
classes of exact solutions in Einstein gravity and (higher dimensional)
modifications, with zero or nonzero torsion, depending on all possible
spacetime coordinates. We consider diagonal and off--diagonal
parametrization of primary and target solutions which are different from
those in \cite{vex1,vpars,vex2,vex3} and other works. In this way we generate
new classes of Einstein spacetimes and modifications and show that the AFDM
encodes various possibilities for generalization.

The second goal is to construct explicit examples of exact solutions as
nonholonomic deformations of the Kerr metric determined by nontrivial
sources and interactions in massive gravity and/or modified $f(R,T)$
gravity, see reviews and original results in Refs. \cite%
{odints1,odints2,capoz,odints3,drg1,drg2,drg3,hr1,hr2,nieu,koyam}. For
non--Hilbert Lagrangians in gravity theories, the functionals $f$ depend on
scalar curvature $R$ (computed, in general, for a linear connection with
nontrivial torsion, or for the Levi--Civita one), on various matter and
effective matter sources for modified gravity theories etc. We provide a
series of exact and/or small parameter-dependent solutions which for small
deformations mimic rotoid Kerr - de Sitter like black holes/ellipsoids
self--consistently imbedded into generic off--diagonal backgrounds of 4/ 6/
8 dimensional spacetimes. With respect to nonholonomic frames and via the
re--definition of generating and integrations functions and coefficients of the
sources, modifications of  Einstein gravity are modelled by effective
polarized cosmological constants and off--diagonal terms in the new classes
of solutions. For certain geometrically well defined conditions, various
effects in massive and $f$--modified gravity can be encoded into vacuum and
non--vacuum,  configurations (exact solutions) with nontrivial effective cosmological constants in GR. In some sense, we can mimic physically important effects in modified gravity effects (for instance,
acceleration of universe, certain dark energy and dark matter locally
anisotropic interactions, effective renormalization of quantum gravity
models, see Refs. \cite{vgrg,vbranef,vepl}) via nonlinear generic
off--diagonal interactions on effective Einstein spaces.  The main question
arising from such models and solutions is whether or not we need to modify
Einstein's gravitational theory, or to try and solve physically important issues in
modern cosmology and quantum gravity by considering only nonlinear and generic
off--diagonal interactions based on the general relativity paradigm.  There
is necessarily additional theoretical and experimental/observational research which is required
in order to analyze and solve these problems. Such directions of research
cannot be developed if we consider only diagonalizable (and rotating ones,
like the Kerr metric) metrics generated by an ansatz with two Killing
symmetries.

The plan of the paper is as follows:\ In section \ref{s2} we provide the
necessary geometric preliminaries on nonholonomic 2+2+2+... splittings of the
spacetime dimensions in GR and MGT. We summarize the key results on the AFDM for
constructing generic off--diagonal solutions in gravity theories depending
on all spacetime coordinates in dimensions 4,5,...,8.

In section \ref{s3} we prove the general decoupling property of the
(modified) Einstein equations which allows us to perform formal integrations
of corresponding systems of nonlinear PDE. The geometric constructions are
performed for the "simplest" case of one Killing symmetry in 4-d and
generalized to non-Killing configurations and for higher dimensions.

Section \ref{s4} is devoted to the theory of nonholonomic deformations of
exact solutions in modified gravity theories containing the Kerr solution as
a "primary" configurations but with target metrics being constructed as exact
solutions in massive gravity and/or $f$--modified gravity. We show how using
the AFDM we can generate as a particular case the Kerr solution. Then we
construct solutions with general off--diagonal deformations of the Kerr
metrics in 4--d massive gravity, provide examples of (non--Einstein) metrics
with nonholonomically induced torsions and study small $f$--modifications of
the Kerr metrics deformed by massive gravity. A separate subsection is
devoted to ellipsoidal 4--d deformations of the Kerr metric resulting in a
target vacuum rotoid or Kerr--de Sitter configurations. Another subsection
is devoted to extra dimensional massive off--diagonal modifications of the
Kerr solutions, for the case of 6--d spacetime with nontrivial cosmological
constant and for 8--d deformations which may model Finsler-like
configurations.

Finally (in section \ref{s5}), we provide our conclusions and speculate on the
physical meaning of the exact solutions constructed using the AFDM for massive
modified gravity theories and how such effects can be modelled by nonlinear
off--diagonal interactions in Einstein gravity.  Some relevant formulae for the coefficients
and sketches of the proofs are presented in the Appendix.

\section{Nonholonomic Frames with 2+2+.... Splitting}

\label{s2} In this section, we state the geometric conventions and outline
the formalism which are necessary for decoupling and integrating the
gravitational field equations in GR and MGTs, see relevant details in \cite%
{vex1,vpars,vex2,vex3}.

\subsection{Geometric preliminaries}

\subsubsection{Conventions}

For (higher dimensional) spacetime geometric models and related exact
solutions on a finite dimensional (pseudo) Riemannian spacetime $\ ^{s}V$,
we consider conventional splitting of dimensions, $\dim V=4+2s=2+2+...+2\geq
4;s\geq 0.$\footnote{%
In a similar form, we can split odd dimensions, for instance, $\dim
V=3+2+...+2$. Here it should be noted that it is not possible to elaborate
any simplified system of notations if we want to integrate in general
explicit form certain systems of PDEs related to higher dimensional
gravitational theories. It is important to distinguish indices and
coordinates corresponding to higher dimensions and nonholonomically
constrained variables.} The anholonomic frame deformation method, AFDM,
allows us to construct exact solutions with arbitrary signatures $(\pm 1,\pm
1,\pm 1,...\pm 1)$ of metrics $\ ^{s}\mathbf{g}$. Let us establish
conventions on (abstract) indices and coordinates $u^{\alpha
_{s}}=(x^{i_{s}},y^{a_{s}}),$ for $s=0,1,2,3,....$ labelling the oriented
number of two dimensional, 2-d, "shells" added to a 4--d spacetime. For $%
s=0, $ we write $u^{\alpha }=(x^{i},y^{a})$ and consider such local systems
of coordinates: {\small
\begin{eqnarray*}
s &=&0:u^{\alpha _{0}}=(x^{i_{0}},y^{a_{0}})=u^{\alpha }=(x^{i},y^{a}), \\
s &=&1:u^{\alpha _{1}}=(x^{\alpha }=u^{\alpha
},y^{a_{1}})=(x^{i},y^{a},y^{a_{1}}), \\
\ s &=&2:u^{\alpha _{2}}=(x^{\alpha _{1}}=u^{\alpha
_{1}},y^{a_{2}})=(x^{i},y^{a},y^{a_{1}},y^{a_{2}}), \\
\ s &=&3:u^{\alpha _{3}}=(x^{\alpha _{2}}=u^{\alpha
_{2}},y^{a_{3}})=(x^{i},y^{a},y^{a_{1}},y^{a_{2}},y^{a_{3}}),...
\end{eqnarray*}%
} when indices run corresponding values $%
i,j,...=1,2;a,b,...=3,4;a_{1},b_{1}...=5,6;a_{2},b_{2}...=7,8;$ $%
a_{3},b_{3}...=9,10,...$ and, for instance, $i_{1},j_{1},...=1,2,3,4;i_{2},$
$j_{2},...$ $=1,2,3,4,5,6;\ i_{3},j_{3},...=1,2,3,4,5,6,7,8;...$ In brief,
we shall write $u=(x,y);$ $\ ^{1}u=(u,\ ^{1}y)=(x,y,\ ^{1}y),\ ^{2}u=(\
^{1}u,\ ^{2}y)=(x,y,\ ^{1}y,\ ^{2}y),...$

Local frames (bases, $e_{\alpha _{s}}$) on $\ ^{s}V$ are denoted in the form
\begin{equation}
e_{\alpha _{s}}=e_{\ \alpha _{s}}^{\underline{\alpha }_{s}}(\ ^{s}u)\partial
/\partial u^{\underline{\alpha }_{s}},  \label{nholfr}
\end{equation}
where partial derivatives are $\partial _{\beta _{s}}:=\partial /\partial
u^{\beta _{s}},$ and indices are underlined if it is necessary to emphasize
that such values are defined with respect to a coordinate frame. In general,
the frames (\ref{nholfr}) are nonholonomic (equivalently, anholonomic, or
non-integrable), $e_{\alpha _{s}}e_{\beta _{s}}-e_{\beta _{s}}e_{\alpha
_{s}}=W_{\alpha _{s}\beta _{s}}^{\gamma _{s}}e_{\gamma _{s}},$ where the
anholonomy coefficients $W_{\alpha _{s}\beta _{s}}^{\gamma _{s}}=W_{\beta
_{s}\alpha _{s}}^{\gamma _{s}}(u)$ vanish for holonomic, i. e. integrable,
configurations. The dual frames are $e^{\alpha _{s}}=e_{\ \underline{\alpha }%
_{s}}^{\ \alpha _{s}}(\ ^{s}u)du^{\underline{\alpha }_{s}}$, which can be
defined from the condition $e^{\alpha _{s}}\rfloor e_{\beta _{s}}=\delta
_{\beta _{s}}^{\alpha _{s}}$ (the 'hook' operator $\rfloor $ corresponds to
the inner derivative and $\delta _{\beta _{s}}^{\alpha _{s}}$ is the
Kronecker symbol).

The conventional $2+2+...$ splitting for a metric is written in the form
\begin{equation}
\ ^{s}\mathbf{g=}g_{\alpha _{s}\beta _{s}}e^{\alpha _{s}}\otimes e^{\beta
_{s}}=g_{\underline{\alpha }_{s}\underline{\beta }_{s}}du^{\underline{\alpha
}_{s}}\otimes du^{\underline{\beta }_{s}},\ s=0,1,2, ...,  \label{metr}
\end{equation}%
where coefficients of the metric transform following the rule%
\begin{equation}
g_{\alpha _{s}\beta _{s}}=e_{\ \alpha _{s}}^{\underline{\alpha }_{s}}e_{\
\beta _{s}}^{\underline{\beta }_{s}}g_{\underline{\alpha }_{s}\underline{%
\beta }_{s}}.  \label{metrtransf}
\end{equation}%
Similar frame transforms can be considered for all tensor objects. We can
not preserve a splitting of dimensions under general frame/coordinate
transforms.

\subsubsection{Nonholonomic splitting with associated N--connections}

To prove the general decoupling property of the Einstein equations and
generalizations/ modifications we have to construct a necessary type of
nonholonomic $2+2+...$ nonholonomic splitting with associated nonlinear
connection (N-connection) structure. Such a splitting is introduced using
nonholonomic distributions:\footnote{%
In modern gravity, it is largely used the so--called ADM
(Arnowit--Deser--Misner) formalism with a 3+1 splitting, see details in \cite%
{misner}. It is not possible to elaborate a technique for a general decoupling of
the gravitational field equations and generating off--diagonal solutions if
we use only nonholonomic frame bases determined by  the "shift" and "lapse"
functions. To construct exact solutions is more convenient to work with a
correspondingly defined non--integrable 2+2+... splitting \cite{vex2,vex3}.}

\begin{enumerate}
\item A N--connection is stated by a Whitney sum
\begin{equation}
\ ^{s}\mathbf{N}:T\ ^{s}\mathbf{V}=h\mathbf{V}\oplus v\mathbf{V}\oplus \
^{1}v\mathbf{V}\oplus \ ^{2}v\mathbf{V}\oplus ...\oplus \ ^{s}v\mathbf{V},
\label{whitney}
\end{equation}%
for a conventional horizontal (h) and vertical (v) "shell by shell"
splitting. We shall write boldface letters for spaces and geometric objects
enabled/adapted to N--connection structure. This defines a local fibered
structure on $\ ^{s}\mathbf{V}$ when the coefficients of N--connection, $%
N_{i_{s}}^{a_{s}},$ for $\ ^{s}\mathbf{N}=N_{i_{s}}^{a_{s}}(\
^{s}u)dx^{i_{s}}\otimes \partial /\partial y^{a_{s}},$ induce a system of
N--adapted local bases, with N-elongated partial derivatives, $\mathbf{e}%
_{\nu _{s}}=(\mathbf{e}_{i_{s}},e_{a_{s}}),$ and cobases with N--adapted
differentials, $\mathbf{e}^{\mu _{s}}=(e^{i_{s}},\mathbf{e}^{a_{s}}).$ On a
4-d $\mathbf{V}$,
\begin{eqnarray}
&&\mathbf{e}_{i}=\frac{\partial }{\partial x^{i}}-\ N_{i}^{a}\frac{\partial
}{\partial y^{a}},\ e_{a}=\frac{\partial }{\partial y^{a}},  \label{nader} \\
&&e^{i}=dx^{i},\mathbf{e}^{a}=dy^{a}+\ N_{i}^{a}dx^{i},  \label{nadif}
\end{eqnarray}%
and on $s\geq 1$ shells,
\begin{eqnarray}
\mathbf{e}_{i_{s}} &=&\frac{\partial }{\partial x^{i_{s}}}-\
N_{i_{s}}^{a_{s}}\frac{\partial }{\partial y^{a_{s}}},\ e_{a_{s}}=\frac{%
\partial }{\partial y^{a_{s}}},  \label{naders} \\
e^{i_{s}} &=&dx^{i_{s}},\mathbf{e}^{a_{s}}=dy^{a_{s}}+\
N_{i_{s}}^{a_{s}}dx^{i_{s}}.  \label{nadifs}
\end{eqnarray}%
The N--adapted operators (\ref{nader}) and (\ref{naders}) define a subclass
of general frame transforms of type (\ref{nholfr}). The corresponding
anholonomy relations
\begin{equation}
\lbrack \mathbf{e}_{\alpha _{s}},\mathbf{e}_{\beta _{s}}]=\mathbf{e}_{\alpha
_{s}}\mathbf{e}_{\beta _{s}}-\mathbf{e}_{\beta _{s}}\mathbf{e}_{\alpha
_{s}}=W_{\alpha _{s}\beta _{s}}^{\gamma _{s}}\mathbf{e}_{\gamma _{s}},
\label{anhrel1}
\end{equation}%
are completely defined by the N--connection coefficients and their partial
derivatives, $W_{i_{s}a_{s}}^{b_{s}}=\partial _{a_{s}}N_{i_{s}}^{b_{s}}$ and
$W_{j_{s}i_{s}}^{a_{s}}=\Omega _{i_{s}j_{s}}^{a_{s}},$ where the curvature
of the N--connection is $\Omega _{i_{s}j_{s}}^{a_{s}}=\mathbf{e}_{j_{s}}\left(
N_{i_{s}}^{a_{s}}\right) -\mathbf{e}_{i_{s}}\left( N_{j_{s}}^{a_{s}}\right)
. $

\item Any metric structure $\ ^{s}\mathbf{g=\{g}_{\alpha _{s}\beta _{s}}%
\mathbf{\}}$ on $\ ^{s}\mathbf{V}$ can be written as a distinguished metric
(d--metric)\footnote{%
geometric objects with coefficients defined with respect to N--adapted
frames are called respectively distinguished metrics, distinguished tensors
etc (in brief, d--metrics, d--tensors etc)}
\begin{eqnarray}
\ \ ^{s}\mathbf{g} &=&\ g_{i_{s}j_{s}}(\ ^{s}u)\ e^{i_{s}}\otimes
e^{j_{s}}+\ g_{a_{s}b_{s}}(\ ^{s}u)\mathbf{e}^{a_{s}}\otimes \mathbf{e}%
^{b_{s}}  \label{dm} \\
&=&g_{ij}(x)\ e^{i}\otimes e^{j}+g_{ab}(u)\ \mathbf{e}^{a}\otimes \mathbf{e}%
^{b}+g_{a_{1}b_{1}}(\ ^{1}u)\ \mathbf{e}^{a_{1}}\otimes \mathbf{e}^{b_{1}}
+....+\ g_{a_{s}b_{s}}(\ ^{s}u)\mathbf{e}^{a_{s}}\otimes \mathbf{e} ^{b_{s}}.
\notag
\end{eqnarray}%
In coordinate frames, a metric (\ref{metr}) is parameterized by generic
off--diagonal matrices%
\begin{equation*}
\ \ \underline{g}_{\alpha \beta }\left(\ u\right) =\left[
\begin{array}{cc}
\ g_{ij}+\ h_{ab}N_{i}^{a}N_{j}^{b} & h_{ae}N_{j}^{e} \\
\ h_{be}N_{i}^{e} & \ h_{ab}%
\end{array}%
\right] ,
\end{equation*}%
%
%
%
%
%
%
%
%
%
%
%
%
%
%
\begin{equation*}
\ \underline{g}_{\alpha _{1}\beta _{1}}\left( \ ^{1}u\right) =\left[
\begin{array}{cc}
\ \underline{g}_{\alpha \beta } & h_{a_{1}e_{1}}N_{\beta _{1}}^{e_{1}} \\
\ h_{b_{1}e_{1}}N_{\alpha _{1}}^{e_{1}} & \ h_{a_{1}b_{1}}%
\end{array}%
\right] ,\ \ \ \underline{g}_{\alpha _{2}\beta _{2}}\left( \ ^{2}u\right) = %
\left[
\begin{array}{cc}
\ \underline{g}_{\alpha _{1}\beta _{1}} & h_{a_{2}e_{2}}N_{\beta
_{1}}^{e_{2}} \\
\ h_{b_{2}e_{2}}N_{\alpha _{1}}^{e_{2}} & \ h_{a_{2}b_{2}}%
\end{array}%
\right] ,\ ...
\end{equation*}%
\begin{equation*}
\ \ \underline{g}_{\alpha _{s}\beta _{s}}\left( \ ^{s}u\right) =\left[
\begin{array}{cc}
\ g_{i_{s}j_{s}}+\ h_{a_{s}b_{s}}N_{i_{s}}^{a_{s}}N_{j_{s}}^{b_{s}} &
h_{a_{s}e_{s}}N_{j_{s}}^{e_{s}} \\
\ h_{b_{s}e_{s}}N_{i_{s}}^{e_{s}} & \ h_{a_{s}b_{s}}%
\end{array}%
\right] .  \label{fansatz}
\end{equation*}
\end{enumerate}

For extra dimensions, such parameterizations are similar to those introduced
in the Kaluza--Klein theory when $y^{a_{s}},s\geq 1,$ are considered as
extra dimension coordinates with cylindrical compactification and $N_{\alpha
}^{e_{s}}(\ ^{s}u)\sim A_{a_{s}\alpha }^{e_{s}}(u)y^{\alpha }$ are for
certain (non) Abelian gauge fields $A_{a_{s}\alpha }^{e_{s}}(u)$. In
general, various parameterizations can be used for warped/trapped
coordinates in brane gravity and modifications of GR, see examples in \cite%
{vp,vt,vsingl1,vgrg}.

\subsubsection{The Levi--Civita and auxiliary N--adapted connections}

There is a subclass of linear connections on $\ ^{s}\mathbf{V}$ which are
adapted to the N--connection splitting (\ref{whitney}). By definition, a
distinguished connection, d--connection, $\mathbf{D}=(hD;vD),\ ^{1}\mathbf{D=%
}(\ ^{1}hD;\ ^{1}vD),...,\ ^{s}\mathbf{D=}(\ ^{s-1}hD;\ ^{s}vD),$ preserves
under parallelism the N--connection structure.\footnote{%
In our works, certain left \textquotedblright
up\textquotedblright\ or \textquotedblright low\textquotedblright\ labels are used in
order to emphasize that certain geometric objects are defined on a
corresponding shell and in terms of a fundamental geometric object. We shall omit such
labels if that does not result in ambiguities.} The coefficients%
\begin{eqnarray}
\mathbf{\Gamma }_{\ \beta \gamma }^{\alpha }
&=&(L_{jk}^{i},L_{bk}^{a};C_{jc}^{i},C_{bc}^{a}),  \notag \\
\mathbf{\Gamma }_{\ \beta _{1}\gamma _{1}}^{\alpha _{1}}&=& (L_{\beta \gamma
}^{\alpha },L_{b_{1}\gamma }^{a_{1}};C_{\beta c_{1}}^{\alpha
},C_{b_{1}c_{1}}^{a_{1}}),\ \mathbf{\Gamma }_{\ \beta _{2}\gamma
_{2}}^{\alpha _{2}} =(L_{\beta _{1}\gamma _{1}}^{\alpha _{1}},L_{b_{2}\gamma
_{1}}^{a_{2}};C_{\beta _{1}c_{2}}^{\alpha _{1}},C_{b_{2}c_{2}}^{a_{2}}),...,
\label{coefd} \\
\mathbf{\Gamma }_{\ \beta _{s}\gamma _{s}}^{\alpha _{s}} &=&(L_{\beta
_{s-1}\gamma _{s-1}}^{\alpha _{s-1}},L_{b_{s}\gamma _{s-1}}^{a_{s}};C_{\beta
_{s-1}c_{s}}^{\alpha _{s-1}},C_{b_{s}c_{s}}^{a_{s}})  \notag
\end{eqnarray}%
of a d--connection $\ ^{s}\mathbf{D=\{D}_{\alpha _{s}}\mathbf{\}}$ can be
computed in N--adapted form with respect to frames (\ref{nader})-- (\ref%
{nadifs}) following equations $\mathbf{D}_{\alpha _{s}}\mathbf{e}_{\beta
_{s}}=\mathbf{\Gamma }_{\ \beta _{s}\gamma _{s}}^{\alpha _{s}}\mathbf{e}%
_{\gamma _{s}}$ and covariant derivatives parameterized in the form
\begin{eqnarray*}
\mathbf{D}_{\alpha } &=&(D_{i};D_{a}),\mathbf{D}_{\alpha _{1}}=(\
^{1}D_{\alpha };D_{a_{1}}),\ \mathbf{D}_{\alpha _{2}} = (\ ^{2}D_{\alpha
_{1}};D_{a_{2}}),...,\mathbf{D}_{\alpha _{s}}=(\ ^{s}D_{\alpha
_{s-1}};D_{a_{s}}), \\
\mbox{ for } hD &=&(L_{jk}^{i},L_{bk}^{a}),vD=(C_{jc}^{i},C_{bc}^{a}), \\
\ ^{1}hD &=&(L_{\beta \gamma }^{\alpha },L_{b_{1}\gamma }^{a_{1}}),\
^{1}vD=(C_{\beta c_{1}}^{\alpha },C_{b_{1}c_{1}}^{a_{1}}),\ \ ^{2}hD =
(L_{\beta _{1}\gamma _{1}}^{\alpha _{1}},L_{b_{2}\gamma _{1}}^{a_{2}}),\
^{2}vD=(C_{\beta _{1}c_{2}}^{\alpha _{1}},C_{b_{2}c_{2}}^{a_{2}}), \\
&&..., \\
\ ^{s}hD &=&(L_{\beta _{s-1}\gamma _{s-1}}^{\alpha _{s-1}},L_{b_{s}\gamma
_{s-1}}^{a_{s}}),\ ^{s}vD=(C_{\beta _{s-1}c_{s}}^{\alpha
_{s-1}},C_{b_{s}c_{s}}^{a_{s}}).
\end{eqnarray*}%
Such coefficients can be computed with respect to mixed subsets of
coordinates and/or N--adapted frames on different shells. It is possible
always to consider such frame transforms when all shell frames are N-adapted
and $\ $%
\begin{equation*}
^{1}D_{\alpha }=\mathbf{D}_{\alpha },\ ^{2}D_{\alpha _{1}}=\mathbf{D}%
_{\alpha _{1}},...\ ^{s}D_{\alpha _{s-1}}=\mathbf{D}_{\alpha _{s-1}}.
\end{equation*}

To perform computations in N--adapted--shell form we can consider a
differential connection 1--form $\mathbf{\Gamma }_{\ \beta _{s}}^{\alpha
_{s}}=\mathbf{\Gamma }_{\ \beta _{s}\gamma _{s}}^{\alpha _{s}}\mathbf{e}%
^{\gamma _{s}}$ and elaborate a differential form calculus with respect to
skew symmetric tensor products of N--adapted frames (\ref{nader})-- (\ref%
{nadifs}). For instance, the torsion $\mathcal{T}^{\alpha _{s}}=\{\mathbf{T}%
_{\ \beta _{s}\gamma _{s}}^{\alpha _{s}}\}$ and curvature $\mathcal{R}%
_{~\beta _{s}}^{\alpha _{s}}=\{\mathbf{\mathbf{R}}_{\ \ \beta _{s}\gamma
_{s}\delta _{s}}^{\alpha _{s}}\}$ d--tensors of $\ \ ^{s}\mathbf{D}$ can be
computed respectively,
\begin{eqnarray}
&&\mathcal{T}^{\alpha _{s}}:=\ ^{s}\mathbf{De}^{\alpha _{s}}=d\mathbf{e}%
^{\alpha _{s}}+\mathbf{\Gamma }_{\ \beta _{s}}^{\alpha _{s}}\wedge \mathbf{e}%
^{\beta _{s}}\   \label{dt} \\
&& \mathcal{R}_{~\beta _{s}}^{\alpha _{s}}:=\ ^{s}\mathbf{D\Gamma }_{\ \beta
_{s}}^{\alpha _{s}}=d\mathbf{\Gamma }_{\ \beta _{s}}^{\alpha _{s}}-\mathbf{%
\Gamma }_{\ \beta _{s}}^{\gamma _{s}}\wedge \mathbf{\Gamma }_{\ \gamma
_{s}}^{\alpha _{s}}=\mathbf{R}_{\ \beta _{s}\gamma _{s}\delta _{s}}^{\alpha
_{s}}\mathbf{e}^{\gamma _{s}}\wedge \mathbf{e}^{\delta _{s}},  \label{dc}
\end{eqnarray}%
see Refs. \cite{vex3} for explicit calculation of the coefficients $\mathbf{R}_{\
\beta _{s}\gamma _{s}\delta _{s}}^{\alpha _{s}}$ in higher dimensions.

For any (pseudo) Riemannian metric $\ \ ^{s}\mathbf{g,}$ we can construct in
standard form the Levi--Civita connection (LC--connection), $\ ^{s}\nabla
=\{\ _{\shortmid }\Gamma _{\ \beta _{s}\gamma _{s}}^{\alpha _{s}}\},$ which
is completely defined by the metric coefficients following two conditions:
This linear connection is metric compatible, $\ ^{s}\nabla (\ ^{s}\mathbf{g)}%
=0,$ and with zero torsion, $\ \ _{\shortmid }\mathcal{T}^{\alpha _{s}}=0$
(see formulas (\ref{dt}) for $\ ^{s}\mathbf{D\rightarrow }\ ^{s}\nabla ).$
Such a linear connection is not a d--connection because it does not preserve
under general coordinate transforms a N--connection splitting.

To elaborate a covariant differential calculus adapted to decomposition (\ref%
{whitney}) we have to introduce a different type of linear connection. This
is the canonical d--connection $\ ^{s}\widehat{\mathbf{D}}$ which is
completely and uniquely defined by a (pseudo) Riemannian metric $\ ^{s}%
\mathbf{g}$ (\ref{fansatz}) for a chosen $\ ^{s}\mathbf{N=\{}%
N_{i_{s}}^{a_{s}}\}$ if and only if $\ \ ^{s}\widehat{\mathbf{D}}(\ ^{s}%
\mathbf{g)}=0$ and the horizontal and vertical torsions are zero, i.e. $h%
\widehat{\mathcal{T}}=\{\widehat{\mathbf{T}}_{\ jk}^{i}\}=0,$ $v\widehat{%
\mathcal{T}}=\{\widehat{\mathbf{T}}_{\ bc}^{a}\}=0,\ ^{1}v\widehat{\mathcal{T%
}}=\{\widehat{\mathbf{T}}_{\ b_{1}c_{1}}^{a_{1}}\}=0,...,\ ^{s}v\widehat{%
\mathcal{T}}=\{\widehat{\mathbf{T}}_{\ b_{s}c_{s}}^{a_{s}}\}=0.$ We \ can
check by straightforward computations that such conditions are satisfied by $%
\ ^{s}\widehat{\mathbf{D}}=\{\widehat{\mathbf{\Gamma }}_{\ \alpha _{s}\beta
_{s}}^{\gamma _{s}}\}$ with coefficients (\ref{coefd}) computed recurrently
\begin{eqnarray}
\widehat{L}_{jk}^{i} &=&\frac{1}{2}g^{ir}\left( \mathbf{e}_{k}g_{jr}+\mathbf{%
e}_{j}g_{kr}-\mathbf{e}_{r}g_{jk}\right) ,  \notag \\
\widehat{L}_{bk}^{a} &=&e_{b}(N_{k}^{a})+\frac{1}{2}h^{ac}\left( \mathbf{e}%
_{k}h_{bc}-h_{dc}\ e_{b}N_{k}^{d}-h_{db}\ e_{c}N_{k}^{d}\right) ,  \notag \\
\widehat{C}_{jc}^{i} &=&\frac{1}{2}g^{ik}e_{c}g_{jk},\ \widehat{C}_{bc}^{a}=%
\frac{1}{2}h^{ad}\left( e_{c}h_{bd}+e_{c}h_{cd}-e_{d}h_{bc}\right) ,
\label{candcon} \\
\widehat{L}_{\beta \gamma }^{\alpha } &=&\frac{1}{2}g^{\alpha \tau }\left(
\mathbf{e}_{\gamma }g_{\beta \tau }+\mathbf{e}_{\beta }g_{\gamma \tau }-%
\mathbf{e}_{\tau }g_{\beta \gamma }\right),  \notag \\
&&  \notag \\
\widehat{L}_{b_{1}\gamma }^{a_{1}} &=&e_{b_{1}}(N_{\gamma }^{a_{1}})+\frac{1%
}{2}h^{a_{1}c_{1}}\left( \mathbf{e}_{\gamma }h_{b_{1}c_{1}}-h_{d_{1}c_{1}}\
e_{b_{1}}N_{\gamma }^{d_{1}}-h_{d_{1}b_{1}}\ e_{c_{1}}N_{\gamma
}^{d_{1}}\right),  \notag \\
\widehat{C}_{\beta c_{1}}^{\alpha } &=&\frac{1}{2}g^{\alpha \tau
}e_{c_{1}}g_{\beta \tau },\ \widehat{C}_{b_{1}c_{1}}^{a_{1}}=\frac{1}{2}%
h^{a_{1}d_{1}}\left(
e_{c_{1}}h_{b_{1}d_{1}}+e_{c_{1}}h_{c_{1}d_{1}}-e_{d_{1}}h_{b_{1}c_{1}}%
\right),  \notag \\
&& ...  \notag \\
&&  \notag \\
\widehat{L}_{\beta _{s-1}\gamma _{s-1}}^{\alpha _{s-1}} &=&\frac{1}{2}%
g^{\alpha _{s-1}\tau _{s-1}}\left( \mathbf{e}_{\gamma _{s-1}}g_{\beta
_{s-1}\tau _{s-1}}+\mathbf{e}_{\beta _{s-1}}g_{\gamma _{s-1}\tau _{s-1}}-%
\mathbf{e}_{\tau _{s-1}}g_{\beta _{s-1}\gamma _{s-1}}\right) ,  \notag \\
\widehat{L}_{b_{s}\gamma _{s-1}}^{a_{s}} &=&e_{b_{s}}(N_{\gamma
_{s-1}}^{a_{s}})+ \frac{1}{2}h^{a_{s}c_{s}}\left( \mathbf{e}_{\gamma
_{s-1}}h_{b_{s}c_{s}}-h_{d_{s}c_{s}}\ e_{b_{s}}N_{\gamma
_{s-1}}^{d_{s}}-h_{d_{s}b_{s}}\ e_{c_{s}}N_{\gamma _{s-1}}^{d_{s}}\right) ,
\notag \\
\widehat{C}_{\beta _{s-1}c_{s}}^{\alpha _{s-1}} &=&\frac{1}{2}g^{\alpha
_{s-1}\tau _{s-1}}e_{c_{s}}g_{\beta _{s-1}\tau _{s-1}},\ \widehat{C}%
_{b_{s}c_{s}}^{a_{s}} = \frac{1}{2}h^{a_{s}d_{s}}\left(
e_{c_{s}}h_{b_{s}d_{s}}+e_{c_{s}}h_{c_{s}d_{s}}-e_{d_{s}}h_{b_{s}c_{s}}%
\right) .  \notag
\end{eqnarray}
The torsion\ d--tensor (\ref{dt}) of $\ ^{s}\widehat{\mathbf{D}}$ is
completely defined by $\ ^{s}\mathbf{g}$ (\ref{fansatz}) for any chosen $\
^{s}\mathbf{N=\{}N_{i_{s}}^{a_{s}}\}$ if the above coefficients (\ref%
{candcon}) are introduced "shell by shell" into formulas
\begin{eqnarray}
\widehat{T}_{\ jk}^{i} &=&\widehat{L}_{jk}^{i}-\widehat{L}_{kj}^{i},\widehat{%
T}_{\ ja}^{i}=\widehat{C}_{jb}^{i},\widehat{T}_{\ ji}^{a}=-\Omega _{\
ji}^{a},\ \widehat{T}_{aj}^{c} = \widehat{L}_{aj}^{c}-e_{a}(N_{j}^{c}),%
\widehat{T}_{\ bc}^{a}=\ \widehat{C}_{bc}^{a}-\ \widehat{C}_{cb}^{a},  \notag
\\
&&....  \label{dtors} \\
\widehat{T}_{\ \beta _{s}\gamma _{s}}^{\alpha _{s}} &=&\widehat{L}_{\ \beta
_{s}\gamma _{s}}^{\alpha _{s}}-\widehat{L}_{\ \gamma _{s}\beta _{s}}^{\alpha
_{s}},\widehat{T}_{\ \beta _{s}b_{s}}^{\alpha _{s}}=\widehat{C}_{\ \beta
_{s}b_{s}}^{\alpha _{s}},\widehat{T}_{\ \beta _{s}\gamma
_{s}}^{a_{s}}=\Omega _{\ \gamma _{s}\beta _{s}}^{a_{s}}.  \notag
\end{eqnarray}%
The N-adapted formulas (\ref{candcon}) and (\ref{dtors}) show that any
coefficient for such objects computed in 4-d can be similarly extended shell
by shell by any value $s=1,2,....$ redefining correspondingly the h- and
v-indices. Hereafter, we shall present coordinate formulas only for $s=0,$
omitting label $s,$ i.e. with $\alpha =(i,a),$ or for some arbitrary
coefficients $\alpha _{s}=(i_{s},a_{s})$ if that will not result in
ambiguities.

Because both linear connections $\ ^{s}\nabla $ and $\ ^{s}\widehat{\mathbf{D%
}}$ are defined by the same metric structure, we can compute a canonical
distortion relation
\begin{equation}
\ ^{s}\nabla =\ ^{s}\widehat{\mathbf{D}}+\ ^{s}\widehat{\mathbf{Z}},
\label{distorsrel}
\end{equation}%
where the distorting tensor $\ ^{s}\widehat{\mathbf{Z}}=\{\widehat{\mathbf{\
Z}}_{\ \beta _{s}\gamma _{s}}^{\alpha _{s}}\}$ is uniquely defined by the
same metric $\ ^{s}\mathbf{g}$ (\ref{fansatz}). The values $\widehat{\mathbf{%
\ Z}}_{\ \beta _{s}\gamma _{s}}^{\alpha _{s}}$ are algebraic combinations of
$\widehat{T}_{\ \beta _{s}\gamma _{s}}^{\alpha _{s}}$ and vanish for zero
torsion. For instance, the GR theory in 4-d can be formulated equivalently
using the connection $\nabla $ and/or $\widehat{\mathbf{D}}$ if the
distorting relation (\ref{distorsrel}) is used \cite{vpars,vex2}. The
nonholonomic variables $(\ ^{s}\mathbf{g}$ (\ref{dm})$\mathbf{,}\ ^{s}%
\mathbf{N,}\ ^{s}\widehat{\mathbf{D}})$ are equivalent to standard ones $(\
^{s}\mathbf{g}$ (\ref{metr})$\mathbf{,}\ ^{s}\nabla ).$ Here we note that $\
^{s}\nabla $ and $\ ^{s}\widehat{\mathbf{D}}$ are not tensor objects and
such connections are subjected to different rules of coordinate transforms.
It is possible to consider frame transforms with certain $\ ^{s}\mathbf{N=\{}%
N_{i_{s}}^{a_{s}}\}$ when the conditions $\ _{\shortmid }\Gamma _{\ \alpha
_{s}\beta _{s}}^{\gamma _{s}}=\widehat{\mathbf{\Gamma }}_{\ \alpha _{s}\beta
_{s}}^{\gamma _{s}}$ are satisfied with respect to some N--adapted frames (%
\ref{nader})-- (\ref{nadifs}) even, in general, $\ ^{s}\nabla \neq \ ^{s}%
\widehat{\mathbf{D}} $ and the corresponding curvature tensors $\
_{\shortmid }R_{\ \beta _{s}\gamma _{s}\delta _{s}}^{\alpha _{s}}\neq
\widehat{\mathbf{R}}_{\ \beta _{s}\gamma _{s}\delta _{s}}^{\alpha _{s}}.$

\subsection{ The Einstein equations in N--adapted variables}

An important motivation to use the linear connection$\ ^{s}\widehat{\mathbf{D%
}}$ is that the Einstein equations written in variables $(\ ^{s}\mathbf{g}$ $%
\mathbf{,}\ ^{s}\mathbf{N,}\ ^{s}\widehat{\mathbf{D}})$ decouple with
respect to N--adapted frames of reference which gives us the possibility to
construct very general classes of solutions, see proofs and examples in \cite%
{vex1,vpars,vex2,vex3,vp,vt,vsingl1,vgrg}. We cannot "see" a general
decoupling property for such nonlinear systems of PDE if we work from the
very beginning with $\ ^{s}\nabla ,$ for instance, in coordinate frames or
with respect to arbitrary nonholonomic ones: The condition of zero torsion, $%
\ \ _{\shortmid }\mathcal{T}^{\alpha _{s}}=0$ states "strong coupling"
conditions between various tensor coefficients in the Einstein equations and
does not allow to decouple the equations.\footnote{%
The condition of decoupling a system of equations to contain, for instance,
only partial derivatives on a coordinate, is different from that of
separation of variables for a function.}

The main idea of the "anholonomic frame deformation method", AFDM, is to use
the data $(\ ^{s}\mathbf{g}$ $\mathbf{,}\ ^{s}\mathbf{N,}\ ^{s}\widehat{%
\mathbf{D}})$ in order to decouple certain gravitational and matter field
equations, then to solve them in very general off-diagonal form, with
possible dependence on all coordinates, and generate exact solutions with
nontrivial nonholonomically induced torsion. Such integral varieties of
solutions depend on a number of arbitrary generating and integration
functions and possible symmetry parameters. This geometric approach can be
applied for constructing exact solutions in various modified gravity
theories with nonlinear effective Lagrangians and nontrivial torsion.
Nevertheless, we can extract "integral subvarieties" of solutions in GR if
at the end (after a class of "generalized" solutions was constructed) we
impose, additionally, the condition of zero torsion (\ref{dtors}). This
constrains the set of admissible generating/integration functions but also
results in generic off--diagonal solutions depending on all coordinates. We can impose certain symmetry/ asymptotic / boundary / Cauchy
conditions in order to determine certain geometrically/physically important
off--diagonal configurations. Following additional assumptions, this can be
related to small parametric off--diagonal, solitonic or other type,
deformations of well known solutions in GR. The goal of this work is to
study possible nonholonomic transformations of the Kerr and several wormhole
metrics into off--diagonal (4-d or higher dimension) exact solutions.

The Ricci d--tensor $Ric=\{\mathbf{R}_{\alpha _{s}\beta _{s}}:=\mathbf{R}_{\
\alpha _{s}\beta _{s}\tau _{s}}^{\tau _{s}}\}$ of a d--connection $\ ^{s}%
\mathbf{D}$ is introduced via a respective contracting of coefficients of
the curvature tensor (\ref{dc}). The explicit formulas for h--/
v--components,
\begin{equation}
\mathbf{R}_{\alpha _{s}\beta _{s}}=\{R_{i_{s}j_{s}}:=R_{\
i_{s}j_{s}k_{s}}^{k_{s}},\ \ R_{i_{1}a_{1}}:=-R_{\
i_{1}k_{1}a_{1}}^{k_{1}},...,\ R_{a_{s}i_{s}}:=R_{\
a_{s}i_{s}b_{s}}^{b_{s}}\},  \label{dricci}
\end{equation}%
are direct recurrent $s$--modifications of those derived in Refs. \cite%
{vex1,vpars,vex2,vex3} (we do not repeat such details in this article).
Contracting such values with the inverse d--metric, with coefficients
computed for the inverse matrix of $\ ^{s}\mathbf{g}$ (\ref{dm}), we define
and compute the scalar curvature of $\ ^{s}\mathbf{D,}$
\begin{eqnarray}
\ ^{s}R &:=&\mathbf{g}^{\alpha _{s}\beta _{s}}\mathbf{R}_{\alpha _{s}\beta
_{s}}=g^{i_{s}j_{s}}R_{i_{s}j_{s}}+h^{a_{s}b_{s}}R_{a_{s}b_{s}}  \notag \\
&=&R+S+\ ^{1}S+...+\ ^{s}S,  \label{rdsc}
\end{eqnarray}%
with respective h-- and v--components of scalar curvature, $R=g^{ij}R_{ij},$
$S=h^{ab}R_{ab},$ $\ ^{1}S=h^{a_{1}b_{1}}R_{a_{1}b_{1}},...,\
^{s}S=h^{a_{s}b_{s}}R_{a_{s}b_{s}}.$

The Einstein d--tensor $\ ^{s}\mathcal{E}=\{\mathbf{E}_{\alpha _{s}\beta
_{s}}\}$ for any data $(\ ^{s}\mathbf{g}$ $\mathbf{,}\ ^{s}\mathbf{N,}\ ^{s}%
\mathbf{D})$ can be defined in standard form,%
\begin{equation}
\mathbf{E}_{\alpha _{s}\beta _{s}}:=\mathbf{R}_{\alpha _{s}\beta _{s}}-\frac{%
1}{2}\mathbf{g}_{\alpha _{s}\beta _{s}}\ ^{s}R.  \label{einstdt}
\end{equation}%
It should be noted that $\ ^{s}\mathbf{D(}\ ^{s}\mathbf{\mathcal{E})}\neq 0\
$ and the d--tensor $\mathbf{R}_{\alpha _{s}\beta _{s}}$ is not symmetric
for a general $\ ^{s}\mathbf{D.}$ Nevertheless, we can always compute, for
instance, $\ ^{s}\widehat{\mathbf{D}}\mathbf{(}\ ^{s}\widehat{\mathbf{%
\mathcal{E}}}\mathbf{)}$ as a unique distortion relation determined by (\ref%
{distorsrel}). This is a consequence of nonholonomic splitting structure (%
\ref{whitney}). It is similar to nonholonomic mechanics when the
conservation laws became more sophisticate when we impose certain
non-integrable constraints on the dynamical equations.

The Einstein equations for a metric $\mathbf{g}_{\beta _{s}\gamma _{s}}$ can
be postulated in standard form using the LC--connection $\ ^{s}\nabla $
(with corresponding Ricci tensor, $\ _{\shortmid }R_{\alpha _{s}\beta _{s}},$
curvature scalar, $\ _{\shortmid }^{s}R,$ and Einstein tensor, $\
_{\shortmid }E_{\alpha _{s}\beta _{s}}),$%
\begin{equation}
\ _{\shortmid }E_{\alpha _{s}\beta _{s}}:=\ _{\shortmid }R_{\alpha _{s}\beta
_{s}}-\frac{1}{2}g_{\alpha _{s}\beta _{s}}\ _{\shortmid }^{s}R=\varkappa \
_{\shortmid }T_{\alpha _{s}\beta _{s}},  \label{einsteq}
\end{equation}%
where $\varkappa $ is the gravitational constant and $\ _{\shortmid
}T_{\alpha _{s}\beta _{s}}$ is the stress--energy tensor for matter fields.
In 4-d, there are well-defined geometric/variational and physically
motivated procedures of constructing $\ _{\shortmid }T_{\alpha _{s}\beta
_{s}}.$ Such values can be similarly (at least geometrically) re--defined
with respect to N--adapted frames using distorting relations (\ref%
{distorsrel}) and introducing extra--dimensions.\footnote{%
We do not need additional field equations for torsion fields like in
Einstein--Cartan, gauge or string gravity theories.}

The gravitational field equations (\ref{einsteq}) can be rewritten
equivalently in N--adapted form for the canonical d--connection $\ ^{s}%
\widehat{\mathbf{D}},$%
\begin{eqnarray}
&& \ ^{s} \widehat{\mathbf{R}}_{\ \beta _{s}\delta _{s}}-\frac{1}{2}\mathbf{g%
}_{\beta _{s}\delta _{s}}\ ^{s}R=\mathbf{\Upsilon }_{\beta _{s}\delta _{s}},
\label{cdeinst} \\
&&\widehat{L}_{a_{s}j_{s}}^{c_{s}}=e_{a_{s}}(N_{j_{s}}^{c_{s}}),\ \widehat{C}%
_{j_{s}b_{s}}^{i_{s}}=0,\ \Omega _{\ j_{s}i_{s}}^{a_{s}}=0,  \label{lcconstr}
\end{eqnarray}%
where the sources $\mathbf{\Upsilon }_{\beta _{s}\delta _{s}}$ are formally
defined in GR but for extra dimensions when $\mathbf{\Upsilon }_{\beta
_{s}\delta _{s}}\rightarrow \varkappa T_{\beta _{s}\delta _{s}}$ \ for $\
^{s}\widehat{\mathbf{D}}\rightarrow \ ^{s}\nabla .$ The solutions of \ (\ref%
{cdeinst}) are found with nonholonomically induced torsion (\ref{dt}). If the
conditions (\ref{lcconstr}) are satisfied, the d-torsion coefficients (\ref%
{dtors}) are zero and we get the LC--connection, i.e. it is possible to
"extract" solutions of the standard Einstein equations. The decoupling
property can be proved in explicit form working with $\ ^{s}\widehat{\mathbf{%
D}}$ and nonholonomic torsion configurations. Having constructed certain
classes of solutions in explicit form, with nonholonomically induced
torsions and depending on various sets of integration and generating
functions and parameters, we can "extract" solutions for $\ ^{s}\nabla $
imposing at the end additional constraints resulting in zero torsion.

\subsection{Nonholonomic massive f(R,T) gravity and extra dimensions}

We shall consider modified gravity theories constructed on dimension shells
derived for the action%
\begin{equation}
S=\frac{1}{16\pi }\int \delta ^{4+2s}u\sqrt{|\mathbf{g}_{\alpha _{s}\beta
_{s}}|}[f(\ ^{s}R,\ ^{s}T)-\frac{\mu _{g}^{2}}{4}\mathcal{U}(\mathbf{g}_{\mu
_{s}\nu _{s}},\mathbf{K}_{\alpha _{s}\beta _{s}})+\ ^{m}L].  \label{act}
\end{equation}%
This generalizes in nonholonomic variables the modified $f(R,T)$ gravity,
see reviews in \cite{odints1,odints2,capoz,odints3}, and the ghost--free
massive gravity (by de Rham, Gabadadze and Tolley, dRGT) \cite%
{drg1,drg2,drg3}. Nontrivial mass terms allow us to solve certain problems
of the bimetric theory by Hassan and Rosen, \cite{hr1,hr2}, with connections to
various recent research in black hole physics and modern cosmology \cite%
{nieu,koyam}, and allows us to model solutions of (\ref{act}) in various
theories with generalized Finsler branes, stochastic processes, Clifford and
phase variables, fractional derivatives etc, see details in Refs. \cite%
{vacarfinslcosm,vfracrf,vbranef,stavr,mavr,castro,calcagni}. \ For instance,
$y^{a_{s}}$--coordinates can be treated as "velocity/momentum" variables, to
model stochastic and fractional processes, or to be considered as "standard"
extra dimensional ones. In this paper, we shall use the units  $\hbar
=c=1$ and the Planck mass $M_{Pl}$ is defined $M_{Pl}^{2}=1/8\pi G$ via
4--d Newton constant $G$ and similar units will be considered for higher
dimensions. We write $\delta ^{4+2s}u$ instead of $d^{4+2s}u$ because N--elongated differentials are used (\ref{nader}) and consider the constant $%
\mu _{g}$ as the mass parameter for gravity (for simplicity, massive gravity
theories will be studied for 4--d spacetimes). The geometric and physical
meaning of the values contained in this formula will be explained below.

The Lagrangian density $\ ^{m}L$ in action (\ref{act}) \ is used for
computing the stress--energy tensor of matter. On nonholonomic
manifolds/bundles such variations can be considered in N--adapted form,
using operators (\ref{nader}) and (\ref{nadif}), on inverse metric d--tensor
(\ref{dm}). For all shells, \ we can compute $\mathbf{T}_{\alpha _{s}\beta
_{s}}=-\frac{2}{\sqrt{|\mathbf{g}_{\mu _{s}\nu _{s}}|}}\frac{\delta (\sqrt{|%
\mathbf{g}_{\mu _{s}\nu _{s}}|}\ ^{m}L)}{\delta \mathbf{g}^{\alpha _{s}\beta
_{s}}}$, when the trace is (by definition) $\ ^{s}T:=\mathbf{g}^{\alpha
_{s}\beta _{s}}\mathbf{T}_{\alpha _{s}\beta _{s}}.$ The functional $f(\
^{s}R,\ ^{s}T)$ modifies the standard Einstein--Hilbert Lagrangian (with a
scalar curvature $R$ usually taken for the Levi--Civita connection $\nabla )$
to that for the modified $f$--gravity in various dimensions but with
dependence on $\ ^{s}R$ and $T.$ For various applications in modern
cosmology, we can assume that
\begin{equation}
\mathbf{T}_{\alpha _{s}\beta _{s}}=(\rho +p)\mathbf{v}_{\alpha _{s}}\mathbf{v%
}_{\beta _{s}}-p\mathbf{g}_{\alpha _{s}\beta _{s}},  \label{emt}
\end{equation}%
for the approximation of perfect fluid matter with the energy density $\rho $
and the pressure $p$. The four--velocity $\mathbf{v}_{\alpha _{s}}$ is
subjected to the conditions $\mathbf{v}_{\alpha _{s}}\mathbf{v}^{\alpha
_{s}}=1$ and $\mathbf{v}^{\alpha _{s}}\widehat{\mathbf{D}}_{\beta _{s}}%
\mathbf{v}_{\alpha _{s}}=0,$ for $\ ^{m}L=-p$ in a corresponding local
N--adapted frame. For simplicity, we can parameterize
\begin{equation}
f(\ ^{s}R,\ ^{s}T)=\ ^{1}f(\ ^{s}R)+\ ^{2}f(\ ^{s}T)  \label{functs}
\end{equation}%
and denote by $\ ^{1}F(\ ^{s}R):=\partial \ ^{1}f(\ ^{s}R)/\partial \ ^{s}R$
and $\ ^{2}F(\ ^{s}T):=\partial \ ^{2}f(\ ^{s}T)/\partial \ ^{s}T.$

A mass term with "gravitational mass" $\mu _{g}$ and potential {\small
\begin{eqnarray}
&& \mathcal{U}/4 =-12+6[\sqrt{\mathcal{S}}]\mathcal{+[S}]\mathcal{-[}\sqrt{%
\mathcal{S}}]^{2}+ \alpha _{3}\{18[\sqrt{\mathcal{S}}]-6[\sqrt{\mathcal{S}}%
]^{2}+[\sqrt{\mathcal{S}}]^{3}+2\mathcal{[S}^{3/2}]-3\mathcal{[S}]([\sqrt{%
\mathcal{S}}]-2)-24\}+  \label{potent} \\
&&\alpha _{4}\{[\sqrt{\mathcal{S}}](24-12\mathcal{[}\sqrt{\mathcal{S}}]-%
\mathcal{[}\sqrt{\mathcal{S}}]^{3})-12[\sqrt{\mathcal{S}}]\mathcal{[S}]+2%
\mathcal{[}\sqrt{\mathcal{S}}]^{2}(3\mathcal{[S}]+2\mathcal{[}\sqrt{\mathcal{%
S}}])+ 3\mathcal{[S}](4-\mathcal{[S}])-8\mathcal{[S}^{3/2}](\sqrt{\mathcal{S}%
}-1)+6\mathcal{[S}^{2}]-24\},  \notag
\end{eqnarray}%
} is considered in (\ref{act}) in addition to the usual $f$--gravity term
(in particular, to the Einstein--Hilbert one). The trace of a shell extended
matrix $\mathcal{S}=(S_{\mu _{s}\nu _{s}})$ is denoted by $\mathcal{[S}%
]:=S_{\ \nu _{s}}^{\nu _{s}}.$ We understand the square root of such a
matrix, $\sqrt{\mathcal{S}}=(\sqrt{\mathcal{S}}_{\ \mu _{s}}^{\nu _{s}}),$
to be a matrix for which $\sqrt{\mathcal{S}}_{\ \alpha _{s}}^{\nu _{s}}\sqrt{%
\mathcal{S}}_{\ \mu _{s}}^{\alpha _{s}}=S_{\ \mu _{s}}^{\nu _{s}}$ and $%
\alpha _{3}$ and $\alpha _{4}$ are free parameters. We use such constants
which transform $\mathcal{U}$ into standard 4--d one for $s=0.$ In works
\cite{drg2,drg3}, see additional arguments in \cite{gratia}), such a
nonlinearly extended Fierz--Pauli type potential was shown to result in a
theory of massive gravity which is seem to be free from ghost--like degrees
of freedom (it takes a special form of total derivative in absence of
dynamics). We emphasize that the potential generating matrix $\mathcal{S}$
is constructed in a special form which results in a d--tensor with shell
decomposition , $\mathbf{K}_{\ \mu _{s}}^{\nu _{s}}=\delta _{\ \mu
_{s}}^{\nu _{s}}-\sqrt{\mathcal{S}}_{\ \mu _{s}}^{\nu _{s}},$ characterizing
metric fluctuations away from a fiducial (flat) 4--d spacetime and possible
extra dimensions, or velocity/momentum type variables.

In 4-d, the coefficients
\begin{equation}
\mathbf{S}_{\ \mu }^{\nu }=\mathbf{g}^{\nu \alpha }\eta _{\overline{\nu }%
\overline{\mu }}\mathbf{e}_{\alpha }s^{\overline{\nu }}\mathbf{e}_{\mu }s^{%
\overline{\mu }},  \label{stuk}
\end{equation}%
with the Minkowski metric $\eta _{\overline{\nu }\overline{\mu }%
}=diag(1,1,1,-1),$ are generated by introducing four scalar St\"{u}kelberg
fields $s^{\overline{\nu }},$ which is necessary for restoring the
diffeomorphism invariance. Using N--adapted shell extended values $\mathbf{g}%
^{\nu _{s}\alpha _{s}}$ and $\mathbf{e}_{\alpha _{s}}$ we can always
transform a tensor $S_{\mu \nu }$ into shell distinguished d--tensor $%
\mathbf{S}_{\mu _{s}\nu _{s}}$ characterizing nonholonomically constrained
fluctuations. This is possible for the values $\mathbf{K}_{\ \mu _{s}}^{\nu
_{s}},\mathbf{S}_{\ \mu _{s}}^{\nu _{s}},\sqrt{\mathcal{S}}_{\ \mu
_{s}}^{\nu _{s}}$ etc even shell extended $s^{\overline{\nu _{s}}}$
transforms as scalar fields under coordinate and frame transforms.

For simplicity, we can consider 4--d variations of the action (\ref{act}) in
N--adapted from for the coefficients of d--metric $\mathbf{g}_{\nu \alpha }$
(\ref{dm}). The corresponding generalized/ effective Einstein equations, for
the $f$--modified massive gravity are
\begin{equation}
\widehat{\mathbf{E}}_{\alpha \beta }=\mathbf{\Upsilon }_{\beta \delta },
\label{efcdeq}
\end{equation}%
where the source encodes three terms of different nature,
\begin{equation}
\mathbf{\Upsilon }_{\beta \delta }=\ ^{ef}\eta \ G\ \mathbf{T}_{\beta \delta
}+\ ^{ef}\mathbf{T}_{\beta \delta }+\mu _{g}^{2}\ ^{K}\mathbf{T}_{\beta
\delta }.  \label{effectsource}
\end{equation}%
The first component is determined by usual matter fields with energy
momentum $\mathbf{T}_{\beta \delta }$ tensor but with effective polarization
of the gravitational constant $\ ^{ef}\eta =[1+\ ^{2}F/8\pi ]/\ ^{1}F.$ The
second term is for the $f$--modifications of the energy--momentum tensor,
\begin{equation}
\ ^{ef}\mathbf{T}_{\beta \delta }=[\frac{1}{2}(\ ^{1}f-\ ^{1}F\ \widehat{R}%
+2p\ ^{2}F+\ ^{2}f)\mathbf{g}_{\beta \delta }-(\mathbf{g}_{\beta \delta }\
\widehat{\mathbf{D}}_{\alpha }\widehat{\mathbf{D}}^{\alpha }-\widehat{%
\mathbf{D}}_{\beta }\widehat{\mathbf{D}}_{\delta })\ ^{1}F]/\ ^{1}F.
\label{efm}
\end{equation}
The mass gravity contribution, i.e. the third term in source is computed as
a dimensionless effective stress--energy tensor
\begin{eqnarray*}
\ ^{K}\mathbf{T}_{\alpha \beta }&:=&\frac{1}{4\sqrt{|\mathbf{g}_{\mu \nu }|}}%
\frac{\delta (\sqrt{|\mathbf{g}_{\mu \nu }|}\ \mathcal{U})}{\delta \mathbf{g}%
^{\alpha \beta }} \\
&=&-\frac{1}{12}\{\ \mathcal{U}\mathbf{g}_{\alpha \beta }/4-2\mathbf{S}%
_{\alpha \beta }+2([\sqrt{\mathcal{S}}]-3)\sqrt{\mathcal{S}}_{\alpha \beta }+
\\
&&\alpha _{3}[3(-6+4\mathcal{[}\sqrt{\mathcal{S}}]+\mathcal{[}\sqrt{\mathcal{%
S}}]^{2}-\mathcal{[S}])\sqrt{\mathcal{S}}_{\alpha \beta }+6(\mathcal{[}\sqrt{%
\mathcal{S}}]-2)\mathbf{S}_{\alpha \beta }-\mathcal{S}_{\alpha \beta
}^{3/2}]- \\
&&\alpha _{4}[24\left( \mathcal{S}_{\alpha \beta }^{2}-([\sqrt{\mathcal{S}}%
]-1)\mathcal{S}_{\alpha \beta }^{3/2}\right) ]+12(2-2[\sqrt{\mathcal{S}}]-%
\mathcal{[S}]+[\sqrt{\mathcal{S}}]^{2})\mathbf{S}_{\alpha \beta }+ \\
&&(24-24[\sqrt{\mathcal{S}}]+12[\sqrt{\mathcal{S}}]^{2}-[\sqrt{\mathcal{S}}%
]^{3}-12[\mathcal{S}]+12[\mathcal{S}][\sqrt{\mathcal{S}}]-8\mathcal{[S}%
^{3/2}])\sqrt{\mathcal{S}}_{\alpha \beta }\}.
\end{eqnarray*}%
The value $\ ^{K}\mathbf{T}_{\alpha \beta }$ encodes a bi--metric
configurations when the second (fiducial) d--metric \textbf{\ }$\mathbf{f}%
_{\alpha \mu }=\eta _{\overline{\nu }\overline{\mu }}\mathbf{e}_{\alpha }s^{%
\overline{\nu }}\mathbf{e}_{\mu }s^{\overline{\mu }}$ is determined by the St%
\"{u}kelberg fields $s^{\overline{\nu }}.$ The potential $\mathcal{U}$ (\ref%
{potent}) defines interactions between $\mathbf{g}_{\mu \nu }$ and $\mathbf{f%
}_{\mu \nu }$ via $\sqrt{\mathcal{S}}_{\ \mu }^{\nu }=\sqrt{\mathbf{g}^{\nu
\mu }\mathbf{f}_{\alpha \nu }}$ and $\mathcal{S}_{\ \mu }^{\nu }:=\mathbf{g}%
^{\nu \mu }\mathbf{f}_{\alpha \nu }.$ We can construct exact solutions in
explicit form and study bi--metric gravity models with $\ ^{K}\mathbf{T}%
_{\alpha \beta }=\ \lambda (x^{k})\ \mathbf{g}_{\alpha \beta },$ which can
be generated by such configurations of $s^{\overline{\nu }}$ when $\mathbf{g}%
_{\mu \nu }=\iota ^{2}(x^{k})\mathbf{f}_{\mu \nu }$ with a possible
nontrivial conformal factor $\iota ^{2}.$ Such nonholonomic configurations
allow us to compute using (\ref{stuk}) a diagonal matrices $\mathcal{S}_{\
\mu }^{\nu }:=\iota ^{-2}\delta _{\ \mu }^{\nu }.$ We can express the
effective polarized anisotropic constant$\ $encoding the contributions of $%
s^{\overline{\nu }}$ as a functional $\lambda \lbrack \iota ^{2}(x^{k})].$

The theories with gravitational field equations (\ref{efcdeq}) are similar
to the Einstein one but for a different metric compatible linear connection,
$\widehat{\mathbf{D}},$ and with nonlinear "gravitationally polarized"
coupling in effective source $\mathbf{\Upsilon }_{\beta \delta }$ (\ref%
{effectsource}). In next sections, we shall prove that such nonlinear
systems of PDE can be integrated in general forms for any N--adapted
parameterizations
\begin{equation}
\mathbf{\Upsilon }_{~\delta }^{\beta }=diag[\mathbf{\Upsilon }_{\alpha }:%
\mathbf{\Upsilon }_{~1}^{1}=\mathbf{\Upsilon }_{~2}^{2}=\Upsilon
(x^{k},y^{3});\mathbf{\Upsilon }_{~3}^{3}=\mathbf{\Upsilon }%
_{~4}^{4}=~^{v}\Upsilon (x^{k})].  \label{source}
\end{equation}%
In particular, we can consider
\begin{equation}
\Upsilon =~^{v}\Upsilon =\Lambda =const,  \label{source1}
\end{equation}%
for an effective cosmological constant $\Lambda ,$ see details in \cite%
{vpars,vex1,vex2,vex3,veym}. It should be noted that $\widehat{\mathbf{D}}%
_{\delta }\ ^{1}F_{\mid \Upsilon =\Lambda }=0$ in (\ref{efm}) if we
prescribe a functional dependence on $\ \widehat{R}=const$ (we have to chose
necessary types of N--coefficients and respective canonical d--connection
structure). For certain general distributions of matter fields and effective
matter, we can prescribe such values for (\ref{source1}) with $\mathbf{T}%
_{\beta \delta }=\check{T}(x^{k})\mathbf{g}_{\beta \delta }$ and $\ ^{s}R=%
\widehat{\Lambda }$ in (\ref{source}), then we can write
\begin{eqnarray}
\Upsilon &=&\widetilde{\Lambda }+\widetilde{\lambda },\mbox{ for }
\widetilde{\lambda }=\mu _{g}^{2}\ \lambda (x^{k}),  \notag \\
\widetilde{\Lambda } &=&\ ^{ef}\eta \ G\ \check{T}(x^{k})+\frac{1}{2}(\
^{1}f(\widehat{\Lambda })-\widehat{\Lambda }\ ^{1}F(\widehat{\Lambda })\
+2p\ ^{2}F(\check{T})+\ ^{2}f(\check{T})),  \notag \\
\ ^{ef}\eta &=&[1+\ ^{2}F(\check{T})/8\pi ]/\ ^{1}F(\widehat{\Lambda }).
\label{source1a}
\end{eqnarray}%
In general, any term may depend on coordinates $x^{i}$ but via
re--definition of generating functions they can be transformed into certain
effective constants. Prescribing the values $\widehat{\Lambda },\check{T},\
\lambda ,p$ and functionals $\ ^{1}f$ and $\ ^{2}f,$ we describe a
nonholonomically constrained matter and effective matter fields dynamics
with respect to N--adapted frames.

All above constructions can be extended to extra shells $s=1,2,...$ via
formal re--definition of indices for higher dimension. Under very general
assumptions, the effective source can be parameterized in the form
\begin{equation}
\mathbf{\Upsilon }_{~\delta _{s}}^{\beta _{s}}=(\ ^{s}\widetilde{\Lambda }+\
^{s}\widetilde{\lambda })\mathbf{\delta }_{~\delta _{s}}^{\beta _{s}}.
\label{source1b}
\end{equation}
This formal diagonal form is fixed with respect to N--adapted frames and
(see next section) for corresponding re--definition of certain generation
functions. Such $(\ ^{s}\widetilde{\Lambda }+\ ^{s}\widetilde{\lambda })$%
--terms encode via nonholonomic constraints and the canonical d--connection $%
\ ^{s}\widehat{\mathbf{D}}$ various physically important information on
modifications of the GR theory by modifications in $f$--functional and/or
massive gravity theories of various dimensions. LC--configurations can be
extracted in all such types of theories by imposing additional constraints
when $\widehat{\mathbf{D}}_{\mathcal{T}=0}\rightarrow \nabla $.

\section{Decoupling \& Integration of (Modified) Einstein Equations}

\label{s3}In this section, we show how the gravitational field equations (%
\ref{cdeinst}) with possible constraints (\ref{lcconstr}), or (\ref{einsteq}%
), can be formally integrated in very general forms for generic
off--diagonal metrics with coefficients depending on all spacetime
coordinates.

\subsection{Off--diagonal configurations with one Killing symmetries}

In the simplest form, the decoupling property can be proven for certain
ansatz with at least one Killing symmetry.

\subsubsection{ Ansatz for metrics, N--connections, and gravitational
polarizations}

Let us consider metrics of type (\ref{dm}) which via frame transformationss (\ref%
{metrtransf}) (for N--adapted transforms, $\mathbf{g}_{\alpha _{s}\beta
_{s}}=e_{\ \alpha _{s}}^{\alpha _{s}^{\prime }}e_{\ \beta _{s}}^{\beta
_{s}^{\prime }}\mathbf{g}_{\alpha _{s}^{\prime }\beta _{s}^{\prime }}$) can
be parameterized in the form\footnote{%
in our former works, we used a quite different system of notation}
\begin{eqnarray}
\ \ _{K}^{s}\mathbf{g} &=&\ g_{i}(x^{k})dx^{i}\otimes
dx^{i}+h_{a}(x^{k},y^{4})\mathbf{e}^{a}\otimes \mathbf{e}^{b}+  \label{ansk}
\\
&&h_{a_{1}}(u^{\alpha },y^{6})\ \mathbf{e}^{a_{1}}\otimes \mathbf{e}%
^{a_{1}}+h_{a_{2}}(u^{\alpha _{1}},y^{8})\ \mathbf{e}^{a_{2}}\otimes \mathbf{%
e}^{b_{2}} +....+\ h_{a_{s}}(\ u^{\alpha _{s-1}},y^{a_{s}})\mathbf{e}%
^{a_{s}}\otimes \mathbf{e}^{a_{s}},  \notag
\end{eqnarray}%
where
\begin{eqnarray*}
\mathbf{e}^{a} &=&dy^{a}+N_{i}^{a}dx^{i},\mbox{\ for \ }%
N_{i}^{3}=n_{i}(x^{k},y^{4}),N_{i}^{4}=w_{i}(x^{k},y^{4}); \\
\mathbf{e}^{a_{1}} &=&dy^{a_{1}}+N_{\alpha }^{a_{1}}du^{\alpha },%
\mbox{\ for
\ }N_{\alpha }^{5}=\ ^{1}n_{\alpha }(u^{\beta },y^{6}),N_{\alpha }^{6}=\
^{1}w_{\alpha }(u^{\beta },y^{6}); \\
\mathbf{e}^{a_{2}} &=&dy^{a_{2}}+N_{\alpha _{1}}^{a_{2}}du^{\alpha _{1}},%
\mbox{\ for \ }N_{\alpha _{1}}^{7}=\ ^{2}n_{\alpha _{1}}(u^{\beta
_{1}},y^{8}),N_{\alpha _{1}}^{8}=\ ^{2}w_{\alpha }(u^{\beta _{1}},y^{8}); \\
&&.... \\
\mathbf{e}^{a_{s}} &=&dy^{a_{s}}+N_{\alpha _{s-1}}^{a_{s}}du^{\alpha _{s-1}},%
\mbox{\ for \ } N_{\alpha _{s-1}}^{4+2s-1}=\ ^{s}n_{\alpha _{1}}(u^{\beta
_{s-1}},y^{4+2s}), N_{\alpha _{1}}^{4+2s}=\ ^{s}w_{\alpha }(u^{\beta
_{s-1}},y^{4+2s}).
\end{eqnarray*}%
Such ansatz contains a Killing vector $\partial /\partial y^{s-1}$ because
the coordinate $y^{s-1}$ is not contained in the coefficients of such
metrics. With respect to coordinate frames, for instance, in $\dim \ ^{s}%
\mathbf{V}=6;\ s=1,u^{\alpha _{1}}=(x^{1},x^{2},y^{3},y^{4},y^{5},y^{6}),$
the metrics (\ref{ansk}) are written in a form similar to that in Figure \ref%
{fig1}.

{\small 
\begin{sidewaysfigure}
\centering
{\scriptsize
\begin{eqnarray*}
&& g_{\alpha _1 \beta _1}=\\
&& \\
&&
\left[
\begin{array}{cccccc}
\begin{array}{c}
g_{1}+(n_{1}^{\ })^{2}h_{3}
+(w_{1}^{\ })^{2}h_{4} \\
+(\ ^{1}n_{1}^{\ })^{2}h_{5}+
(\ ^{1}w_{1}^{\ })^{2}h_{6}
\end{array}
&
\begin{array}{c}
n_{1}n_{2}h_{3}+
w_{1}w_{2}h_{4}+ \\
\ ^{1}n_{1}^{\ }\ ^{1}n_{2}^{\ }h_{5}+
\ ^{1}w_{1}^{\ }\ ^{1}w_{2}^{\ }h_{6}%
\end{array}
&
\begin{array}{c}
n_1 h_3+\\ \ ^{1}n_{1}^{\ }\ ^{1}n_{3}^{\ }h_{5}+
\ ^{1}w_{1}^{\ }\ ^{1}w_{3}^{\ }h_{6}%
\end{array}
&
\begin{array}{c}
w_1 h_4+\\ \ ^{1}n_{1}^{\ }\ ^{1}n_{4}^{\ }h_{5}+
\ ^{1}w_{1}^{\ }\ ^{1}w_{4}^{\ }h_{6}%
\end{array}
& \ ^{1}n_{1}^{\ }h_{5} & \ ^{1}w_{1}^{\ }h_{6} \\
&  &  &  &  &  \\
\begin{array}{c}
n_{1}n_{2}h_{3}+
w_{1}w_{2}h_{4}+ \\
\ ^{1}n_{1}^{\ }\ ^{1}n_{2}^{\ }h_{5}+
\ ^{1}w_{1}^{\ }\ ^{1}w_{2}^{\ }h_{6}%
\end{array}
&
\begin{array}{c}
g_{2}+(n_{2}^{\ })^{2}h_{3} +(w_{2}^{\ })^{2}h_{4} \\
+(\ ^{1}n_{2}^{\ })^{2}h_{5}+ (\ ^{1}w_{2}^{\ })^{2}h_{6}%
\end{array}
&
\begin{array}{c}
n_2 h_3+ \\ \ ^{1}n_{2}^{\ }\ ^{1}n_{3}^{\ }h_{5}+
\ ^{1}w_{2}^{\ }\ ^{1}w_{3}^{\ }h_{6}%
\end{array}
&
\begin{array}{c}
w_2 h_4+ \\ \ ^{1}n_{2}^{\ }\ ^{1}n_{4}^{\ }h_{5}+
\ ^{1}w_{2}^{\ }\ ^{1}w_{4}^{\ }h_{6}%
\end{array}
& \ ^{1}n_{2}^{\ }h_{5} & \ ^{1}w_{2}^{\ }h_{6} \\
&  &  &  &  &  \\
\begin{array}{c}
n_1 h_3 + \\ \ ^{1}n_{1}^{\ }\ ^{1}n_{3}^{\ }h_{5}
+\ ^{1}w_{1}^{\ }\ ^{1}w_{3}^{\ }h_{6}%
\end{array}
&
\begin{array}{c}
n_2 h_3 + \\ \ ^{1}n_{2}^{\ }\ ^{1}n_{3}^{\ }h_{5}
+\ ^{1}w_{2}^{\ }\ ^{1}w_{3}^{\ }h_{6}%
\end{array}
&
\begin{array}{c}
h_{3}+(\ ^{1}n_{3}^{\ })^{2}h_{5} +(\ ^{1}w_{3}^{\ })^{2}h_{6}%
\end{array}
& \begin{array}{c}
\ ^{1}n_{3}^{\ }\ ^{1}n_{4}^{\ }h_{5}+
\ ^{1}w_{3}^{\ }\ ^{1}w_{4}^{\ }h_{6}%
\end{array} & \ ^{1}n_{3}^{\ }h_{5} & \ ^{1}w_{3}^{\ }h_{6} \\
&  &  &  &  &  \\
\begin{array}{c}
w_1 h_4 + \\ \ ^{1}n_{1}^{\ }\ ^{1}n_{4}^{\ }h_{5}+ \ ^{1}w_{1}^{\ }\ ^{1}w_{4}^{\ }h_{6}%
\end{array}
&
\begin{array}{c}
w_2 h_4 + \\ \ ^{1}n_{2}^{\ }\ ^{1}n_{4}^{\ }h_{5}+
\ ^{1}w_{2}^{\ }\ ^{1}w_{4}^{\ }h_{6}%
\end{array}
& \begin{array}{c}
\ ^{1}n_{3}^{\ }\ ^{1}n_{4}^{\ }h_{5}+
\ ^{1}w_{3}^{\ }\ ^{1}w_{4}^{\ }h_{6}%
\end{array} &
\begin{array}{c}
h_{4}+(\ ^{1}n_{4}^{\ })^{2}h_{5} +(\ ^{1}w_{4}^{\ })^{2}h_{6}%
\end{array}
& \ ^{1}n_{4}^{\ }h_{5} & \ ^{1}w_{4}^{\ }h_{6} \\
&  &  &  &  &  \\
\ ^{1}n_{1}^{\ }h_{5} & \ ^{1}n_{2}^{\ }h_{5} & \ ^{1}n_{3}^{\ }h_{5} & \
^{1}n_{4}^{\ }h_{5} & h_{5} & 0 \\
&  &  &  &  &  \\
\ ^{1}w_{1}^{\ }h_{6} & \ ^{1}w_{2}^{\ }h_{6} & \ ^{1}w_{3}^{\ }h_{6} & \
^{1}w_{4}^{\ }h_{6} & 0 & h_{6}%
\end{array}%
\right]
 \label{odm}
\end{eqnarray*}%
}
\label{fig1}
\caption{Generic off--diagonal metrics with respect to coordinate frames in 6-d spaces}
\end{sidewaysfigure}
}


We note that nonholonomic 2+2+... parameterizations of type (\ref{fansatz})
prescribe certain algebraic symmetries of metrics both with respect to
N--adapted and/or coordinate frames. For instance, a splitting 3+3+3+ ...
may contain more complex topological configurations but to integrate the
Einstein gravitational equations in such cases is not possible for general
"non--Killing" ansatz.

In a more general context, a d--metric (\ref{ansk}) can be a result of
nonholonomic deformations of some "primary" geometric/physical data into
certain "target" data,
\begin{equation*}
\mbox{[ primary ]}(\ _{\circ }^{s}\mathbf{g,}\ _{\circ }^{s}\mathbf{N,}\
_{\circ }^{s}\widehat{\mathbf{D}})\ \rightarrow \mbox{[ target ]}(\ _{\eta
}^{s}\mathbf{g}=\ ^{s}\mathbf{\mathbf{g},}\ _{\eta }^{s}\mathbf{N}=\ ^{s}%
\mathbf{\mathbf{N},}\ _{\eta }^{s}\widehat{\mathbf{D}}=\ ^{s}\widehat{%
\mathbf{D}}).
\end{equation*}%
In this work we shall consider that the values labeled by "$\circ "$ may
define, or not, exact solutions in a gravity theory. The metrics with "$\eta
$" will be constrained always  to define a solution of
gravitational field equations (\ref{cdeinst}), or (\ref{einsteq}). For
simplicity, we shall use prime ansatz of type
\begin{eqnarray}
\ \ _{\circ }^{s}\mathbf{g} &=&\ \mathring{g}_{i}(x^{k})dx^{i}\otimes dx^{i}+%
\mathring{h}_{a}(x^{k},y^{4})\mathbf{\mathring{e}}^{a}\otimes \mathbf{%
\mathring{e}}^{b}+\epsilon _{a_{1}}\ dy^{a_{1}}\otimes \ dy^{a_{1}}+....+\
\epsilon _{a_{s}}dy^{a_{s}}\otimes \ dy^{a_{s}},  \notag \\
\mathbf{\mathring{e}}^{a} &=&dy^{a}+\mathring{N}_{i}^{a}(x^{k},y^{4})dx^{i},%
\mbox{ with }\mathring{N}_{i}^{3}=\mathring{n}_{i},\mathring{N}_{i}^{4}=%
\mathring{w}_{i},  \label{ansprime}
\end{eqnarray}%
where the constants $\epsilon _{a_{s}}$  take  values $+1$ and/or $-1$ which
depends on the signature of the higher dimensional spacetime and on $(\mathring{g}_{i},\mathring{h}%
_{a};\mathring{N}_{i}^{a}).$ Such an ansatz may define, for instance, a Kerr
black hole (or a wormhole) solution trivially imbedded into a $4+2s$
spacetime if the corresponding values of the coefficients are constructed
respectively for different type solutions of the gravitational field equations. We choose the
target metric ansatz (\ref{ansk}) as
\begin{eqnarray}
g_{\alpha _{s}} &=&\eta _{\alpha _{s}}(u^{\beta _{s}})\mathring{g}_{\alpha
_{s}};N_{i_{s}}^{a_{s}}=\ _{\eta }N_{i_{s}}^{a_{s}}(u^{\beta
_{s-1}},y^{4+2s})  \label{etad} \\
n_{i} &=&\eta _{i}^{3}\mathring{n}_{i},w_{i}=\eta _{i}^{4}\mathring{w}_{i},%
\mbox{ not summation on i};  \notag
\end{eqnarray}%
with so--called gravitational "polarization" functions and extra dimensional
N-coefficients, $\eta _{\alpha _{s}},\eta _{i}^{a}$ and$\ _{\eta
}N_{i_{s}}^{a_{s}}.$ In order to consider the limits
\begin{equation*}
(\ _{\eta }^{s}\mathbf{g,}\ _{\eta }^{s}\mathbf{N,}\ _{\eta }^{s}\widehat{%
\mathbf{D}})\rightarrow (\ _{\circ }^{s}\mathbf{g,}\ _{\circ }^{s}\mathbf{N,}%
\ _{\circ }^{s}\widehat{\mathbf{D}}),\mbox{ for }\varepsilon \rightarrow 0,
\end{equation*}%
depending on a small parameter $\varepsilon ,0\leq \varepsilon \ll 1,$ we
shall introduce "small" polarizations of type $\eta =1+\varepsilon \chi
(u...)$ and $_{\eta }N_{i_{s}}^{a_{s}}=\varepsilon n_{i_{s}}^{a_{s}}(u...).$

It should be noted that if a target d--metric (\ref{ansk}) is generated by a
nonholonomic deformation with nontrivial $\eta $- , or $\chi , $-functions,
it contains both "old" geometric/physical information on a prime metric (\ref%
{ansprime}) and additional data for a new class of exact solutions.

\subsubsection{ Ricci d--tensors and N--adapted sources}

Let us consider an ansatz (\ref{ansk}) with $\partial _{4}h_{a}\neq
0,\partial _{6}h_{a_{1}}\neq 0,...,\partial _{2s}h_{a_{s}}\neq 0,$\footnote{%
we can construct more special classes of solutions if such conditions are
not satisfied; for simplicity, we suppose that via frame transforms it is
always possible to introduce necessary type parameterizations for d--metrics}
when the partial derivatives are denoted, for instance, $\partial
_{1}h=\partial h/\partial x^{1},$ $\partial _{4}h=\partial h/\partial y^{4},$
and $\partial _{44}h=\partial ^{2}h/\partial y^{4}\partial y^{4}$. A tedious
computation of the coefficients of the canonical d--connection $\widehat{%
\mathbf{\Gamma }}_{\ \alpha _{s}\beta _{s}}^{\gamma _{s}}$(\ref{candcon})
and then of corresponding non-trivial coefficients of the Ricci d--tensor $%
\mathbf{\hat{R}}_{\alpha _{s}\beta _{s}}$ (\ref{dricci}), see similar
details in \cite{vex1,vpars,vex2,vex3}, results in such nontrivial values:
\begin{eqnarray}
\widehat{R}_{1}^{1} &=&\widehat{R}_{2}^{2}=-\frac{1}{2g_{1}g_{2}}[\partial
_{11}g_{2}-\frac{(\partial _{1}g_{1})(\partial _{1}g_{2})}{2g_{1}}-\frac{%
\left( \partial _{1}g_{2}\right) ^{2}}{2g_{2}}+\partial _{22}g_{1}-\frac{%
(\partial _{2}g_{1})(\partial _{2}g_{2})}{2g_{2}}-\frac{\left( \partial
_{2}g_{1}\right) ^{2}}{2g_{1}}],  \label{equ1} \\
\widehat{R}_{3}^{3} &=&\widehat{R}_{4}^{4}=-\frac{1}{2h_{3}h_{4}}[\partial
_{44}h_{3}-\frac{\left( \partial _{4}h_{3}\right) ^{2}}{2h_{3}}-\frac{%
(\partial _{4}h_{3})(\partial _{4}h_{4})}{2h_{4}}],  \label{equ2} \\
\widehat{R}_{3k} &=&\frac{h_{3}}{2h_{4}}\partial _{44}n_{k}+\left( \frac{%
h_{3}}{h_{4}}\partial _{4}h_{4}-\frac{3}{2}\partial _{4}h_{3}\right) \frac{%
\partial _{4}n_{k}}{2h_{4}},  \label{equ3} \\
\widehat{R}_{4k} &=&\frac{w_{k}}{2h_{3}}[\partial _{44}h_{3}-\frac{\left(
\partial _{4}h_{3}\right) ^{2}}{2h_{3}}-\frac{(\partial _{4}h_{3})(\partial
_{4}h_{4})}{2h_{4}}]+\frac{\partial _{4}h_{3}}{4h_{3}}(\frac{\partial
_{k}h_{3}}{h_{3}}+\frac{\partial _{k}h_{4}}{h_{4}})-\frac{\partial
_{k}(\partial _{4}h_{3})}{2h_{3}},  \label{equ4}
\end{eqnarray}%
and, on shells $s=1,2,...$,
\begin{eqnarray}
\widehat{R}_{5}^{5} &=&\widehat{R}_{6}^{6}=-\frac{1}{2h_{5}h_{6}}[\partial
_{66}h_{5}-\frac{\left( \partial _{6}h_{5}\right) ^{2}}{2h_{5}}-\frac{%
(\partial _{6}h_{5})(\partial _{6}h_{6})}{2h_{6}}],  \label{equ5} \\
\widehat{R}_{5\tau } &=&\frac{h_{5}}{2h_{6}}\partial _{66}\ ^{1}n_{\tau
}+\left( \frac{h_{5}}{h_{6}}\partial _{6}h_{6}-\frac{3}{2}\partial
_{6}h_{5}\right) \frac{\partial _{6}\ ^{1}n_{\tau }}{2h_{6}},  \label{equ6}
\\
\widehat{R}_{6\tau } &=&\frac{\ ^{1}w_{\tau }}{2h_{5}}[\partial _{66}h_{5}-%
\frac{\left( \partial _{6}h_{5}\right) ^{2}}{2h_{5}}-\frac{(\partial
_{6}h_{5})(\partial _{6}h_{6})}{2h_{6}}]+\frac{\partial _{6}h_{5}}{4h_{5}}(%
\frac{\partial _{\tau }h_{5}}{h_{5}}+\frac{\partial _{\tau }h_{6}}{h_{6}})-%
\frac{\partial _{\tau }(\partial _{6}h_{5})}{2h_{5}},  \label{equ7}
\end{eqnarray}%
when $\tau =1,2,3,4;$
\begin{eqnarray}
\widehat{R}_{7}^{7} &=&\widehat{R}_{8}^{8}=-\frac{1}{2h_{7}h_{8}}[\partial
_{88}h_{7}-\frac{\left( \partial _{8}h_{7}\right) ^{2}}{2h_{7}}-\frac{%
(\partial _{8}h_{7})(\partial _{8}h_{8})}{2h_{6}}],  \notag \\
\widehat{R}_{7\tau } &=&\frac{h_{7}}{2h_{8}}\partial _{88}\ ^{2}n_{\tau
_{1}}+\left( \frac{h_{7}}{h_{8}}\partial _{8}h_{8}-\frac{3}{2}\partial
_{8}h_{7}\right) \frac{\partial _{8}\ ^{2}n_{\tau _{1}}}{2h_{7}},  \notag \\
\widehat{R}_{8\tau _{1}} &=&\frac{\ ^{2}w_{\tau _{1}}}{2h_{7}}[\partial
_{88}h_{7}-\frac{\left( \partial _{8}h_{7}\right) ^{2}}{2h_{7}}-\frac{%
(\partial _{8}h_{7})(\partial _{8}h_{8})}{2h_{8}}]+\frac{\partial _{8}h_{7}}{%
4h_{7}}(\frac{\partial _{\tau _{1}}h_{7}}{h_{7}}+\frac{\partial _{\tau
_{1}}h_{8}}{h_{8}})-\frac{\partial _{\tau _{1}}(\partial _{8}h_{7})}{2h_{7}},
\label{equ4d}
\end{eqnarray}%
when $\tau _{1}=1,2,3,4,5,6.$ Similar formulas can be written recurrently
for arbitrary finite extra dimensions.

Using the above formulas, we can compute the Ricci scalar (\ref{rdsc}) for $\
^{s}\widehat{\mathbf{D}}$ (for simplicity, we consider $s=1),$ $\ ^{s}%
\widehat{R}=2(\widehat{R}_{1}^{1}+\widehat{R}_{3}^{3}+\widehat{R}_{5}^{5}).$
There are certain N--adapted symmetries of the Einstein d--tensor (\ref%
{einstdt}) for the ansatz (\ref{ansk}), $\widehat{E}_{1}^{1}=\widehat{E}%
_{2}^{2}=-(\widehat{R}_{3}^{3}+\widehat{R}_{5}^{5}),\widehat{E}_{3}^{3}=%
\widehat{E}_{4}^{4}=-(\widehat{R}_{1}^{1}+\widehat{R}_{5}^{5}),\widehat{E}%
_{5}^{5}=\widehat{E}_{6}^{6}=-(\widehat{R}_{1}^{1}+\widehat{R}_{3}^{3})$. In
a similar form, we find symmetries for $s=2:$%
\begin{eqnarray*}
\widehat{E}_{1}^{1} &=&\widehat{E}_{2}^{2}=-(\widehat{R}_{3}^{3}+\widehat{R}%
_{5}^{5}+\widehat{R}_{7}^{7}),\widehat{E}_{3}^{3}=\widehat{E}_{4}^{4}=-(%
\widehat{R}_{1}^{1}+\widehat{R}_{5}^{5}+\widehat{R}_{7}^{7}), \\
\widehat{E}_{5}^{5} &=&\widehat{E}_{6}^{6}=-(\widehat{R}_{1}^{1}+\widehat{R}%
_{3}^{3}+\widehat{R}_{7}^{7}),\widehat{E}_{7}^{7}=\widehat{E}_{8}^{8}=-(%
\widehat{R}_{1}^{1}+\widehat{R}_{3}^{3}+\widehat{R}_{5}^{5}).
\end{eqnarray*}

We search for solutions of the nonholonomic Einstein equations (\ref{equ1}%
)--(\ref{equ4d}) with nontrivial $\Lambda $--sources written in the form
\begin{eqnarray}
\widehat{R}_{1}^{1} &=&\widehat{R}_{2}^{2}=-\Lambda (x^{k}),\ \widehat{R}%
_{3}^{3}=\widehat{R}_{4}^{4}=-\ ^{v}\Lambda (x^{k},y^{4}),  \label{sourc1} \\
\widehat{R}_{5}^{5} &=&\widehat{R}_{6}^{6}=-\ _{1}^{v}\Lambda (u^{\beta
},y^{6}),\ \widehat{R}_{7}^{7}=\widehat{R}_{8}^{8}=-\ _{2}^{v}\Lambda
(u^{\beta _{1}},y^{8}).  \notag
\end{eqnarray}%
Similar equations can be written recurrently for arbitrary finite extra
dimensions. This constrains us to define such N--adapted frame transformations
when the sources $\mathbf{\Upsilon }_{\beta _{s}\delta _{s}}$ in (\ref%
{cdeinst}) are parameterized
\begin{eqnarray*}
\mathbf{\Upsilon }_{1}^{1} &=&\mathbf{\Upsilon }_{2}^{2}=\ ^{v}\Lambda +\
_{1}^{v}\Lambda +\ _{2}^{v}\Lambda ,\mathbf{\Upsilon }_{3}^{3}=\mathbf{%
\Upsilon }_{4}^{4}=\Lambda +\ _{1}^{v}\Lambda +\ _{2}^{v}\Lambda , \\
\mathbf{\Upsilon }_{5}^{5} &=&\mathbf{\Upsilon }_{6}^{6}=\Lambda +\
^{v}\Lambda +\ _{2}^{v}\Lambda ,\mathbf{\Upsilon }_{7}^{7}=\mathbf{\Upsilon }%
_{8}^{8}=\Lambda +\ ^{v}\Lambda +\ _{1}^{v}\Lambda .
\end{eqnarray*}%
For certain models of extra dimensional gravity, we can write $\ _{1}^{v}\Lambda
=\ _{2}^{v}\Lambda =\ ^{\circ }\Lambda =const.$ Re--defining the generating
functions (see below) for non--vacuum configurations, we can always
introduce such effective sources.

\subsubsection{Decoupling of gravitational field equations}

Introducing the ansatz (\ref{ansk}) for $\ g_{i}(x^{k})=\epsilon _{i}e^{\psi
(x^{k})}$ with nonzero $\partial _{4}\phi ,\partial _{4}h_{a},$ $\partial
_{6}\ ^{1}\phi ,\partial _{6}h_{a_{1}},\partial _{8}\ ^{2}\phi ,\partial
_{8}h_{a_{2}}$ in (\ref{equ1})--(\ref{equ4d}) with respective sources,
we obtain this system of PDEs:
\begin{equation}
\epsilon _{1}\partial _{11}\psi +\epsilon _{2}\partial _{22}\psi =2\Lambda
(x^{k}),  \label{e1}
\end{equation}%
\begin{eqnarray}
(\partial _{4}\phi )(\partial _{4}h_{3}) &=&2h_{3}h_{4}\ ^{v}\Lambda
(x^{k},y^{4}),\   \label{e2} \\
\partial _{44}n_{i}+\gamma \partial _{4}n_{i} &=&0,  \label{e3} \\
\beta w_{i}-\alpha _{i} &=&0,\   \label{e4}
\end{eqnarray}%
\begin{eqnarray}
(\partial _{6}\ ^{1}\phi )(\partial _{6}h_{5}) &=&2h_{5}h_{6}\
_{1}^{v}\Lambda (u^{\beta },y^{6}),  \label{e2aa} \\
\partial _{66}\ ^{1}n_{\tau }+\ ^{1}\gamma \partial _{6}\ ^{1}n_{\tau } &=&0,
\label{e3aa} \\
\ ^{1}\beta \ ^{1}w_{\tau }-\ ^{1}\alpha _{\tau } &=&0,\   \label{e4aa}
\end{eqnarray}%
\begin{eqnarray}
(\partial _{6}\ ^{2}\phi )(\partial _{6}h_{7}) &=&2h_{7}h_{8}\
_{2}^{v}\Lambda (u^{\beta _{1}},y^{8}),  \notag \\
\partial _{88}\ ^{2}n_{\tau _{1}}+\ ^{2}\gamma \partial _{8}\ ^{2}n_{\tau
_{1}} &=&0,  \notag \\
\ ^{2}\beta \ ^{2}w_{\tau _{1}}-\ ^{2}\alpha _{\tau _{1}} &=&0,\
\label{e4dd}
\end{eqnarray}%
\begin{equation*}
\mbox{ (similar equations can be written recurrently  for arbitrary finite
extra dimensions),}
\end{equation*}%
where the coefficients are defined respectively
\begin{eqnarray}
\phi  &=&\ln \left\vert \frac{\partial _{4}h_{3}}{\sqrt{|h_{3}h_{4}|}}%
\right\vert ,  \label{ca1} \\
&&\gamma :=\partial _{4}(\ln \frac{|h_{3}|^{3/2}}{|h_{4}|}),\ \ \alpha _{i}=%
\frac{\partial _{4}h_{3}}{2h_{3}}\partial _{i}\phi ,\ \beta =\frac{\partial
_{4}h_{3}}{2h_{3}}\partial _{4}\phi ,  \label{c1}
\end{eqnarray}%
\begin{eqnarray}
\ ^{1}\phi  &=&\ln \left\vert \frac{\partial _{6}h_{5}}{\sqrt{|h_{5}h_{6}|}}%
\right\vert ,  \label{ca2} \\
&&\ ^{1}\gamma :=\partial _{6}(\ln \frac{|h_{5}|^{3/2}}{|h_{6}|}),\
^{1}\alpha _{\tau }=\frac{\partial _{6}h_{5}}{2h_{5}}\partial _{\tau }\
^{1}\phi ,\ ^{1}\beta =\frac{\partial _{6}h_{5}}{2h_{5}}\partial _{\tau }\
^{1}\phi ,  \label{c2}
\end{eqnarray}%
\begin{eqnarray*}
\ ^{2}\phi  &=&\ln \left\vert \frac{\partial _{8}h_{7}}{\sqrt{|h_{7}h_{8}|}}%
\right\vert , \\
&&\ ^{2}\gamma :=\partial _{8}(\ln \frac{|h_{7}|^{3/2}}{|h_{8}|}),\ \
^{2}\alpha _{\tau _{1}}=\frac{\partial _{8}h_{7}}{2h_{7}}\partial _{\tau
_{1}}\ ^{2}\phi ,\ \ ^{2}\beta =\frac{\partial _{8}h_{7}}{2h_{7}}\partial
_{\tau _{1}}\ ^{2}\phi ,
\end{eqnarray*}%
and similarly for extra shells.

The equations (\ref{e1})-- (\ref{e4dd}) reflect a very important decoupling
property of the (generalized) Einstein equations with respect to the
corresponding N--adapted frames. In explicit form, such formulas can be
obtained for metrics with at least one Killing symmetry (the constructions
can be generalized for non--Killing configurations). Let us explain in brief
the decoupling property for 4--d configurations following such steps:

\begin{enumerate}
\item The equation (\ref{e1}) is just a 2-d Laplace, or d'Alambert one
(depending on prescribed signature), which can be solved for any value $%
\Lambda (x^{k}).$

\item The equation (\ref{e2}) contains only the partial derivative $\partial
_{4}$ and is related to the formula for the coefficient (\ref{ca1})
for the values $h_{3}(x^{i},y^{4}),$ $h_{4}(x^{i},y^{4})$ and $\phi
(x^{i},y^{4})$ and source $\ ^{v}\Lambda (x^{k},y^{4}).$ Prescribing any two
such functions, we can define (by integrating with respect to $y^{4})$ the other two such
functions.

\item Using $h_{3}$ and $\phi $ in the previous point, we can compute the
coefficients $\alpha _{i}$ and $\beta ,$ see (\ref{c1}), which allows us to
define $n_{i}$ from the algebraic equations (\ref{e3}).

\item Having computed the coefficient $\gamma $ (\ref{c1}), the
N--connection coefficients $w_{i}$ can be defined after two integrations with respect to  $%
y^{4}$ in (\ref{e4}).
\end{enumerate}

The procedure 2-4 can be repeated step by step on the other shells for higher
dimensions. We have to add the corresponding dependencies on the extra dimensional
coordinates and additional partial derivatives. For instance, the equation (%
\ref{e2aa}) and formula (\ref{ca2}) with partial derivative $\partial _{6}$
involves the functions $h_{5}(x^{i},y^{a},y^{6}),$ $h_{6}(x^{i},y^{a},y^{6})$ and $%
\ ^{1}\phi (x^{i},y^{a},y^{6})$ $\ $and source $\ _{1}^{v}\Lambda (u^{\beta
},y^{6}).$ We can compute any two such functions integrating with respect to  $y^{6}$ if the
two other ones are prescribed.  In a similar form, we follow the steps in points 3 and 4
with $\ ^{1}\alpha _{\tau },\ ^{1}\beta ,\ ^{1}\gamma ,$ see (\ref{c2}), and
compute the higher order N--connection coefficients $\ ^{1}n_{\tau }$ and $\
^{1}w_{\tau }.$

\subsubsection{ Integration of (modified) Einstein equations by generating
functions and effective sources}

The system of nonlinear PDEs (\ref{e1})-- (\ref{e4dd}) can be integrated in
general forms for any finite dimension $\dim \ ^{s}\mathbf{V}\geq 4.$

\paragraph{ 4--d non--vacuum configurations:}

The coefficients $g_{i}=\epsilon _{i}e^{\psi (x^{k})}$ are defined by
solutions of the corresponding Laplace/ d'Alambert equation (\ref{e1}).

We can solve (\ref{e2}) and (\ref{ca1}) for any $\partial _{4}\phi \neq
0,h_{a}\neq 0$ and $\ ^{v}\Lambda \neq 0$ if we re-write the equations as
\begin{equation}
\ h_{3}h_{4}=(\partial _{4}\phi )(\partial _{4}h_{3})/2\ ^{v}\Lambda
\mbox{
and }|h_{3}h_{4}|=(\partial _{4}h_{3})^{2}e^{-2\phi },  \label{eq4bb}
\end{equation}%
for any nontrivial source $\ ^{v}\Lambda .$ Inserting the first equation
into the second one, we find
\begin{equation}
|\partial _{4}h_{3}|=\frac{\partial _{4}(e^{-2\phi })}{4|\ ^{v}\Lambda |}=%
\frac{\partial _{4}[\Phi ^{2}]}{2|\ ^{v}\Lambda |},  \label{aux01}
\end{equation}%
for $\Phi :=e^{\phi }$. This formula can be integrated with respect to $y^{4},$ which
results in

\begin{equation*}
h_{3}[\Phi ,\ ^{v}\Lambda ]=\ ^{0}h_{3}(x^{k})+\frac{\epsilon _{3}\epsilon
_{4}}{4}\int dy^{4}\frac{\partial _{4}(\Phi ^{2})}{\ ^{v}\Lambda },
\end{equation*}%
where $\ ^{0}h_{3}=\ ^{0}h_{3}(x^{k})$ is an integration function and $%
\epsilon _{3},\epsilon _{4}=\pm 1.$ To find $h_{4}$ we can use the first
equation (\ref{eq4bb}) and write
\begin{equation}
h_{4}[\Phi ,\ ^{v}\Lambda ]=\frac{(\partial _{4}\phi )}{\ ^{v}\Lambda }%
\partial _{4}(\ln \sqrt{|h_{3}|})=\frac{1}{2\ ^{v}\Lambda }\frac{\partial
_{4}\Phi }{\Phi }\frac{\partial _{4}h_{3}}{h_{3}}.  \label{h4aux}
\end{equation}%
These formulas for $h_{a}$ can be simplified if we introduce an "effective"
cosmological constant $\widetilde{\Lambda }=const\neq 0$ and re--define the
generating function $\Phi \rightarrow \tilde{\Phi},$ for which $\frac{%
\partial _{4}[\Phi ^{2}]}{\ ^{v}\Lambda }=\frac{\partial _{4}[\tilde{\Phi}%
^{2}]}{\ \tilde{\Lambda}},$ i.e.%
\begin{equation}
\Phi ^{2}=\widetilde{\Lambda }^{-1}\int dy^{4}(\ ^{v}\Lambda )\partial _{4}(%
\tilde{\Phi}^{2})\mbox{
and }\tilde{\Phi}^{2}=\widetilde{\Lambda }\int dy^{4}(\ ^{v}\Lambda
)^{-1}\partial _{4}(\Phi ^{2}).  \label{rescgf}
\end{equation}%
Introducing the integration function$\ ^{0}h_{3}(x^{k})$ and $\epsilon _{3}$
and $\epsilon _{4}$ in $\Phi $ and, respectively, in $\ ^{v}\Lambda ,$ we can
express%
\begin{equation}
h_{3}[\tilde{\Phi},\widetilde{\Lambda }]=\frac{\tilde{\Phi}^{2}}{4\widetilde{%
\Lambda }}\mbox{ and
}h_{4}[\tilde{\Phi},\widetilde{\Lambda }]=\frac{(\partial _{4}\tilde{\Phi}%
)^{2}}{\Xi },  \label{solha}
\end{equation}%
where $\Xi =\int dy^{4}(\ ^{v}\Lambda )\partial _{4}(\tilde{\Phi}^{2}).$ We
can work for convenience with two couples of generating data, $(\Phi ,\
^{v}\Lambda )$ and $(\tilde{\Phi},\ \tilde{\Lambda}),$ related by formulas (%
\ref{rescgf}).

Using the values $h_{a}$ (\ref{solha}), we compute the coefficients $\alpha
_{i},\beta $ and $\gamma $ from (\ref{c1}). The resulting solutions for
N--coefficients can be expressed recurrently,
\begin{eqnarray}
n_{k} &=&\ _{1}n_{k}+\ _{2}n_{k}\int dy^{4}h_{4}/(\sqrt{|h_{3}|})^{3}=\
_{1}n_{k}+\ _{2}\widetilde{n}_{k}\int dy^{4}(\partial _{4}\tilde{\Phi})^{2}/%
\tilde{\Phi}^{3}\Xi ,  \notag \\
w_{i} &=&\partial _{i}\phi /\partial _{4}\phi =\partial _{i}\Phi /\partial
_{4}\Phi ,  \label{solhn}
\end{eqnarray}%
where $\ _{1}n_{k}(x^{i})$ and $\ _{2}n_{k}(x^{i}),$ or $_{2}\widetilde{n}%
_{k}(x^{i})=8\ _{2}n_{k}(x^{i})|\widetilde{\Lambda }|^{3/2},$ are
integration functions. The quadratic line elements determined by
coefficients (\ref{solha})-(\ref{solhn}) are parameterized in the form
\begin{eqnarray}
ds_{4dK}^{2} &=&g_{\alpha \beta }(x^{k},y^{4})du^{\alpha }du^{\beta
}=\epsilon _{i}e^{\psi (x^{k})}(dx^{i})^{2}+  \label{qnk4d} \\
&&\frac{\tilde{\Phi}^{2}}{4\widetilde{\Lambda }}\left[ dy^{3}+\left( \
_{1}n_{k}+_{2}\widetilde{n}_{k}\int dy^{4}\frac{(\partial _{4}\tilde{\Phi}%
)^{2}}{\tilde{\Phi}^{3}\Xi }\right) dx^{k}\right] ^{2}+\frac{(\partial _{4}%
\tilde{\Phi})^{2}}{\Xi }\ \left[ dy^{4}+\frac{\partial _{i}\Phi }{\partial
_{4}\Phi }dx^{i}\right] ^{2}.  \notag
\end{eqnarray}
This line element defines a family of generic off--diagonal solutions with
Killing symmetry in $\partial /\partial y^{3}$ of the 4--d Einstein
equations (\ref{sourc1}) for the canonical d--connection $\ \widehat{\mathbf{%
D}}$ (the label $4dK$ is for "nonholonomic 4-d Killing solutions). We can
verify by straightforward computations of the corresponding anholonomy
coefficients $W_{\alpha \beta }^{\gamma }$ in (\ref{anhrel1}) that such
values are not get zero if arbitrary generating function $\phi $ and
integration ones ($\ ^{0}h_{a},_{1}n_{k}$ and $\ _{2}n_{k})$ are considered.

\paragraph{ 4--d vacuum configurations:}

The limits to the off--diagonal solutions with $\ \Lambda =\ ^{v}\Lambda =0$ can
be not smooth because, for instance, we have multiples of $(\ ^{v}\Lambda
)^{-1} $ in the coefficients of (\ref{qnk4d}). For the ansatz (\ref{ansk}), we
can analyze solutions when the nontrivial coefficients of the Ricci
d--tensor (\ref{equ1})--(\ref{equ4d}) are zero. The first equation is a
typical example of 2--d wave, or Laplace, equation. We can express such
solutions in a similar form $g_{i}=\epsilon _{i}e^{\psi (x^{k},\Lambda
=0)}(dx^{i})^{2}.$

There are three classes of off--diagonal metrics which result in zero
coefficients (\ref{equ2})--(\ref{equ4d}).

\begin{itemize}
\item In the first case, we can impose the condition $\partial
_{4}h_{3}=0,h_{3}\neq 0,$ which results only in one nontrivial equation
(derived from (\ref{equ3})),%
\begin{equation*}
\partial _{44}n_{k}+\partial _{4}n_{k}\ \partial _{4}\ln |h_{4}|=0,
\end{equation*}%
where $h_{4}(x^{i},y^{4})\neq 0$ and $w_{k}(x^{i},y^{4})$ are arbitrary
functions. If $\partial _{4}h_{4}=0,$ we must take $\partial _{44}n_{k}=0.$
For $\partial _{4}h_{4}\neq 0,$ we get
\begin{equation}
n_{k}=\ _{1}n_{k}+\ _{2}n_{k}\int dy^{4}/h_{4}  \label{wsol}
\end{equation}%
with integration functions $\ _{1}n_{k}(x^{i})$ and $\ _{2}n_{k}(x^{i}).$
The corresponding quadratic line element is of the type {\small
\begin{equation}
ds_{v1}^{2} =\epsilon _{i}e^{\psi (x^{k},\Lambda =0)}(dx^{i})^{2}+\
^{0}h_{3}(x^{k})[dy^{3}+ (\ _{1}n_{k}(x^{i})+ \ _{2}n_{k}(x^{i})\int
dy^{4}/h_{4}) dx^{i}]^{2} +
h_{4}(x^{i},y^{4})[dy^{4}+w_{i}(x^{k},y^{4})dx^{i}]^{2}.  \label{vs1}
\end{equation}
}

\item In the second case, $\partial _{4}h_{3}\neq 0$ and $\partial
_{4}h_{4}\neq 0.$ We can solve (\ref{equ2}) and/or (\ref{e2}) in a
self--consistent form for $\ ^{v}\Lambda =0$ if $\partial _{4}\phi =0$ for
coefficients (\ref{ca1}) and (\ref{c1}). For $\phi =\phi _{0}=const,$ we can
consider arbitrary functions $w_{i}(x^{k},y^{4})$ because $\beta =\alpha
_{i}=0$ for such configurations. The condition (\ref{ca1}) is satisfied by
any
\begin{equation}
h_{4}=\ ^{0}h_{4}(x^{k})(\partial _{4}\sqrt{|h_{3}|})^{2},  \label{h34vacuum}
\end{equation}
where $\ ^{0}h_{3}(x^{k})$ is an integration function and $%
h_{3}(x^{k},y^{4}) $ is any generating function. The coefficients $n_{k}$
can be found from (\ref{equ3}), see (\ref{wsol}). Such a family of vacuum
metrics is described by
\begin{eqnarray}
ds_{v2}^{2} &=&\epsilon _{i}e^{\psi (x^{k},\Lambda
=0)}(dx^{i})^{2}+h_{3}(x^{i},y^{4})[dy^{3}+(\ _{1}n_{k}(x^{i})+\
_{2}n_{k}(x^{i}) \int dy^{4}/h_{4})dx^{i}]^{2}+  \label{vs2} \\
&& \ ^{0}h_{4}(x^{k})(\partial _{4}\sqrt{|h_{3}|}%
)^{2}[dy^{4}+w_{i}(x^{k},y^{4})dx^{i}]^{2}.  \notag
\end{eqnarray}

\item In the third case, $\partial _{4}h_{3}\neq 0$ but $\partial
_{4}h_{4}=0.$ The equation (\ref{equ2}) transforms into $\partial _{44}h_{3}-%
\frac{\left( \partial _{4}h_{3}\right) ^{2}}{2h_{3}}=0$, when the general
solution is $h_{3}(x^{k},y^{4})=\left[ c_{1}(x^{k})+c_{2}(x^{k})y^{4}\right]
^{2}$, with generating functions $c_{1}(x^{k}),c_{2}(x^{k})$, and $h_{4}=\
^{0}h_{4}(x^{k}).$ For $\phi =\phi _{0}=const,$ we can take any values $%
w_{i}(x^{k},y^{4})$ because $\beta =\alpha _{i}=0.$ The coefficients $n_{i}$
are found from (\ref{equ3}) and/or, equivalently, from (\ref{e3}) with $\gamma
=\frac{3}{2}\partial _{4}|h_{3}|.$ We obtain
\begin{equation*}
n_{i}=\ _{1}n_{i}(x^{k})+\ _{2}n_{i}(x^{k})\int dy^{4}|h_{3}|^{-3/2}=\
_{1}n_{i}(x^{k})+\ _{2}\widetilde{n}_{i}(x^{k})\left[
c_{1}(x^{k})+c_{2}(x^{k})y^{4}\right] ^{-2},
\end{equation*}%
with integration functions $\ _{1}n_{i}(x^{k})$ and $\ _{2}n_{i}(x^{k}),$ or
re--defined $\ \ _{2}\widetilde{n}_{i}=-\ _{2}n_{i}/2c_{2}.$ The quadratic
line element for this class of solutions for vacuum metrics is described by
{\small
\begin{eqnarray}
ds_{v3}^{2} &=&\epsilon _{i}e^{\psi (x^{k},\Lambda =0)}(dx^{i})^{2}+\left[
c_{1}(x^{k})+c_{2}(x^{k})y^{4}\right] ^{2}[dy^{3}+(\ _{1}n_{i}(x^{k})+\ _{2}%
\widetilde{n}_{i}(x^{k})\left[ c_{1}(x^{k})+c_{2}(x^{k})y^{4}\right]
^{-2})dx^{i}]^{2}  \notag \\
&&+\ ^{0}h_{4}(x^{k})[dy^{4}+w_{i}(x^{k},y^{4})dx^{i}]^{2}.  \label{vs3}
\end{eqnarray}%
}
\end{itemize}

Finally, we note that such solutions have nontrivial induced torsions (%
\ref{dtors}).

\paragraph{Extra dimensional non--vacuum solutions:}

The solutions for higher dimensions can be constructed in a certain fashion
which are similar to the 4--d ones using new classes of generating and
integration functions with dependencies on extra dimension coordinates. For
instance, we can generate solutions of the system (\ref{e2aa})--(\ref{e4aa})
with coefficients (\ref{ca2}) and (\ref{c2}) following a formal analogy when
$\partial _{4}\rightarrow \partial _{6},\phi (x^{k},y^{4})\rightarrow \
^{1}\phi (u^{\tau },y^{6}),\ ^{v}\Lambda (x^{k},y^{4})\rightarrow \
_{1}^{v}\Lambda (u^{\tau },y^{6})...$ and associate values $\ ^{1}\tilde{\Phi%
}(u^{\tau },y^{6})$ and $\ ^{1}\widetilde{\Lambda }$ as we considered in the
previous paragraph.

The extra--dimensional coefficients are computed%
\begin{equation*}
h_{5}[\ ^{1}\tilde{\Phi},\ ^{1}\widetilde{\Lambda }]=\frac{\ ^{1}\tilde{\Phi}%
^{2}}{4\ ^{1}\widetilde{\Lambda }}\mbox{ and
}h_{6}[\ ^{1}\tilde{\Phi}]=\frac{(\partial _{6}\ ^{1}\tilde{\Phi})^{2}}{\
^{1}\Xi },
\end{equation*}%
for $\ ^{1}\Xi =\int dy^{6}(\ _{1}^{v}\Lambda )\partial _{6}(\ ^{1}\tilde{%
\Phi}^{2})$ and, for N--coefficients,
\begin{eqnarray*}
\ ^{1}n_{\tau } &=&\ _{1}^{1}n_{\tau }+\ _{2}^{1}n_{\tau }\int dy^{6}h_{6}/(%
\sqrt{|h_{5}|})^{3}=\ _{1}^{1}n_{k}+\ _{2}^{1}\widetilde{n}_{k}\int
dy^{6}(\partial _{6}\ ^{1}\tilde{\Phi})^{2}/(\ ^{1}\tilde{\Phi})^{3}\
^{1}\Xi , \\
\ ^{1}w_{\tau } &=&\partial _{\tau }\ ^{1}\phi /\partial _{6}\ ^{1}\phi
=\partial _{\tau }\ ^{1}\Phi /\partial _{6}\ ^{1}\Phi ,
\end{eqnarray*}%
where $\ ^{0}h_{a_{1}}=\ ^{0}h_{a_{1}}(u^{\tau }),$ $\ _{1}^{1}n_{k}(u^{\tau
})$ and $\ _{2}^{1}n_{k}(u^{\tau }),$ are integration functions.

A general class of quadratic line elements in 6--d spacetimes can be
parameterized in the form {\small
\begin{equation}
ds_{6dK}^{2}=ds_{4dK}^{2}+\frac{\ ^{1}\tilde{\Phi}^{2}}{4\ ^{1}\widetilde{%
\Lambda }}\left[ dy^{5}+\left( \ _{1}^{1}n_{k}+\ _{2}^{1}\widetilde{n}%
_{k}\int dy^{6}\frac{(\partial _{6}\ ^{1}\tilde{\Phi})^{2}}{(\ ^{1}\tilde{%
\Phi})^{3}\ ^{1}\Xi }\right) du^{\tau }\right] ^{2}+\ \frac{(\partial _{6}\
^{1}\tilde{\Phi})^{2}}{\ ^{1}\Xi }\left[ dy^{6}+\frac{\partial _{\tau }\ \
^{1}\Phi }{\partial _{6}\ \ ^{1}\Phi }du^{\tau }\right] ^{2},  \label{qnk6d}
\end{equation}%
} where $ds_{4dK}^{2}$ is given by formula (\ref{qnk4d}) and $\tau =1,2,3,4.$
This quadratic line element has a Killing symmetry in $\partial _{5}$ (in
N--adapted frames, the metric does not depend on $y^{5}$).

Extending the constructions to the shell $s=2$ with $\partial
_{6}\rightarrow \partial _{8},\ ^{1}\phi (u^{\tau },y^{6})\rightarrow \
^{2}\phi (u^{\tau _{1}},y^{8}),\ _{1}^{v}\Lambda (u^{\tau
},y^{6})\rightarrow \ _{2}^{v}\Lambda (u^{\tau _{1}},y^{8})...,\ ^{2}\tilde{%
\Phi}(u^{\tau _{1}},y^{8}),\ ^{2}\widetilde{\Lambda }$, where $\tau
_{1}=1,2,...,$ $5,6,$ we generate off--diagonal solutions in 8--d gravity,
{\small
\begin{equation}
ds_{8dK}^{2}=ds_{6dK}^{2}+\frac{\ ^{2}\tilde{\Phi}^{2}}{4\ ^{2}\widetilde{%
\Lambda }}\left[ dy^{7}+\left( \ _{1}^{2}n_{k}+\ _{2}^{2}\widetilde{n}%
_{k}\int dy^{8}\frac{(\partial _{8}\ ^{2}\tilde{\Phi})^{2}}{(\ ^{2}\tilde{%
\Phi})^{3}\ ^{2}\Xi }\right) du^{\tau _{1}}\right] ^{2}+\ \frac{(\partial
_{8}\ ^{2}\tilde{\Phi})^{2}}{\ ^{2}\Xi }\left[ dy^{8}+\frac{\partial _{\tau
_{1}}\ \ ^{2}\Phi }{\partial _{8}\ \ ^{2}\Phi }du^{\tau _{1}}\right] ^{2},
\label{qnk8d}
\end{equation}%
} where $ds_{6dK}^{2}$ is given by (\ref{qnk6d}), $\ ^{2}\Xi =\int dy^{8}(\
_{2}^{v}\Lambda )\partial _{8}(\ ^{2}\tilde{\Phi}^{2}),$ and corresponding
integration/generating functions $\ ^{0}h_{a_{2}}(u^{\tau _{1}});a_{2}=7,8;\
_{1}n_{\tau _{1}}(u^{\tau _{1}})$ and $\ _{2}n_{\tau _{1}}(u^{\tau _{1}})$
are integration functions.

Using \ 2+2+... symmetries of off--diagonal parameterizations (\ref{odm}),
we can construct exact solutions for arbitrary finite dimension of extra
dimensional spacetime $\ ^{s}\mathbf{V.}$

\paragraph{ Extra dimensional vacuum solutions:}

The off--diagonal solutions (\ref{qnk4d}), (\ref{qnk6d}), (\ref{qnk8d}),...
have been constructed for nontrivial sources $\ ^{v}\Lambda (x^{k}, y^{4}),$
$\ _{1}^{v}\Lambda (u^{\tau },y^{6}),$ $\ _{2}^{v}\Lambda (u^{\tau
},y^{8}),...$ In a similar manner, we can generate vacuum configurations
with effective zero cosmological constants by extending to higher dimensions
the 4-d vacuum metrics of type $ds_{v1}^{2}$ (\ref{vs1}), $ds_{v2}^{2}$ (\ref%
{vs2}), $ds_{v3}^{2}$ (\ref{vs3}) etc. It is possible to generate solutions
when the sources for (\ref{sourc1}) are zero on some shells and nonzero for
other ones.

We provide here an example of quadratic line element for 6--d gravity
derived as a $s=1$ generalization of (\ref{vs2}). For such solutions, $%
\partial _{4}h_{a}\neq 0,\partial _{6}h_{a_{1}}\neq 0,...$ and $\phi =\phi
_{0}=const,$ $\ ^{1}\phi =\ ^{1}\phi _{0}=const,...$
\begin{eqnarray}
ds_{v2s3}^{2} &=&\epsilon _{i}e^{\psi (x^{k},\Lambda
=0)}(dx^{i})^{2}+h_{3}(x^{i},y^{4})[dy^{3}+\left( \ _{1}n_{k}(x^{i})+\
_{2}n_{k}(x^{i})\int dy^{4}/h_{4}\right) dx^{i}]^{2}+  \label{qe6dvacuum} \\
&&\ ^{0}h_{4}(x^{k})(\partial _{4}\sqrt{|h_{3}|}%
)^{2}[dy^{4}+w_{i}(x^{k},y^{4})dx^{i}]^{2}+h_{5}(u^{\tau
},y^{6})[dy^{5}+\left( \ _{1}^{1}n_{\lambda }(u^{\tau })+\
_{2}^{1}n_{\lambda }(u^{\tau })\int dy^{6}/h_{6}\right) du^{\lambda }]^{2}
\notag \\
&&+\ ^{0}h_{6}(u^{\tau })(\partial _{6}\sqrt{|h_{5}|})^{2}[dy^{6}+\
^{1}w_{\lambda }(u^{\tau },y^{6})du^{\lambda }]^{2},  \notag
\end{eqnarray}%
where $\ ^{0}h_{3}(x^{k}),\ ^{0}h_{5}(u^{\tau }),\ _{1}n_{k}(x^{i}),\
_{2}n_{k}(x^{i}),\ _{1}^{1}n_{\lambda }(u^{\tau }),\ _{2}^{1}n_{\lambda
}(u^{\tau })$ are integration functions. \ The values $h_{4}(x^{k},y^{4})$
and $h_{6}(u^{\tau },y^{6})$ are any generating functions. We can consider
arbitrary functions $w_{i}(x^{k},y^{4})$ and $\ ^{1}w_{\lambda }(u^{\tau
},y^{6})$ because, respectively, $\beta =\alpha _{i}=0$ and $\ ^{1}\beta =\
^{1}\alpha _{\tau }=0$ for such configurations, see formulas (\ref{ca1}), (%
\ref{c1}) and (\ref{ca2}), (\ref{c2}).

\subsubsection{Coefficients of metrics as generating functions}

For nontrivial sources $\ ^{v}\Lambda (x^{k},y^{4}),$ $\ _{1}^{v}\Lambda
(u^{\tau },y^{6}),$ $\ _{2}^{v}\Lambda (u^{\tau },y^{8}),...$ , we can
prescribe respectively $h_{3},h_{5}$ and $h_{7}$ (with nonzero $\partial
_{4}h_{3},\partial _{6}h_{5}$ and $\partial _{8}h_{7}$) as generating
functions. Let us perform such constructions in explicit form for $s=0.$
Using formula (\ref{aux01}), we find (up to an integration function
depending on $x^{i}$) that
\begin{equation}
\Phi ^{2}=2\varepsilon _{\Phi }\int dy^{4}\ ^{v}\Lambda \ \partial _{4}h_{3},
\label{aux02}
\end{equation}%
where $\varepsilon _{\Phi }=\pm 1$ in order to have $\Phi ^{2}>0.$
Inserting this value into (\ref{h4aux}), we express $h_{4}$ in terms of $\
^{v}\Lambda $ and $h_{3},$%
\begin{equation*}
h_{4}[\ ^{v}\Lambda ,\ h_{3}]=\varepsilon _{4}(\partial _{4}h_{3})^{2}/2\
^{v}\Lambda h_{3}\int dy^{4}(\ ^{v}\Lambda h_{3}),\ \varepsilon _4=\pm 1.
\end{equation*}%
The N--connection coefficients are computed following the formulas in (\ref{solhn})
with $\Phi \lbrack \ ^{v}\Lambda ,\ h_{3}]$ expressed in the form (\ref{aux02}),%
\begin{equation*}
w_{i}[\ ^{v}\Lambda ,\ h_{3}]=\frac{\partial _{i}\Phi }{\partial _{4}\Phi }=%
\frac{\partial _{i}\Phi ^{2}}{\partial _{4}\Phi ^{2}}=\frac{\int
dy^{4}\partial _{i}|\ ^{v}\Lambda \partial _{4}h_{3}|}{|\ ^{v}\Lambda
\partial _{4}h_{3}|},
\end{equation*}%
and
\begin{equation*}
n_{k}[\ ^{v}\Lambda ,\ h_{3}]=\ _{1}n_{k}+\ _{2}n_{k}\int dy^{4}\frac{%
(\partial _{4}h_{3})^{2}}{\ ^{v}\Lambda (\sqrt{|h_{3}|})^{5}%
\int_{0}^{y^{4}}dy^{4^{\prime }}(\ ^{v}\Lambda h_{3})},
\end{equation*}%
where $\varepsilon _{4}/2$ is included in $n_{2}.$

We can use for $s=1$ and $s=2$ certain formulas similar to (\ref{aux02}),
\begin{equation*}
\ ^{1}\Phi ^{2}=2\varepsilon _{\ ^{1}\Phi }\int dy^{6}\ _{1}^{v}\Lambda \
\partial _{6}h_{5}\mbox{ and }\ ^{2}\Phi ^{2}=2\varepsilon _{\ ^{2}\Phi
}\int dy^{8}\ _{2}^{v}\Lambda \ \partial _{8}h_{7},\ \varepsilon _{\
^{1}\Phi }=\pm 1, \varepsilon _{\ ^{2}\Phi }=\pm 2.
\end{equation*}%
The solutions (\ref{qnk4d}), (\ref{qnk6d}) and (\ref{qnk8d}) are
respectively re--parameterized as
\begin{eqnarray*}
ds_{4dK}^{2} &=&\epsilon _{i}e^{\psi (x^{k})}(dx^{i})^{2}+h_{3}\left[
dy^{3}+\left( \ _{1}n_{k}+\ _{2}n_{k}\int dy^{4}\frac{(\partial
_{4}h_{3})^{2}}{\ ^{v}\Lambda (\sqrt{|h_{3}|})^{5}\int_{0}^{y^{4}}dy^{4^{%
\prime }}(\ ^{v}\Lambda h_{3})}\right) dx^{k}\right] ^{2} \\
&&+\varepsilon _{4}\frac{(\partial _{4}h_{3})^{2}}{2\ ^{v}\Lambda h_{3}\int
dy^{4}(\ ^{v}\Lambda h_{3})}\ \left[ dy^{4}+\frac{\int dy^{4}\partial _{i}|\
^{v}\Lambda \partial _{4}h_{3}|}{|\ ^{v}\Lambda \partial _{4}h_{3}|}dx^{i}%
\right] ^{2},
\end{eqnarray*}%
\begin{eqnarray*}
ds_{6dK}^{2} &=&ds_{4dK}^{2}+h_{5}\left[ dy^{5}+\left( \ _{1}^{1}n_{\tau }+\
_{2}^{1}n_{\tau }\int dy^{6}\frac{(\partial _{6}h_{5})^{2}}{\
_{1}^{v}\Lambda (\sqrt{|h_{5}|})^{5}\int_{0}^{y^{6}}dy^{6^{\prime }}(\
_{1}^{v}\Lambda h_{5})}\right) du^{\tau }\right] ^{2} \\
&&+\varepsilon _{6}\frac{(\partial _{6}h_{5})^{2}}{2\ _{1}^{v}\Lambda
h_{5}\int dy^{6}(\ _{1}^{v}\Lambda h_{5})}\left[ dy^{6}+\frac{\int
dy^{6}\partial _{\tau }|\ _{1}^{v}\Lambda \partial _{6}h_{5}|}{|\
_{1}^{v}\Lambda \partial _{6}h_{5}|}du^{\tau }\right] ^{2},
\end{eqnarray*}%
and
\begin{eqnarray*}
ds_{8dK}^{2} &=&ds_{6dK}^{2}+h_{7}\left[ dy^{7}+\left( \ _{1}^{2}n_{\tau
_{1}}+\ _{2}^{2}n_{\tau _{1}}\int dy^{8}\frac{(\partial _{8}h_{7})^{2}}{\
_{2}^{v}\Lambda (\sqrt{|h_{7}|})^{5}\int_{0}^{y^{8}}dy^{8^{\prime }}(\
_{2}^{v}\Lambda h_{7})}\right) du^{\tau _{1}}\right] ^{2} \\
&&+\ \varepsilon _{8}\frac{(\partial _{8}h_{7})^{2}}{2\ _{2}^{v}\Lambda
h_{7}\int dy^{8}(\ _{2}^{v}\Lambda h_{7})}\left[ dy^{8}+\frac{\int
dy^{8}\partial _{\tau _{1}}|\ _{2}^{v}\Lambda \partial _{8}h_{7}|}{|\
_{2}^{v}\Lambda \partial _{8}h_{7}|}du^{\tau _{1}}\right] ^{2}.
\end{eqnarray*}

We can introduce effective cosmological constants via re--definition of the
generating functions of the type (\ref{rescgf}) when $(\Phi ,\ ^{v}\Lambda
)\rightarrow (\tilde{\Phi},\widetilde{\Lambda }),(\ ^{1}\Phi ,\
_{1}^{v}\Lambda )\rightarrow (\ ^{1}\tilde{\Phi},\ _{1}\widetilde{\Lambda })$
and $(\ ^{2}\Phi ,\ _{2}^{v}\Lambda )\rightarrow (\ ^{2}\tilde{\Phi},\ _{2}%
\widetilde{\Lambda }).$ For such parameterizations, the coefficients of the
metrics depend explicitly on $\tilde{\Phi},\ ^{1}\tilde{\Phi}$ and $\ ^{2}%
\tilde{\Phi}.$ Finally, we note that such formulas can be similarly
generalized for higher dimensions with shells $s=3,4 ...$.

\subsubsection{ The Levi--Civita conditions}

\label{sslc}All solutions constructed in previous sections define certain
subclasses of generic off--diagonal metrics (\ref{ansk}) for canonical
d--connections $\ ^{s}\widehat{\mathbf{D}}$ and nontrivial nonholonomically
induced d--torsion coefficients $\widehat{\mathbf{T}}_{\ \alpha _{s}\beta
_{s}}^{\gamma _{s}}\ $ (\ref{dtors}). Such a torsion vanishes for a subclass
of nonholonomic distributions with necessary types of parameterizations of
the generating and integration functions and sources. In explicit form, we
construct LC--configurations by imposing additional constraints, shell by
shell, on the d--metric and N--connection coefficients. By straightforward
computations (see details in Refs. \cite{vex1,vpars,vex2,vex3}, and Appendix %
\ref{zt}), we can verify that if in N--adapted frames
\begin{eqnarray}
\mbox{ for }s &=&0:\ \partial _{4}w_{i}=\mathbf{e}_{i}\ln \sqrt{|\ h_{4}|},%
\mathbf{e}_{i}\ln \sqrt{|\ h_{3}|}=0,\partial _{i}w_{j}=\partial _{j}w_{i}%
\mbox{ and }\partial _{4}n_{i}=0;  \notag \\
s &=&1:\ \partial _{6}\ ^{1}w_{\alpha }=\ ^{1}\mathbf{e}_{\alpha }\ln \sqrt{%
|\ h_{6}|},\ ^{1}\mathbf{e}_{\alpha }\ln \sqrt{|\ h_{5}|}=0,\partial
_{\alpha }\ ^{1}w_{\beta }=\partial _{\beta }\ ^{1}w_{\alpha }\mbox{ and }%
\partial _{6}\ ^{1}n_{\gamma }=0;  \label{zerot} \\
s &=&2:\ \partial _{8}\ ^{2}w_{\alpha _{1}}=\ ^{2}\mathbf{e}_{\alpha
_{1}}\ln \sqrt{|\ h_{8}|},\ ^{2}\mathbf{e}_{\alpha _{1}}\ln \sqrt{|\ h_{7}|}%
=0,\partial _{\alpha _{1}}\ ^{2}w_{\beta _{1}}=\partial _{\beta _{1}}\
^{2}w_{\alpha _{1}}\mbox{ and }\partial _{8}\ ^{2}n_{\gamma _{1}}=0;  \notag
\end{eqnarray}%
(similar equations can be written recurrently for arbitrary finite extra
dimensions) then the torsion coefficients become zero. For $n$--coefficients,
such conditions are satisfied if $\ _{2}n_{k}(x^{i})=0$ and $\partial _{i}\
_{1}n_{j}(x^{k})=\partial _{j}\ _{1}n_{i}(x^{k});\ _{2}^{1}n_{\alpha
}(u^{\beta })=0$ and $\partial _{\gamma }\ _{1}^{1}n_{\tau }(u^{\beta
})=\partial _{\tau }\ _{1}^{1}n_{\gamma }(u^{\beta });\ _{2}^{2}n_{\alpha
_{1}}(u^{\beta _{1}})=0$ and $\partial _{\gamma _{1}}\ _{1}^{2}n_{\tau
_{1}}(u^{\beta _{1}})=\partial _{\tau _{1}}\ _{1}^{2}n_{\gamma
_{1}}(u^{\beta _{1}})$ etc. The explicit form of the solutions of the constraints on
$w_{k}$ derived from (\ref{zerot}) depend on the class of vacuum or
non--vacuum metrics we try to construct.

Let us show how we can satisfy the LC--conditions (\ref{zerot}) for $s=0.$
We note that such nonholonomic constraints cannot be solved in explicit
form for arbitrary data $(\Phi ,\ ^{v}\Lambda ),$ or $(\tilde{\Phi},\ \tilde{%
\Lambda}),$ and all types of nonzero integration functions $\
_{1}n_{j}(x^{k})$ and $\ _{2}n_{k}(x^{i})=0.$ Nevertheless, certain general
classes of solutions can be written in explicit form if via coordinate and
frame transformations we can fix $_{2}n_{k}(x^{i})=0$ $\ $and $\
_{1}n_{j}(x^{k})=\partial _{j}n(x^{k})$ for a function $n(x^{k}).$ Then we
use the property that
\begin{equation*}
\mathbf{e}_{i}\Phi =(\partial _{i}-w_{i}\partial _{4})\Phi \equiv 0
\end{equation*}%
for any $\Phi $ if $w_{i}=\partial _{i}\Phi /\partial _{4}\Phi ,$ see (\ref%
{solhn}). For any functional $H[\Phi ],$ one has the equality
\begin{equation*}
\mathbf{e}_{i}H=(\partial _{i}-w_{i}\partial _{4})H=\frac{\partial H}{%
\partial \Phi }(\partial _{i}-w_{i}\partial _{4})\Phi \equiv 0.
\end{equation*}%
We can restrict our construction to a subclass of generating data $(\Phi ,\
^{v}\Lambda )$ and $(\tilde{\Phi},\ \tilde{\Lambda})$ which are related via formulas (%
\ref{rescgf}) when $H=\tilde{\Phi}[\Phi ]$ is a functional which allows us
to generate LC--configurations in explicit form. Using $h_{3}[\tilde{\Phi}]=%
\tilde{\Phi}^{2}/4\widetilde{\Lambda }$ (\ref{solha}) for $H= $ $\tilde{\Phi}%
=\ln \sqrt{|\ h_{3}|}$, we satisfy the second condition, $\mathbf{e}_{i}\ln
\sqrt{|\ h_{3}|}=0,$ in (\ref{zerot}) for $s=0.$

In the second step, we solve firstly the condition in (\ref{zerot}), for $s=0.$
Taking the derivative $\partial _{4}$ of $\ w_{i}=\partial _{i}\Phi
/\partial _{4}\Phi $ (\ref{solhn}), we obtain%
\begin{equation}
\partial _{4}w_{i}=\frac{(\partial _{4}\partial _{i}\Phi )(\partial _{4}\Phi
)-(\partial _{i}\Phi )\partial _{4}\partial _{4}\Phi }{(\partial _{4}\Phi
)^{2}}=\frac{\partial _{4}\partial _{i}\Phi }{\partial _{4}\Phi }-\frac{%
\partial _{i}\Phi }{\partial _{4}\Phi }\frac{\partial _{4}\partial _{4}\Phi
}{\partial _{4}\Phi }.  \label{fder}
\end{equation}%
If $\Phi =\check{\Phi},$ for which
\begin{equation}
\partial _{4}\partial _{i}\check{\Phi}=\partial _{i}\partial _{4}\check{\Phi}%
,  \label{explcond}
\end{equation}
and using (\ref{fder}), we compute $\partial _{4}w_{i}=\mathbf{e}_{i}\ln
|\partial _{4}\Phi |$. For $h_{4}[\Phi ,\ ^{v}\Lambda ]$ (\ref{h4aux}), $%
\mathbf{e}_{i}\ln \sqrt{|\ h_{4}|}=\mathbf{e}_{i} [ \ln |\partial _{4}\Phi
|-\ln \sqrt{|\ ^{v}\Lambda |}]$, were we used the conditions (\ref{explcond}%
) and the property $\mathbf{e}_{i}\check{\Phi}=0.$ Using the last two
formulas, we can obtain $\partial _{4}w_{i}=\mathbf{e}_{i}\ln \sqrt{|\ h_{4}|%
}$ if $\mathbf{e}_{i}\ln \sqrt{|\ ^{v}\Lambda |}=0.$ This is possible for $\
^{v}\Lambda =const,$ or if $\ ^{v}\Lambda $ can be expressed as a functional
$\ ^{v}\Lambda (x^{i},y^{4})=\ ^{v}\Lambda \lbrack \check{\Phi}].$

Finally, we note that the third condition for $s=0$, $\partial
_{i}w_{j}=\partial _{j}w_{i},$ see (\ref{zerot}), holds for any $\check{A}=%
\check{A}(x^{k},y^{4})$ for which $w_{i}=\check{w}_{i}=\partial _{i}\check{%
\Phi}/\partial _{4}\check{\Phi}=\partial _{i}\check{A}.$

Following similar considerations for other shells' generating functions
\begin{eqnarray}
s=1: &&\ ^{1}\Phi =\ ^{1}\check{\Phi}(u^{\tau },y^{6}),\partial _{6}\partial
_{\tau }\ ^{1}\check{\Phi}=\partial _{\tau }\partial _{6}\ ^{1}\check{\Phi};
\label{expconda} \\
&&\partial _{\alpha }\ ^{1}\check{\Phi}/\partial _{6}\ ^{1}\check{\Phi}%
=\partial _{\alpha }\ ^{1}\check{A};\ _{1}^{1}n_{\tau }=\partial _{\tau }\
^{1}n(u^{\beta });  \notag \\
s=2: &&\ ^{2}\Phi =\ ^{2}\check{\Phi}(u^{\tau _{1}},y^{8}),\partial
_{8}\partial _{\tau _{1}}\ ^{2}\check{\Phi}=\partial _{\tau _{1}}\partial
_{8}\ ^{2}\check{\Phi};  \notag \\
&&\partial _{\alpha _{1}}\ ^{2}\check{\Phi}/\partial _{8}\ ^{2}\check{\Phi}%
=\partial _{\alpha _{2}}\ ^{2}\check{A};\ _{1}^{2}n_{\tau _{1}}=\partial
_{\tau _{1}}\ ^{2}n(u^{\beta _{1}});  \notag
\end{eqnarray}%
(similar formulas can be written recurrently for arbitrary extra shells); we
can construct quadratic line elements for LC--configurations
\begin{eqnarray}
ds_{8dK}^{2} &=&\epsilon _{i}e^{\psi (x^{k})}(dx^{i})^{2}+\frac{\ (\tilde{%
\Phi}[\check{\Phi}])^{2}}{4\widetilde{\Lambda }}\left[ dy^{3}+(\partial
_{i}\ n)dx^{i}\right] ^{2}+\frac{(\partial _{4}\tilde{\Phi}[\check{\Phi}%
])^{2}}{\Xi (\tilde{\Phi}[\check{\Phi}])}\left[ dy^{4}+(\partial _{i}\
\check{A})dx^{i}\right] ^{2}  \notag \\
&&+\frac{(\ ^{1}\tilde{\Phi}[\ ^{1}\check{\Phi}])^{2}}{4\ ^{1}\widetilde{%
\Lambda }}\left[ dy^{5}+(\partial _{\tau }\ ^{1}n)du^{\tau }\right] ^{2}+\
\frac{(\partial _{6}\ ^{1}\tilde{\Phi}[\ ^{1}\check{\Phi}])^{2}}{\ ^{1}\Xi
(\ ^{1}\tilde{\Phi}[\ ^{1}\check{\Phi}])}\ \left[ dy^{6}+(\partial _{\tau }\
^{1}\check{A})du^{\tau }\right] ^{2}  \label{qellcs} \\
&&+\frac{(\ ^{2}\tilde{\Phi}[\ ^{2}\check{\Phi}])^{2}}{4\ ^{2}\widetilde{%
\Lambda }}\left[ dy^{7}+(\partial _{\tau _{1}}\ ^{2}n)du^{\tau _{1}}\right]
^{2}+\ \frac{(\partial _{8}\ ^{2}\tilde{\Phi}[\ ^{2}\check{\Phi}])^{2}}{\
^{2}\Xi (\ ^{2}\tilde{\Phi}[\ ^{2}\check{\Phi}])}\ \left[ dy^{8}+(\partial
_{\tau _{1}}\ ^{2}\check{A})du^{\tau _{1}}\right] ^{2}.  \notag
\end{eqnarray}%
In these formulas, the generating functions are functionals of "inverse hat"
values, when
\begin{eqnarray*}
\check{\Phi}^{2} &=&\widetilde{\Lambda }^{-1}\int dy^{4}(\ ^{v}\Lambda
)\partial _{4}(\tilde{\Phi}^{2})\mbox{
and }\tilde{\Phi}^{2}=\widetilde{\Lambda }\int dy^{4}(\ ^{v}\Lambda
)^{-1}\partial _{4}(\check{\Phi}^{2}); \\
\ ^{1}\check{\Phi}^{2} &=&(\ ^{1}\widetilde{\Lambda })^{-1}\int dy^{6}(\
_{1}^{v}\Lambda )\partial _{6}(\ ^{1}\tilde{\Phi}^{2})\mbox{
and }\ ^{1}\tilde{\Phi}^{2}=\ ^{1}\widetilde{\Lambda }\int dy^{6}(\
_{1}^{v}\Lambda )^{-1}\partial _{6}(\ ^{1}\check{\Phi}^{2}); \\
\ ^{2}\check{\Phi}^{2} &=&(\ ^{2}\widetilde{\Lambda })^{-1}\int dy^{8}(\
_{2}^{v}\Lambda )\partial _{8}(\ ^{2}\tilde{\Phi}^{2})\mbox{
and }\ ^{2}\tilde{\Phi}^{2}=\ ^{2}\widetilde{\Lambda }\int dy^{8}(\
_{2}^{v}\Lambda )^{-1}\partial _{8}(\ ^{2}\check{\Phi}^{2}).
\end{eqnarray*}%
We can compute the values $\Xi (\tilde{\Phi}[\check{\Phi}]),$ $\ ^{1}\Xi (\
^{1}\tilde{\Phi}[\ ^{1}\check{\Phi}])$ and $\ ^{2}\Xi (\ ^{2}\tilde{\Phi}[\
^{2}\check{\Phi}])$ as in (\ref{qnk8d}).

The torsions for such non--vacuum exact solutions (\ref{qellcs}) generated
by the respective data $(\ ^{s}\mathbf{\check{g},}\ ^{s}\mathbf{\check{N},}\ ^{s}%
\mathbf{\check{\nabla}})$ are zero, which is different from the class of
exact solutions (\ref{qnk8d}) with nontrivial canonical d--torsions (\ref%
{dtors}) and completely determined by arbitrary data $(\ ^{s}\mathbf{g,}\ ^{s}%
\mathbf{N,}\ ^{s}\widehat{\mathbf{D}})$ with Killing symmetry on $\partial
_{7}.$

\subsection{ Non--Killing configurations}

The off--diagonal integral varieties of solutions of gravitational field
equations constructed in the previous section possess for any shell $s\geq 0$ at
least one Killing vector symmetry on $\partial /\partial y^{a_{s}-1}$ when
the metrics do not depend on coordinate $y^{a_{s}-1}$ in a class of
N--adapted frames. There are two general possibilities to generate
"non--Killing" configurations: 1) performing a formal embedding into higher
dimensional vacuum spacetimes and/or via 2) "vertical" conformal nonholonomic
deformations.

\subsubsection{Embedding into a higher dimension vacuum}

We analyze a subclass of off--diagonal metrics for 6--d spaces which via
nonholonomic constraints and re--parameterizations transform into 4--d
non--Killing vacuum solutions. Let us consider certain geometric data $%
\Lambda =\ ^{v}\Lambda =\ _{1}^{v}\Lambda =0$ and $h_{3}=\epsilon
_{3},h_{5}=\epsilon _{5},n_{k}=0$ and $\ ^{1}n_{\alpha }=0$ with a 2-d $h$%
--metric $\epsilon _{i}e^{\psi (x^{k},\Lambda =0)}(dx^{i})^{2}.$ The
coefficients of the Ricci d--tensor are zero (see formulas (\ref{equ1})-(\ref%
{equ4}) and (\ref{equ5})-(\ref{equ7})). Here we note that one cannot use the
equations (\ref{e1})-(\ref{e4aa}) derived for $\partial _{4}h_{3}\neq 0,$ $%
\partial _{6}h_{5}\neq 0$ etc which does not allow, for instance, the values $%
h_{3}=\epsilon _{3},h_{5}=\epsilon _{5},$ for any nontrivial data $%
h_{4}(x^{i},y^{4}),w_{k}(x^{i},y^{4});$ $h_{6}(x^{i},y^{4},y^{6}),\
^{1}w_{k}(x^{i},y^{4})$, $\ ^{1}w_{4}(x^{i},y^{4},y^{6}).$ Such values can
be considered as generating functions for the vacuum quadratic line elements%
\begin{eqnarray}
ds_{6\rightarrow 4}^{2} &=&\epsilon _{i}e^{\psi (x^{k},\Lambda
=0)}(dx^{i})^{2}+\epsilon _{3}(dy^{3})^{2}+h_{4}(dy^{4}+w_{k}dx^{k})^{2}
\label{6to4} \\
&&+\epsilon _{5}(dy^{5})^{2}+h_{6}(dy^{6}+\ ^{1}w_{k}dx^{k}+\
^{1}w_{4}dy^{4})^{2}.  \notag
\end{eqnarray}%
In general, this class of vacuum 6-d metrics have a nonzero
nonholonomically induced d--torsion (\ref{dtors}). Such solutions do not
consist obligatory of a subclass of vacuum solutions (\ref{qe6dvacuum}) when $%
h_{3}\rightarrow \epsilon _{3}$ and $h_{5}\rightarrow \epsilon _{5};$ the
conditions $\partial _{4}h_{3}\neq 0$ and $\partial _{6}h_{5}\neq 0$
restrict the class of possible generating functions $h_{4}$ and $h_{6}.$ If
we fix from the very beginning certain configurations with $\partial
_{4}h_{3}=0$ and $\partial _{6}h_{5}=0,$ we can consider $h_{4},h_{6}$ and $%
w_{k},\ ^{1}w_{k},\ ^{1}w_{4}$ as independent generating functions.

If the coefficients in (\ref{6to4}) are subjected additionally to the
constraints (\ref{zerot}) for $s=0$ and $s=1,$ we generate
LC--configurations. We can follow a formal procedure which is similar to
that outlined in section \ref{sslc}. The conditions $\mathbf{e}_{i}\ln \sqrt{%
|\ h_{3}|}=0$ and $\ ^{1}\mathbf{e}_{\alpha }\ln \sqrt{|\ h_{5}|}=0$ are
satisfied respectively for any constant $h_{3}=\epsilon _{3}$ and $%
h_{5}=\epsilon _{5}.$ Let us show how we can restrict the class of
generating functions in order to obtain solutions for which
\begin{eqnarray}
\partial _{4}w_{i}(x^{i},y^{4}) &=&\mathbf{e}_{i}\ln \sqrt{|\
h_{4}(x^{i},y^{4})|},\partial _{i}w_{j}=\partial _{j}w_{i},\mbox{ and }\
\label{zerota} \\
\partial _{6}\ ^{1}w_{\alpha }(x^{i},y^{4},y^{6}) &=&\ ^{1}\mathbf{e}%
_{\alpha }\ln \sqrt{|\ h_{6}(x^{i},y^{4},y^{6})|},\partial _{\alpha }\
^{1}w_{\beta }=\partial _{\beta }\ ^{1}w_{\alpha }.  \notag
\end{eqnarray}%
We emphasize that the above N--adapted formulas do not depend on $y^{3}$ and $%
y^{5}.$ Prescribing any values of $\ h_{4}$ and $\ h_{6}$ we can find
LC--admissible $w$--coefficients solving respective systems of first order
partial derivative equations in (\ref{zerota}). In general, such solutions
are defined by nonholonomic configurations, i.e. in "non--explicit" form. If
all values $h_{4}[\check{\Phi}],h_{6}[\ ^{1}\check{\Phi}]$ and $w_{k}[\check{%
\Phi}],\ ^{1}w_{k}[\ ^{1}\check{\Phi}],\ ^{1}w_{4}[\ ^{1}\check{\Phi}]$ are
respectively determined by $\check{\Phi}(x^{i},y^{4})$ and $\ ^{1}\check{\Phi%
}(x^{i},y^{4},y^{6})$ satisfying conditions of type (\ref{explcond}) and (%
\ref{expconda}) (but $h_{3}$ and $h_{5}$ are not functionals of type (\ref%
{solha})), we can solve the equations (\ref{zerota}) in explicit form. Let
us chose any generating functions $\check{\Phi}$ and $\ ^{1}\check{\Phi},$
consider any functionals $h_{4}[\check{\Phi}],h_{6}[\ ^{1}\check{\Phi}]$ and
compute%
\begin{eqnarray}
w_{i} &=&\check{w}_{i}=\partial _{i}\check{\Phi}/\partial _{4}\check{\Phi}%
=\partial _{i}\check{A}\mbox{ and }  \label{data4c} \\
\ ^{1}w_{i} &=&\ ^{1}\check{w}_{i}=\partial _{i}\ ^{1}\check{\Phi}/\partial
_{6}\ ^{1}\check{\Phi}=\partial _{i}\ ^{1}\check{A},\ ^{1}w_{4}=\ ^{1}\check{%
w}_{4}=\partial _{4}\ ^{1}\check{\Phi}/\partial _{6}\ ^{1}\check{\Phi}%
=\partial _{4}\ ^{1}\check{A},  \notag
\end{eqnarray}%
for some $\check{A}(x^{i},y^{4})$ and $\ ^{1}\check{A}(x^{i},y^{4},y^{6})$
which are necessary for $\partial _{i}w_{j}=\partial _{j}w_{i}$ and $%
\partial _{\alpha }\ ^{1}w_{\beta }=\partial _{\beta }\ ^{1}w_{\alpha }.$
Considering functional derivatives of type (\ref{fder}) and N--coefficients
of the type in (\ref{data4c}) when $H[\check{\Phi}]=\ln \sqrt{|\ h_{4}|}$ and $\
^{1}H[\ ^{1}\check{\Phi}]=\ln \sqrt{|\ h_{6}|},$ we can satisfy the
LC--conditions (\ref{zerota}).

Putting together the above formulas, we construct a subclass of metrics of (\ref%
{6to4}) determined by generic off--diagonal metrics as solutions of 6--d
vacuum Einstein equations,
\begin{eqnarray}
ds_{6\rightarrow 4}^{2} &=&\epsilon _{i}e^{\psi (x^{k},\Lambda
=0)}(dx^{i})^{2}+\epsilon _{3}(dy^{3})^{2}+h_{4}[\check{\Phi}%
](dy^{4}+\partial _{k}\check{A}dx^{k})^{2}  \label{6to4lc} \\
&&+\epsilon _{5}(dy^{5})^{2}+h_{6}[\ ^{1}\check{\Phi}](dy^{6}+\partial _{k}\
^{1}\check{A}\ dx^{k}+\partial _{4}\ ^{1}\check{A}\ dy^{4})^{2}.  \notag
\end{eqnarray}%
We note that in this quadratic line element the terms $\epsilon
_{3}(dy^{3})^{2}$ and $\epsilon _{5}(dy^{5})^{2}$ are used for trivial extensions
from 4-d to 6--d. Re--defining the coordinate $y^{6}\rightarrow y^{3},$ we
generate vacuum solutions in 4--d gravity with metrics (\ref{6to4lc})
depending on all four coordinates $x^{i},y^{3}$ and $y^{4}.$ The anholonomy
coefficients (\ref{anhrel1}) are not zero and such metrics cannot be
diagonalized by coordinate transformations. This class of 4--d vacuum spacetimes
do not possess, in general, Killing symmetries.

\subsubsection{"Vertical" conformal nonholonomic deformations}

There is another possibility to generate off--diagonal solutions depending
on all spacetime coordinates and, in general, with nontrivial sources of the
type in (\ref{sourc1}), see details and proofs in Ref. \cite{vex3}. By
straightforward computations, we can check that any metric
\begin{eqnarray}
\mathbf{g} &=&g_{i}(x^{k})dx^{i}\otimes dx^{i}+\omega ^{2}(u^{\alpha
})h_{a}(x^{k},y^{4})\mathbf{e}^{a}\otimes \mathbf{e}^{a}+  \label{ans1} \\
&&\ ^{1}\omega ^{2}(u^{\alpha _{1}})h_{a_{1}}(u^{\alpha },y^{6})\mathbf{e}%
^{a_{1}}\otimes \mathbf{e}^{a_{1}}+\ ^{2}\omega ^{2}(u^{\alpha
_{2}})h_{a_{2}}(u^{\alpha _{1}},y^{8})\mathbf{e}^{a_{2}}\otimes \mathbf{e}%
^{a_{2}},  \notag
\end{eqnarray}%
with the conformal $v$--factors subjected to the conditions
\begin{eqnarray}
\mathbf{e}_{k}\omega  &=&\partial _{k}\omega +n_{k}\partial _{3}\omega
+w_{k}\partial _{4}\omega =0,  \label{vconfc} \\
\ ^{1}\mathbf{e}_{\beta }\ ^{1}\omega  &=&\partial _{\beta }\ ^{1}\omega +\
^{1}n_{\beta }\partial _{5}\ ^{1}\omega +\ ^{1}w_{\beta }\partial _{6}\
^{1}\omega =0,  \notag \\
\ ^{2}\mathbf{e}_{\beta _{1}}\ ^{2}\omega  &=&\partial _{\beta _{1}}\
^{2}\omega +\ ^{2}n_{\beta _{1}}\partial _{7}\ ^{2}\omega +\ ^{2}w_{\beta
_{1}}\partial _{8}\ ^{2}\omega =0,  \notag
\end{eqnarray}%
(similar equations can be written recurrently for arbitrary finite extra
dimensions) does not change the Ricci d--tensor (\ref{equ1})--(\ref{equ4d}).
Any class of solutions considered in this section can be generalized to
non--Killing configurations using nonholonomic "vertical" conformal
transforms.

In 4--d, the ansatz (\ref{ans1}) can be parameterized with respect to
coordinate frames in a form with nontrivial $\omega ^{2}(u^{\alpha })$ which
is different from that given in Figure \ref{fig1},
\begin{equation}
g_{\underline{\alpha }\underline{\beta }}=\left[
\begin{array}{cccc}
g_{1}+\omega ^{2}(n_{1}^{\ 2}h_{3}+w_{1}^{\ 2}h_{4}) & \omega
^{2}(n_{1}n_{2}h_{3}+w_{1}w_{2}h_{4}) & \omega ^{2}n_{1}h_{3} & \omega
_{1}^{2}w_{1}h_{4} \\
\omega ^{2}(n_{1}n_{2}h_{3}+w_{1}w_{2}h_{4}) & g_{2}+\omega ^{2}(n_{2}^{\
2}h_{3}+w_{2}^{\ 2}h_{4}) & \omega ^{2}n_{2}h_{3} & \omega ^{2}w_{2}h_{4} \\
\omega ^{2}n_{1}h_{3} & \omega ^{2}n_{2}h_{3} & \omega ^{2}h_{3} & 0 \\
\omega ^{2}w_{1}h_{4} & \omega ^{2}w_{2}h_{4} & 0 & \omega ^{2}h_{4}%
\end{array}%
\right] .  \label{ans1a}
\end{equation}
A general metric $g_{\alpha \beta }(u^{\gamma })$ can be parameterized in
the form (\ref{ans1a}) if there are any geometrically and physically
well--defined frame transformations $g_{\alpha \beta }=e_{\ \alpha }^{\underline{%
\alpha }}e_{\ \beta }^{\underline{\beta }}g_{\underline{\alpha }\underline{%
\beta }}.$ For certain given values $g_{\alpha \beta }$ and $g_{\underline{%
\alpha }\underline{\beta }}$ (in GR, there are 6 + 6 independent
components), we have to solve a system of quadratic algebraic equation in
order to determine 16 coefficients $e_{\ \alpha }^{\underline{\alpha }},$ up
to a fixed coordinate system. We have to fix such nonholonomic 2+2 splitting
and partitions on manifolds when the algebraic equations have real
nondegenerate solutions.

Finally, we note that we can consider generic off--diagonal coordinate
decompositions which are similar to (\ref{ans1a}) but with dependencies on
all coordinates for higher order shells.

\section{Nonholonomic Deformations \& the Kerr Metric}

\label{s4} In this section, we show how using the AFDM formalism the Kerr
solution can be constructed as a particular case when corresponding types of
generating and integration functions are prescribed. We provide a series of
new classes of solutions when the metrics are nonholonomically deformed into
general or ellipsoidal stationary configurations in four dimensional gravity
and/or extra dimensions. Explicit examples are studied of generic
off--diagonal metrics encoding interactions in massive gravity, $f$%
--modifications and nonholonomically induced torsion effects. We find such
nonholonomic constraints when modified massive, and zero mass, gravitational
effects can be modelled by nonlinear off--diagonal interactions in GR.

\subsection{Generating the Kerr vacuum solution}

Let us consider the ansatz%
\begin{equation*}
ds_{[0]}^{2}=Y^{-1}e^{2h}(d\rho ^{2}+dz^{2})-\rho
^{2}Y^{-1}dt^{2}+Y(d\varphi +Adt)^{2}
\end{equation*}%
parameterized in terms of three functions $(h,Y,A)$ on coordinates $(\rho
,z). $ We obtain the Kerr solution of the vacuum Einstein equations in 4--d,
for rotating black holes, if we chose
\begin{eqnarray*}
Y &=&\frac{1-(p\widehat{x}_{1})^{2}-(q\widehat{x}_{2})^{2}}{(1+p\widehat{x}%
_{1})^{2}+(q\widehat{x}_{2})^{2}},\ A=2M\frac{q}{p}\frac{(1-\widehat{x}%
_{2})(1+p\widehat{x}_{1})}{1-(p\widehat{x}_{1})-(q\widehat{x}_{2})}, \\
e^{2h} &=&\frac{1-(p\widehat{x}_{1})^{2}-(q\widehat{x}_{2})^{2}}{p^{2}[(%
\widehat{x}_{1})^{2}+(\widehat{x}_{2})^{2}]},\ \rho ^{2}=M^{2}(\widehat{x}%
_{1}^{2}-1)(1-\widehat{x}_{2}^{2}),\ z=M\widehat{x}_{1}\widehat{x}_{2},
\end{eqnarray*}%
where $M=const$ and $\rho =0$ consists of the horizon $\widehat{x}_{1}=0$
and the "north / south" segments of the rotation axis, $\widehat{x}%
_{2}=+1/-1.$ Such a metric can be written in the form (\ref{ansprime}),%
\begin{equation}
ds_{[0]}^{2}=(dx^{1})^{2}+(dx^{2})^{2}-\rho ^{2}Y^{-1}(\mathbf{e}^{3})^{2}+Y(%
\mathbf{e}^{4})^{2},  \label{kerr1}
\end{equation}%
if the coordinates $x^{1}(\widehat{x}_{1},\widehat{x}_{2})$ and $x^{2}(%
\widehat{x}_{1},\widehat{x}_{2})$ are defined for any%
\begin{equation*}
(dx^{1})^{2}+(dx^{2})^{2}=M^{2}e^{2h}(\widehat{x}_{1}^{2}-\widehat{x}%
_{2}^{2})Y^{-1}\left( \frac{d\widehat{x}_{1}^{2}}{\widehat{x}_{1}^{2}-1}+%
\frac{d\widehat{x}_{2}^{2}}{1-\widehat{x}_{2}^{2}}\right)
\end{equation*}%
and $y^{3}=t+\widehat{y}^{3}(x^{1},x^{2}),y^{4}=\varphi +\widehat{y}%
^{4}(x^{1},x^{2},t),$ when
\begin{equation*}
\mathbf{e}^{3}=dt+(\partial _{i}\widehat{y}^{3})dx^{i},\mathbf{e}%
^{4}=dy^{4}+(\partial _{i}\widehat{y}^{4})dx^{i},
\end{equation*}%
for some functions $\widehat{y}^{a},$ $a=3,4,$ with $\partial _{t}\widehat{y}%
^{4}=-A(x^{k}).$

For many purposes, the Kerr metric was written in the so--called
Boyer--Linquist coordinates $(r,\vartheta ,\varphi ,t),$ for $r=m_{0}(1+p%
\widehat{x}_{1}),\widehat{x}_{2}=\cos \vartheta .$ The parameters $p,q$ are
related to the total black hole mass, $m_{0}$ (it should be not confused
with the parameter $\mu _{g}$ in massive gravity) and the total angular
momentum, $am_{0},$ for the asymptotically flat, stationary and axisymmetric
Kerr spacetime. The formulas $m_{0}=Mp^{-1}$ and $a=Mqp^{-1}$ when $%
p^{2}+q^{2}=1$ implies $m_{0}^{2}-a^{2}=M^{2}$ (see monographs \cite%
{heusler,kramer,misner} for the standard methods and bibliography on
stationary black hole solutions; we note here that the coordinates $\widehat{%
x}_{1},\widehat{x}_{2}$ correspond respectively to $x,y$ from chapter 4 of
the first book). In such variables, the vacuum solution (\ref{kerr1}) can be
written%
\begin{eqnarray}
ds_{[0]}^{2} &=&(dx^{1^{\prime }})^{2}+(dx^{2^{\prime }})^{2}+\overline{A}(%
\mathbf{e}^{3^{\prime }})^{2}+(\overline{C}-\overline{B}^{2}/\overline{A})(%
\mathbf{e}^{4^{\prime }})^{2},  \label{kerrbl} \\
\mathbf{e}^{3^{\prime }} &=&dt+d\varphi \overline{B}/\overline{A}%
=dy^{3^{\prime }}-\partial _{i^{\prime }}(\widehat{y}^{3^{\prime }}+\varphi
\overline{B}/\overline{A})dx^{i^{\prime }},\mathbf{e}^{4^{\prime
}}=dy^{4^{\prime }}=d\varphi ,  \notag
\end{eqnarray}%
for any coordinate functions
\begin{equation*}
x^{1^{\prime }}(r,\vartheta ),\ x^{2^{\prime }}(r,\vartheta ),\ y^{3^{\prime
}}=t+\widehat{y}^{3^{\prime }}(r,\vartheta ,\varphi )+\varphi \overline{B}/%
\overline{A},y^{4^{\prime }}=\varphi ,\ \partial _{\varphi }\widehat{y}%
^{3^{\prime }}=-\overline{B}/\overline{A},
\end{equation*}
for which $(dx^{1^{\prime }})^{2}+(dx^{2^{\prime }})^{2}=\Xi \left( \Delta
^{-1}dr^{2}+d\vartheta ^{2}\right) $, and the coefficients are%
\begin{eqnarray}
\overline{A} &=&-\Xi ^{-1}(\Delta -a^{2}\sin ^{2}\vartheta ),\overline{B}%
=\Xi ^{-1}a\sin ^{2}\vartheta \left[ \Delta -(r^{2}+a^{2})\right] ,  \notag
\\
\overline{C} &=&\Xi ^{-1}\sin ^{2}\vartheta \left[ (r^{2}+a^{2})^{2}-\Delta
a^{2}\sin ^{2}\vartheta \right] ,\mbox{ and }  \notag \\
\Delta &=&r^{2}-2m_{0}+a^{2},\ \Xi =r^{2}+a^{2}\cos ^{2}\vartheta .
\label{kerrcoef}
\end{eqnarray}

The quadratic linear elements (\ref{kerr1}) (or (\ref{kerrbl})) with prime
data
\begin{eqnarray}
\mathring{g}_{1} &=&1,\mathring{g}_{2}=1,\mathring{h}_{3}=-\rho ^{2}Y^{-1},%
\mathring{h}_{4}=Y,\mathring{N}_{i}^{a}=\partial _{i}\widehat{y}^{a},
\label{dkerr} \\
(\mbox{ or \ }\mathring{g}_{1^{\prime }} &=&1,\mathring{g}_{2^{\prime }}=1,%
\mathring{h}_{3^{\prime }}=\overline{A},\mathring{h}_{4^{\prime }}=\overline{%
C}-\overline{B}^{2}/\overline{A},  \notag \\
\mathring{N}_{i^{\prime }}^{3} &=&\mathring{n}_{i^{\prime }}=-\partial
_{i^{\prime }}(\widehat{y}^{3^{\prime }}+\varphi \overline{B}/\overline{A}),%
\mathring{N}_{i^{\prime }}^{4}=\mathring{w}_{i^{\prime }}=0)  \notag
\end{eqnarray}%
define solutions of the vacuum Einstein equations parameterized in the form (\ref%
{cdeinst}) and (\ref{lcconstr}) with zero sources. Here we note that we have
to consider a correspondingly N--adapted system of coordinates instead of
the "standard" prolate spherical, or Boyer--Linquist ones because
parameterizations with the data (\ref{dkerr}) are most convenient for a
straightforward application of the AFDM. Following such an approach, we can
generalize the solutions in order to get dependencies of the coefficients on
more than two coordinates, with non--Killing configurations and/or extra
dimensions.

In some sense, the Kerr vacuum solution in GR consists a "degenerate" case
of the 4--d off--diagonal vacuum solutions determined by primary metrics with
the data (\ref{dkerr}) when the diagonal coefficients depend only on two
"horizontal" N--adapted coordinates and the off--diagonal terms are induced
by rotation frames.

\subsection{Deformations of Kerr metrics in 4--d massive gravity}

Let us consider the coefficients (\ref{dkerr}) for the Kerr metric as the
data for a prime metric $\mathbf{\mathring{g}}$ (in general, it may be, or
not, an exact solution of the Einstein or other modified gravitational
equations, or any fiducial metric). Our goal is to construct nonholonomic
deformations,
\begin{equation*}
(\mathbf{\mathring{g}},\mathbf{\mathring{N},\ }^{v}\mathring{\Upsilon}=0,%
\mathring{\Upsilon}=0)\rightarrow (\widetilde{\mathbf{g}},\widetilde{\mathbf{%
N}}\mathbf{,\ }^{v}\widetilde{\Upsilon }=\widetilde{\lambda },\widetilde{%
\Upsilon }=\widetilde{\lambda }),\widetilde{\lambda }=const\neq 0,
\end{equation*}%
see sources (\ref{source1b}) for the shell $s=0$ and (\ref{source1a}). The
main condition is that the target metric $\mathbf{g}$ positively defines a
generic off--diagonal solution of field equations in 4--d massive gravity.
The N--adapted deformations of coefficients of the metrics, frames and sources
are parameterized in the form
\begin{eqnarray}
&&[\mathring{g}_{i},\mathring{h}_{a},\mathring{w}_{i},\mathring{n}%
_{i}]\rightarrow \lbrack \widetilde{g}_{i}=\widetilde{\eta }_{i}\mathring{g}%
_{i},\widetilde{h}_{3}=\widetilde{\eta }_{3}\mathring{h}_{3},\widetilde{h}%
_{4}=\widetilde{\eta }_{4}\mathring{h}_{4},\widetilde{w}_{i}=\mathring{w}%
_{i}+\ ^{\eta }w_{i},n_{i}=\mathring{n}_{i}+\ ^{\eta }n_{i}],  \notag \\
&&\ \widetilde{\Upsilon }=\widetilde{\lambda },\ ^{v}\hat{\Upsilon}%
(x^{k^{\prime }})=\ ^{v}\Lambda =\mu _{g}^{2}\ \lambda (x^{k^{\prime }})%
\mathring{h}_{4}^{-1},\widetilde{\Lambda }=\mu _{g}^{2}\ \widetilde{\lambda }%
,\tilde{\Phi}^{2}=\exp [2\varpi (x^{k^{\prime }},y^{4})]\ \mathring{h}_{3},
\label{ndefbm}
\end{eqnarray}%
where the values \ $\widetilde{\eta }_{a},\widetilde{w}_{i},\tilde{n}_{i}$
and $\varpi $ are functions of three coordinates $(x^{k^{\prime }},y^{4})$
and $\widetilde{\eta }_{i}(x^{k})$ depend only on h--coordinates. The prime
data $\mathring{g}_{i},\mathring{h}_{a},\mathring{w}_{i},\mathring{n}_{i}$
are given by coefficients depending only on $(x^{k}).$

In terms of $\eta $--functions (\ref{etad}) resulting in $h_{a}^{\ast }\neq
0 $ and $g_{i}=c_{i}e^{\psi {(x^{k})}},$ the solutions of type (\ref{qnk4d})
with $\widetilde{\Lambda }\rightarrow \widetilde{\lambda }$ and $\
_{2}n_{k^{\prime }}=0$ (we use "primed" coordinates and prime Kerr data (\ref%
{kerrbl}) and (\ref{dkerr})) can be re--written in the form%
\begin{eqnarray}
ds^{2} &=&e^{\psi (x^{k^{\prime }})}[(dx^{1^{\prime }})^{2}+(dx^{2^{\prime
}})^{2}] -  \label{nvlcmgs} \\
&& \frac{e^{2\varpi }}{4\mu _{g}^{2}\ |\widetilde{\lambda }|}\overline{A}%
[dy^{3^{\prime }}+\left( \partial _{k^{\prime }}\ ^{\eta }n(x^{i^{\prime
}})-\partial _{k^{\prime }}(\widehat{y}^{3^{\prime }}+\varphi \overline{B}/%
\overline{A})\right) dx^{k^{\prime }}]^{2}+\frac{(\varpi ^{\ast })^{2}}{\
\mu _{g}^{2}\ \lambda (x^{k^{\prime }})}(\overline{C}-\overline{B}^{2}/%
\overline{A})[d\varphi +(\partial _{i^{\prime }}\ ^{\eta }\widetilde{A}%
)dx^{i^{\prime }}]^{2},  \notag
\end{eqnarray}%
for
\begin{equation*}
\Xi =\int dy^{4}(\ ^{v}\Lambda )\partial _{4}(\tilde{\Phi}^{2})=\mu
_{g}^{2}\ \lambda (x^{k^{\prime }})\mathring{h}_{4}^{-1}\tilde{\Phi}^{2},
\end{equation*}%
with $\tilde{\Phi}^{2}/\mathring{h}_{4}$ parameterized using formulas (\ref%
{ndefbm}).\footnote{%
Hereafter we shall consider that we can approximate $\lambda (x^{k^{\prime
}})\simeq \widetilde{\lambda }=const.$} The gravitational polarizations $%
(\eta _{i},\eta _{a})$ and N--coefficients $(n_{i},w_{i})$ are computed
following formulas\
\begin{eqnarray*}
e^{\psi (x^{k})} &=&\widetilde{\eta }_{1^{\prime }}=\widetilde{\eta }%
_{2^{\prime }},\ \widetilde{\eta }_{3^{\prime }}=\frac{e^{2\varpi }}{4\mu
_{g}^{2}\ |\widetilde{\lambda }|},\ \widetilde{\eta }_{4^{\prime }}=\frac{%
(\varpi ^{\ast })^{2}}{\ \mu _{g}^{2}\ \lambda (x^{k^{\prime }})}, \\
w_{i^{\prime }} &=&\mathring{w}_{i^{\prime }}+\ ^{\eta }w_{i^{\prime
}}=\partial _{i^{\prime }}(\ ^{\eta }\widetilde{A}[\varpi ]),\ n_{k^{\prime
}}=\mathring{n}_{k^{\prime }}+\ ^{\eta }n_{k^{\prime }}=\partial _{k^{\prime
}}(-\widehat{y}^{3^{\prime }}+\varphi \overline{B}/\overline{A}+\ ^{\eta }n),
\end{eqnarray*}%
where $\ ^{\eta }\widetilde{A}(x^{k},y^{4})$ is introduced via formulas and
assumptions similar to (\ref{expconda}), for $s=1,$ and $\psi ^{\bullet
\bullet }+\psi ^{\prime \prime }=2\ \mu _{g}^{2}\ \lambda (x^{k^{\prime }}).$
For N--coefficients, the parameterizations are used (\ref{solhn}) with
$\check{\Phi}=\exp [\varpi (x^{k^{\prime }},y^{4})]\sqrt{\ |\mathring{h}%
_{3^{\prime }}|}$, when $\mathring{h}_{3^{\prime }}\mathring{h}_{4^{\prime
}}=\overline{A}\overline{C}-\overline{B}^{2}$ and
\begin{equation*}
w_{i^{\prime }}=\mathring{w}_{i^{\prime }}+\ ^{\eta }w_{i^{\prime
}}=\partial _{i^{\prime }}(\ e^{\varpi }\sqrt{|\overline{A}\overline{C}-%
\overline{B}^{2}|})/\ \varpi ^{\ast }e^{\varpi }\sqrt{|\overline{A}\overline{%
C}-\overline{B}^{2}|}=\partial _{i^{\prime }}\ ^{\eta }\widetilde{A}.
\end{equation*}%
We can take any function $\ ^{\eta }n(x^{k})$ and put $\lambda =const\neq 0$
using the corresponding re--definitions of coordinates and generating functions.

The solutions (\ref{nvlcmgs}) are valid for stationary LC--configurations
determined by off--diagonal massive gravity effects on Kerr black holes when
the new class of spacetimes have a Killing symmetry in $\partial /\partial
y^{3^{\prime }}$ and a generic dependence on three (from maximally four)
coordinates, $(x^{i^{\prime }}(r,\vartheta ),\varphi ). $ Off--diagonal
modifications are possible even for very small values of the mass parameter $%
\ \mu _{g}.$ The solutions depend on the type of generating function $\varpi
(x^{i^{\prime }},\varphi )$ we have to fix in order to satisfy certain
experimental/observational data in certain fixed systems of
reference/coordinates. Various data can be re--parameterized for an
effective $\lambda =const\neq 0.$ In such variables, we can mimic stationary
massive gravity effects by off--diagonal configurations in GR with
integration parameters which should be also fixed by imposing additional assumptions
on the symmetries of the interactions (for instance, to have an ellipsoid
configuration, see section \ref{4dellipsc} and details and discussion on
parametric Killing symmetries in Refs. \cite{ger1,ger2,vpars}).

\subsubsection{Nonholonomically induced torsion and massive gravity}

If we do not impose the LC--conditions (\ref{lcconstr}), a nontrivial source
$\ ^{\mu }\widetilde{\Lambda }=\mu _{g}^{2}\ \widetilde{\lambda }$ from
massive gravity induces stationary configuration with nontrivial d--torsion (%
\ref{dtors}). The torsion coefficients are determined by metrics of the type (%
\ref{qnk4d}) with $\widetilde{\Lambda }\rightarrow \widetilde{\lambda }$ and
parameterizations of coefficients and coordinates distinguishing the prime
data for a Kerr metric (\ref{dkerr}). Such solutions can be written in the
form {\small
\begin{eqnarray}
ds^{2} &=&e^{\psi (x^{k^{\prime }})}[(dx^{1^{\prime }})^{2}+(dx^{2^{\prime
}})^{2}]-\frac{\Phi ^{2}}{4\mu _{g}^{2}\ |\widetilde{\lambda }|}\overline{A}%
[dy^{3^{\prime }}+\left( \ _{1}n_{k^{\prime }}(x^{i^{\prime }})+\
_{2}n_{k^{\prime }}(x^{i^{\prime }})\frac{4\mu _{g}(\Phi ^{\ast })^{2}}{\Phi
^{5}}-\partial _{k^{\prime }}(\widehat{y}^{3^{\prime }}+\varphi \overline{B}/%
\overline{A})\right) dx^{k^{\prime }}]^{2}  \notag \\
&&+\frac{(\partial _{\varphi }\Phi )^{2}}{\ \mu _{g}^{2}\ \lambda
(x^{k^{\prime }})\Phi ^{2}}(\overline{C}-\overline{B}^{2}/\overline{A}%
)[d\varphi +\frac{\partial _{i^{\prime }}\Phi }{\partial _{\varphi }\Phi }%
dx^{i^{\prime }}]^{2},  \label{ofindtmg}
\end{eqnarray}%
} where we use a generating function $\Phi (x^{i^{\prime }},\varphi )$
instead of $e^{\varpi }$ and consider nonzero values of $\
_{2}n_{k}(x^{i^{\prime }}).$ We can see that nontrivial stationary
off--diagonal torsion effects may result in additional effective rotations
proportional to $\mu _{g}$ if the integration function $\ _{2}n_{k}\neq 0.$
Considering two different classes of off--diagonal solutions (\ref{ofindtmg}%
) and (\ref{nvlcmgs}), we can study if a massive gravity theory is described in terms of an
induced torsion or characterized by additional nonholonomic constraints as
in GR (with zero torsion).

It should be noted that configurations of the type (\ref{ofindtmg}) can be
constructed in various theories with noncommutative, brane,
extra--dimension, warped and trapped brane type variables in string, or
Finsler like and/or Ho\v{r}ava--Lifshits theories \cite%
{vex1,vp,vt,vgrg,vbranef} when nonholonomically induced torsion effects play
a substantial role. Those classes of solutions were constructed for
different sets of interactions constants and, for instance, for propagating
Schwarzschild and/or ellipsoid type configurations on Tau\ NUT backgrounds
etc. The off--diagonal deformations and effective polarizations of the
coefficients of the metrics correspond to a prime Kerr metric and are related to target
configuration in massive gravity.

\subsubsection{Small $f$--modifications of Kerr metrics and massive gravity}

Using the AFDM, we can construct off--diagonal solutions for superposition
of $f$--modified and massive gravity interactions. Such nonlinear effects
can be distinguished in explicit form if we consider for additional $f$%
--deformations, for instance, a "prime" solution for massive gravity/
effectively modelled in GR with source $\ ^{\mu }\Lambda =\mu _{g}^{2}\
\lambda (x^{k^{\prime }}),$ or re--defined to $\ ^{\mu }\tilde{\Lambda}=\mu
_{g}^{2}\ \tilde{\lambda}=const.$ Adding a "small" value $\ \widetilde{%
\Lambda }$ determined by $f$--modifications, we work in N--adapted frames
with an effective source $\Upsilon =\widetilde{\Lambda }+\widetilde{\lambda }
$ (see formulas (\ref{source1a}) and (\ref{source1b})). As a result, we
construct a class of off--diagonal solutions in modified $f$--gravity
generated from the Kerr black hole solution as a result of two nonholonomic
deformations
\begin{equation*}
(\mathbf{\mathring{g}},\mathbf{\mathring{N},\ }^{v}\mathring{\Upsilon}=0,%
\mathring{\Upsilon}=0)\rightarrow (\widetilde{\mathbf{g}},\widetilde{\mathbf{%
N}},\ ^{v}\widetilde{\Upsilon }=\widetilde{\lambda },\widetilde{\Upsilon }=%
\widetilde{\lambda })\rightarrow (\ ^{\varepsilon }\mathbf{g},\
^{\varepsilon }\mathbf{N,\Upsilon =\varepsilon \ }\widetilde{\Lambda }+\
^{\mu }\tilde{\Lambda},\mathbf{\Upsilon =\varepsilon \ }\widetilde{\Lambda }%
+\ ^{\mu }\tilde{\Lambda}),
\end{equation*}%
when the target data $\mathbf{g=}\ ^{\varepsilon }\mathbf{g}$ and$\ \mathbf{%
N=}\ ^{\varepsilon }\mathbf{N}$ depend on a small parameter $\varepsilon ,$ $%
0<\varepsilon \ll 1.$ For simplicity, we restrict our considerations for
solutions when $|\mathbf{\varepsilon \ }\widetilde{\Lambda }|\ll |\ ^{\mu }%
\tilde{\Lambda}|,$ i.e. consider that $f$--modifications in N--adapted
frames are much smaller than massive gravity effects (in a similar from, we
can analyze nonlinear interactions with $|\mathbf{\varepsilon \ }\widetilde{%
\Lambda }|\gg |\ ^{\mu }\tilde{\Lambda}|).$ The corresponding N--adapted
transforms are parameterized as
\begin{eqnarray}
&&[\mathring{g}_{i},\mathring{h}_{a},\mathring{w}_{i},\mathring{n}%
_{i}]\rightarrow  \label{def2} \\
&&[g_{i}=(1+\varepsilon \chi _{i})\widetilde{\eta }_{i}\mathring{g}%
_{i},h_{3}=(1+\varepsilon \chi _{3})\widetilde{\eta }_{3}\mathring{h}%
_{3},h_{4}=(1+\varepsilon \chi _{4})\widetilde{\eta }_{4}\mathring{h}_{4},\
^{\varepsilon }w_{i}=\mathring{w}_{i}+\widetilde{w}_{i}+\varepsilon
\overline{w}_{i},\ ^{\varepsilon }n_{i}=\mathring{n}_{i}+\tilde{n}%
_{i}+\varepsilon \overline{n}_{i}];  \notag \\
&&\mathbf{\Upsilon =}\ ^{\mu }\tilde{\Lambda}(1+\varepsilon \ \widetilde{%
\Lambda }/\ ^{\mu }\tilde{\Lambda});\ \ \ ^{\varepsilon }\tilde{\Phi}=\tilde{%
\Phi}(x^{k},\varphi )[1+\varepsilon \ \ ^{1}\tilde{\Phi}(x^{k},\varphi )/%
\tilde{\Phi}(x^{k},\varphi )]=\exp [\ \ ^{\varepsilon }\varpi (x^{k},\varphi
)],  \notag
\end{eqnarray}%
leading to a 4--d LC--configuration with d--metric
\begin{equation*}
ds_{4\varepsilon dK}^{2}=\epsilon _{i}(1+\varepsilon \chi _{i})e^{\psi
(x^{k})}(dx^{i})^{2}+\frac{\ \ \ ^{\varepsilon }\tilde{\Phi}^{2}}{4\ \mathbf{%
\Upsilon }}\left[ dy^{3}+(\partial _{i}\ n)dx^{i}\right] ^{2}+\ \frac{%
(\partial _{\varphi }\ \ ^{\varepsilon }\tilde{\Phi})^{2}}{\ \mathbf{%
\Upsilon }\ \ ^{\varepsilon }\tilde{\Phi}^{2}}\left[ dy^{4}+(\partial _{i}\
\ ^{\varepsilon }\check{A})dx^{i}\right] ^{2},
\end{equation*}%
for $\partial _{i}\ \ ^{\varepsilon }\check{A}=\partial _{i}\ \
^{\varepsilon }\check{A}+\varepsilon \partial _{i}\ \ ^{1}\check{A}$
determined by $\ ^{\varepsilon }\tilde{\Phi}=\tilde{\Phi}+\varepsilon \ ^{1}%
\tilde{\Phi}$ following conditions in (\ref{data4c}). The values labeled by "$%
\circ $" and "$\widetilde{}$" are taken from (\ref{ndefbm}) (for simplicity,
we omit priming of indices). The $\chi $- and $w$--values (
corresponding to a re--definition of coefficients; for simplicity, we consider $%
\varepsilon \overline{n}_{i}=0$) have to be computed to define $\varepsilon $%
--deformed LC--configurations, see formulas (\ref{zerot}) for $s=0$, as
solutions of the system (\ref{sourc1}) in the form (\ref{e1})--(\ref{e4})
for a source $\mathbf{\Upsilon =\ ^{\mu }\tilde{\Lambda}+\ }\varepsilon
\widetilde{\Lambda }.$

The deformations (\ref{def2}) of the off--diagonal solutions (\ref{nvlcmgs})
result in a new class of $\varepsilon $--deformed solutions with%
\begin{eqnarray}
\chi _{1} &=&\chi _{2}=\chi ,\mbox{ for }\partial _{11}\chi +\epsilon
_{2}\partial _{22}\chi =2\widetilde{\Lambda };  \label{edefcel} \\
\chi _{3} &=&2\ ^{1}\tilde{\Phi}/\tilde{\Phi}-\mathbf{\ }\widetilde{\Lambda }%
/\ ^{\mu }\tilde{\Lambda},\ \chi _{4}=2\partial _{4}\ ^{1}\tilde{\Phi}/%
\tilde{\Phi}-2\ ^{1}\tilde{\Phi}/\tilde{\Phi}-\widetilde{\Lambda }/\ ^{\mu }%
\tilde{\Lambda},  \notag \\
\overline{w}_{i} &=&(\frac{\partial _{i}\ ^{1}\tilde{\Phi}}{\partial _{i}%
\tilde{\Phi}}-\frac{\partial _{4}\ ^{1}\tilde{\Phi}}{\partial _{4}\tilde{\Phi%
}})\frac{\partial _{i}\tilde{\Phi}}{\partial _{4}\tilde{\Phi}}=\partial
_{i}\ \ ^{1}\check{A},\overline{n}_{i}=0,  \notag
\end{eqnarray}%
where there is not summation on index "$i"$ in the last formula and $%
\mathring{h}_{3^{\prime }}\mathring{h}_{4^{\prime }}=\overline{A}\overline{C}%
-\overline{B}^{2}.$ Such nonholonomic deformations are determined
respectively by two generating functions $\tilde{\Phi}=e^{\varpi }$ and $\
^{1}\tilde{\Phi}$ and two sources $\ ^{\mu }\tilde{\Lambda}$ and $\widetilde{%
\Lambda }.$ Putting all this together, we construct an off--diagonal generalization
of the Kerr metric via "main" massive gravity terms and additional $\varepsilon $%
--parametric $f$--modifications,
\begin{eqnarray}
ds^{2} &=&e^{\psi (x^{k^{\prime }})}(1+\varepsilon \chi (x^{k^{\prime
}}))[(dx^{1^{\prime }})^{2}+(dx^{2^{\prime }})^{2}]-  \notag \\
&&\frac{e^{2\varpi }}{4|\ ^{\mu }\tilde{\Lambda}|}\overline{A}[1+\varepsilon
(2e^{-\varpi }\ ^{1}\tilde{\Phi}-\mathbf{\ }\widetilde{\Lambda }/\ ^{\mu }%
\tilde{\Lambda})][dy^{3^{\prime }}+\left( \partial _{k^{\prime }}\ ^{\eta
}n(x^{i^{\prime }})-\partial _{k^{\prime }}(\widehat{y}^{3^{\prime
}}+\varphi \overline{B}/\overline{A})\right) dx^{k^{\prime }}]^{2}+
\label{nvlcmgse} \\
&&\frac{(\varpi ^{\ast })^{2}}{\ \ ^{\mu }\tilde{\Lambda}}(\overline{C}-%
\overline{B}^{2}/\overline{A})[1+\varepsilon (2e^{-\varpi }\partial _{4}\
^{1}\tilde{\Phi}-2e^{-\varpi }\ ^{1}\tilde{\Phi}-\widetilde{\Lambda }/\
^{\mu }\tilde{\Lambda})][d\varphi +(\partial _{i^{\prime }}\ \widetilde{A}%
+\varepsilon \partial _{i^{\prime }}\ \ ^{1}\check{A})dx^{i^{\prime }}]^{2}.
\notag
\end{eqnarray}

We can consider $\varepsilon $--deformations of the type (\ref{def2}) for (\ref%
{ofindtmg}), which allows us to generate new classes of off--diagonal
solutions with nonholonomically induced torsion determined both by massive
and $f$--modifications of GR. Such a spacetime cannot be modelled as an
effective one with anisotropic polarizations in GR.

\subsection{ Ellipsoidal 4--d deformations of the Kerr metric}

\label{4dellipsc}We provide some examples how the Kerr primary data (\ref%
{dkerr}) is nonholonomically deformed into target generic off--diagonal
solutions of vacuum and non--vacuum Einstein equations for the canonical
d--connection and/or the Levi--Civita connection.

\subsubsection{ Vacuum ellipsoidal configurations}

Let us construct a class of parametric solutions with such nonholonomic
constraints on the coefficients given by  (\ref{ofindtmg}) which transform the metrics
into effective 4--d vacuum LC--configurations of the type (\ref{vs2}). This
defines a model when $f$--modifications compensate massive gravity
deformations of a Kerr solution, with $\mathbf{\Upsilon =\ ^{\mu }\tilde{%
\Lambda}+\ }\varepsilon \widetilde{\Lambda }=0,$ and result in ellipsoidal
off--diagonal configurations in GR, where $\varepsilon =-\ ^{\mu }\tilde{%
\Lambda}\mathbf{/}\widetilde{\Lambda }\ll 1$ can be considered as an
eccentricity parameter. We find solutions for $\varepsilon $--deformations
into vacuum solutions. The ansatz for the target metrics is of the type
\begin{eqnarray}
ds^{2} &=&e^{\psi (x^{k^{\prime }})}(1+\varepsilon \chi (x^{k^{\prime
}}))[(dx^{1^{\prime }})^{2}+(dx^{2^{\prime }})^{2}]  \label{vacum4del} \\
&&-\frac{e^{2\varpi }}{4\mu _{g}^{2}|\ \widetilde{\lambda }|}\overline{A}%
[1+\varepsilon \chi _{3^{\prime }}][dy^{3^{\prime }}+\left( \partial
_{k^{\prime }}\ ^{\eta }n(x^{i^{\prime }})-\partial _{k^{\prime }}(\widehat{y%
}^{3^{\prime }}+\varphi \overline{B}/\overline{A})\right) dx^{k^{\prime
}}]^{2}  \notag \\
&&+\frac{(\partial _{4}\varpi )^{2}\eta _{4^{\prime }}}{\ \mu _{g}^{2}\
\widetilde{\lambda }}(\overline{C}-\overline{B}^{2}/\overline{A}%
)[1+\varepsilon \chi _{4^{\prime }}][d\varphi +(\partial _{i^{\prime }}\
\widetilde{A}+\varepsilon \partial _{i^{\prime }}\ \ ^{1}\check{A}%
)dx^{i^{\prime }}]^{2},  \notag
\end{eqnarray}%
when the prime metrics (\ref{nvlcmgs}) are obtained for $\eta _{4^{\prime
}}=1.$ The condition (\ref{ca1}) for $\phi =const,$ i.e. for vacuum
off--diagonal configurations, when $h_{4^{\prime }}=\ ^{0}h_{4^{\prime
}}(\partial _{4}\sqrt{|h_{3^{\prime }}|})^{2}$ (\ref{h34vacuum}), is
satisfied for $\eta _{4^{\prime }}=\overline{A}\sqrt{|\overline{B}^{2}-%
\overline{C}\overline{A}|}e^{2\varpi }$. For terms proportional to $%
\varepsilon ),$ we compute $\chi _{4^{\prime }}=(\partial _{4}\varpi
)^{-1}(1+e^{-\varpi }\chi _{3^{\prime }})$, where $\varpi (r,\vartheta
,\varphi )$ and $\chi _{3^{\prime }}(r,\vartheta ,\varphi )$ are generating
functions. We can consider as generating functions for N--coefficients any $%
\widetilde{A}(r,\vartheta ,\varphi )$ and $\ \ ^{1}\check{A}(r,\vartheta
,\varphi ),$ which for $w_{i^{\prime }}=\partial _{i^{\prime }}(\ \widetilde{%
A}+\varepsilon \ ^{1}\check{A})$ solve the LC--conditions. The
LC--conditions $\mathbf{e}_{i}\ln \sqrt{|\ h_{3}|}=0,$ $\partial
_{i}w_{j}=\partial _{j}w_{i}$ for $s=0,$ see (\ref{zerot}) can be satisfied
if we parameterize
\begin{equation*}
w_{i^{\prime }}=\partial _{i^{\prime }}\ ^{\varepsilon }\Phi /\partial
_{\varphi }\ ^{\varepsilon }\Phi =\partial _{i^{\prime }}(\ \widetilde{A}%
+\varepsilon \ ^{1}\check{A}),
\end{equation*}%
for $\ ^{\varepsilon }\Phi =\exp (\varpi +\varepsilon \chi _{3^{\prime }})$,
see discussions related to (\ref{zerota}) and (\ref{data4c}). Because $%
h_{4^{\prime }}$ for (\ref{vacum4del}) can be approximated up to $%
\varepsilon ^{2}$ to be a functional on $\ ^{\varepsilon }\Phi ,$ we can
satisfy for certain classes of generating functions $\ ^{\varepsilon }\Phi
=\ ^{\varepsilon }\tilde{\Phi}=\ ^{\varepsilon }\check{\Phi},$ see (\ref%
{explcond}), the conditions $\partial _{\varphi }w_{i^{\prime }}=\mathbf{e}%
_{i^{\prime }}\ln \sqrt{|\ h_{4}|}.$

We can chose such a generating function $\chi _{3^{\prime }},$ when the
constraint $h_{3^{\prime }}=0$ defines a stationary rotoid configuration
(different from to the ergo sphere for the Kerr solutions): Prescribing
\begin{equation}
\chi _{3^{\prime }}=2\underline{\zeta }\sin (\omega _{0}\varphi +\varphi
_{0}),  \label{chi3prim}
\end{equation}%
for constant parameters $\underline{\zeta },\omega _{0}$ and $\varphi _{0},$
and introducing the values
\begin{eqnarray*}
\overline{A}(r,\vartheta )[1+\varepsilon \chi _{3^{\prime }}(r,\vartheta
,\varphi )] &=&\widehat{A}(r,\vartheta ,\varphi )=-\Xi ^{-1}(\widehat{\Delta
}-a^{2}\sin ^{2}\vartheta ), \\
\widehat{\Delta }(r,\varphi ) &=&r^{2}-2m(\varphi )+a^{2},
\end{eqnarray*}%
as $\varepsilon $--deformations of Kerr coefficients (\ref{kerrcoef}), we
get an effective "anisotropically polarized" mass
\begin{equation}
m(\varphi )=m_{0}/\left( 1+\varepsilon \underline{\zeta }\sin (\omega
_{0}\varphi +\varphi _{0})\right) .  \label{polarm}
\end{equation}%
The condition $h_{3}=0,$ i.e. $\ ^{\varphi }\Delta (r,\varphi ,\varepsilon
)=a^{2}\sin ^{2}\vartheta ,$ results in an ellipsoidal "deformed horizon" $%
r(\vartheta ,\varphi )=m(\varphi )+\left( m^{2}(\varphi )-a^{2}\sin
^{2}\vartheta \right) ^{1/2}$. For $a=0$, this is just the parametric
formula for an ellipse with eccentricity $\varepsilon ,$%
\begin{equation*}
r_{+}=\frac{2m_{0}}{1+\varepsilon \underline{\zeta }\sin (\omega _{0}\varphi
+\varphi _{0})}.
\end{equation*}

If the anholonomy coefficients (\ref{anhrel1}) computed for (\ref{vacum4del}%
) are not trivial for such $w_{i}$ and $\ n_{k}=\ _{1}n_{k},$ the generated
solutions cannot be diagonalized via coordinate transformations. The
corresponding 4--d spacetimes have one Killing symmetry with respect to $\partial
/\partial y^{3^{\prime }}.$ For small $\varepsilon ,$ the singularity at $%
\Xi =0$ is "hidden" under ellipsoidal deformed horizons if $m_{0}\geq a.$
Both the event horizon, $r_{+}=m(\varphi
)+\left( m^{2}(\varphi )-a^{2}\sin ^{2}\vartheta \right) ^{1/2}$, and the
Cauchy horizon, $r_{-}=m(\varphi )-\left( m^{2}(\varphi )-a^{2}\sin
^{2}\vartheta \right) ^{1/2}$, are $\varphi $--deformed and are effectively embedded into an
off--diagonal background determined by the N--coefficients. In some sense, such
configurations determine Kerr-like black hole solutions with additional
dependencies on the variable $\varphi $ of certain diagonal and off--diagonal
coefficients of the metric.  For $a=0,$ but $\varepsilon \neq 0,$ we get
ellipsoidal deformations of the Schwarzschild black holes (see \cite{vex1}
and references therein on the stability and interpretation of such solutions
with both commutative and/or noncommutative parameters). Such an
interpretation is not possible for "non-small" $N$--deformations of the Kerr
metric. In general, it is not clear what physical importance such
target exact solutions may have even if they may be defined to preserve the Levi--Civita
configurations.

The eccentricity $\varepsilon =-\mathbf{\widetilde{\lambda }/}\widetilde{%
\Lambda }\ll 1$ depends both on massive gravity and $f$--modifications
encoded into effective cosmological constants. We proved that via
nonholonomic deformations it is possible to transform non--vacuum solutions
with an effective locally anisotropically cosmological constant into effective
off--diagonal vacuum configurations in GR. If the generating functions are
prescribed to possess necessarily certain type of smooth conditions, the solutions are
similar to certain Kerr black holes with ellipsoidal $\varepsilon $%
--deformed horizons and embedded self--consistently into non--trivial
off--diagonal vacuum configurations. Polarizations of such vacuums encode
massive gravity contributions and $f$--modifications.

\subsubsection{Ellipsoid Kerr -- de Sitter configurations}

We construct a subclass of solutions (\ref{nvlcmgse}) with rotoid
configurations if we constrain $\chi _{3}$ appearing in the $\varepsilon $%
--deformations in (\ref{edefcel}) to be of the form
\begin{equation*}
\chi _{3}=2\ ^{1}\tilde{\Phi}/\tilde{\Phi}-\mathbf{\ }\widetilde{\Lambda }/\
^{\mu }\tilde{\Lambda}=2\underline{\zeta }\sin (\omega _{0}\varphi +\varphi
_{0})
\end{equation*}%
which is similar to (\ref{chi3prim}). Expressing $\ ^{1}\tilde{\Phi}%
=e^{\varpi }[\mathbf{\ }\widetilde{\Lambda }/2\ ^{\mu }\tilde{\Lambda}+%
\underline{\zeta }\sin (\omega _{0}\varphi +\varphi _{0})],$ for $\tilde{\Phi%
}=e^{\varpi },$ we generate a class of generic off--diagonal metrics associated with the
ellipsoid Kerr -- de Sitter configurations {\small
\begin{eqnarray}
&&ds^{2}=e^{\psi (x^{k^{\prime }})}(1+\varepsilon \chi (x^{k^{\prime
}}))[(dx^{1^{\prime }})^{2}+(dx^{2^{\prime }})^{2}]-  \notag \\
&&\frac{e^{2\varpi }}{4|\ ^{\mu }\tilde{\Lambda}|}\overline{A}%
[1+2\varepsilon \underline{\zeta }\sin (\omega _{0}\varphi +\varphi
_{0})][dy^{3^{\prime }}+\left( \partial _{k^{\prime }}\ ^{\eta
}n(x^{i^{\prime }})-\partial _{k^{\prime }}(\widehat{y}^{3^{\prime
}}+\varphi \overline{B}/\overline{A})\right) dx^{k^{\prime }}]^{2}+
\label{elkdscon} \\
&&\frac{(\varpi ^{\ast })^{2}}{\ ^{\mu }\tilde{\Lambda}}(\overline{C}-%
\overline{B}^{2}/\overline{A})[1+\varepsilon (\partial _{4}\varpi \mathbf{\ }%
\widetilde{\Lambda }/\widetilde{\lambda }+2\partial _{4}\varpi \underline{%
\zeta }\sin (\omega _{0}\varphi +\varphi _{0})+2\omega _{0}\mathbf{\ }%
\underline{\zeta }\cos (\omega _{0}\varphi +\varphi _{0}))][d\varphi
+(\partial _{i^{\prime }}\ \widetilde{A}+\varepsilon \partial _{i^{\prime
}}\ \ ^{1}\check{A})dx^{i^{\prime }}]^{2}.  \notag
\end{eqnarray}%
}

Such metrics have a Killing symmetry in $\partial /\partial y^{3}$ and
are completely defined by a generating function $\varpi (x^{k^{\prime }},\varphi
)$ and the sources $\ ^{\mu }\tilde{\Lambda}=\mu _{g}^{2}\ \lambda $ and $\
\widetilde{\Lambda }.$ They define $\varepsilon $--deformations of Kerr --
de Sitter black holes into ellipsoid configurations with effective
(polarized) cosmological constants determined by the constants in massive
gravity and $f$--modifications. If the LC--conditions are satisfied, such
metrics can be modelled in GR.

\subsection{Extra dimension off--diagonal (non) massive modifications of the
Kerr solutions}

Various classes of generic off--diagonal deformations of the Kerr metric
into higher dimensional exact solutions can be constructed. The explicit
geometric and physical properties depend on the type of additional
generating and integration functions and (non) vacuum configurations and
(non) zero sources we consider. Let us analyze a series of 6--d and 8--d
solutions encoding possible higher dimensional interactions with effective
cosmological constants, nontrivial massive gravity contributions, $f$%
--modifications and certain analogies to Finsler gravity models.

\subsubsection{6--d deformations with nontrivial cosmological constant}

Off--diagonal extra dimensional gravitational interactions modify a Kerr
metric for any nontrivial cosmological constant in 6--d.\footnote{%
In a similar form we can generalize the constructions in 8--d gravity.} Such
higher dimensional Kerr -- de Sitter configurations can be generated by
nonholonomic deformations $(\mathbf{\mathring{g}},\mathbf{\mathring{N},\ }%
^{v}\mathring{\Upsilon}=0,\mathring{\Upsilon}=0)\rightarrow (\widetilde{%
\mathbf{g}},\widetilde{\mathbf{N}}\mathbf{,\ }^{v}\widetilde{\Upsilon }%
=\Lambda ,\widetilde{\Upsilon }=\Lambda ,\mathbf{\ }^{v_{1}}\widetilde{%
\Upsilon }=\Lambda )$. The solutions are not stationary, are characterized by a
Killing symmetry in $\partial /\partial y^{5}$ and can be parameterized in
the form%
\begin{eqnarray}
ds^{2} &=&e^{\psi (x^{k^{\prime }})}[(dx^{1^{\prime }})^{2}+(dx^{2^{\prime
}})^{2}] -\frac{e^{2\varpi }}{4\Lambda }\overline{A}[dy^{3^{\prime }}+\left(
\partial _{k^{\prime }}\ ^{\eta }n(x^{i^{\prime }})-\partial _{k^{\prime }}(%
\widehat{y}^{3^{\prime }}+\varphi \overline{B}/\overline{A})\right)
dx^{k^{\prime }}]^{2}+  \label{6dks} \\
&&\frac{(\partial _{\varphi }\varpi )^{2}}{\ \Lambda }(\overline{C}-%
\overline{B}^{2}/\overline{A})[d\varphi +(\partial _{i^{\prime }}\ ^{\eta }%
\widetilde{A})dx^{i^{\prime }}]^{2} +\frac{\ ^{1}\tilde{\Phi}^{2}}{4\
\Lambda }\left[ dy^{5}+(\partial _{\tau }\ ^{1}n)du^{\tau }\right] ^{2}+\
\frac{(\partial _{6}\ ^{1}\tilde{\Phi})^{2}}{\ \Lambda \ ^{1}\tilde{\Phi}^{2}%
}\left[ dy^{6}+(\partial _{\tau }\ ^{1}\check{A})du^{\tau }\right] ^{2}.
\notag
\end{eqnarray}%
The "primary" data $\overline{A},\overline{B},\overline{C}$ is described by (%
\ref{kerrcoef}) and the generating functions
\begin{eqnarray*}
\varpi &=&\varpi (x^{k^{\prime }},\varphi ),\ ^{1}\tilde{\Phi}(u^{\beta
},y^{6})=\ ^{1}\tilde{\Phi}(x^{k^{\prime }},t,\varphi ,y^{6});\ ^{\eta }n=\
^{\eta }n(x^{i^{\prime }}), \\
\ ^{1}n &=&\ ^{1}n(u^{\beta },y^{6});\ ^{\eta }\widetilde{A}=\ ^{\eta }%
\widetilde{A}(x^{k^{\prime }},\varphi ),\ ^{1}\check{A}=\ ^{1}\check{A}%
(u^{\beta },y^{6}),
\end{eqnarray*}%
subjected to the LC--conditions and integrability conditions.

We can "extract" ellipsoid configurations for a subclass of metrics with
"additional" $\varepsilon $--deformations,
\begin{eqnarray*}
ds^{2} &=&e^{\psi (x^{k^{\prime }})}[(dx^{1^{\prime }})^{2}+(dx^{2^{\prime
}})^{2}]-  \notag \\
&& \frac{e^{2\varpi }}{4\Lambda }\overline{A}[1+2\varepsilon \underline{%
\zeta }\sin (\omega _{0}\varphi +\varphi _{0})] [dy^{3^{\prime }}+\left(
\partial _{k^{\prime }}\ ^{\eta }n(x^{i^{\prime }})-\partial _{k^{\prime }}(%
\widehat{y}^{3^{\prime }}+\varphi \overline{B}/\overline{A})\right)
dx^{k^{\prime }}]^{2}+ \\
&& \frac{(\partial _{\varphi }\varpi )^{2}}{\Lambda }(\overline{C}-\overline{%
B}^{2}/\overline{A}) [1+\varepsilon (2\partial _{4}\varpi \underline{\zeta }%
\sin (\omega _{0}\varphi +\varphi _{0})+2\omega _{0}\underline{\zeta }\cos
(\omega _{0}\varphi +\varphi _{0}))][d\varphi +(\partial _{i^{\prime }}\
^{\eta }\widetilde{A})dx^{i^{\prime }}]^{2} \\
&&+\frac{\ ^{1}\tilde{\Phi}^{2}}{4\ \Lambda }\left[ dy^{5}+(\partial _{\tau
}\ ^{1}n)du^{\tau }\right] ^{2}+\ \frac{(\partial _{6}\ ^{1}\tilde{\Phi})^{2}%
}{\ \Lambda \ ^{1}\tilde{\Phi}^{2}}\left[ dy^{6}+(\partial _{\tau }\ ^{1}%
\check{A})du^{\tau }\right] ^{2}.
\end{eqnarray*}
For small values of eccentricity $\varepsilon ,$ such metrics describe
"slightly" deformed Kerr black holes embedded self--consistently into a
generic off--diagonal 6--d spacetime. In general, extra dimensions are not
compactified. Nevertheless, imposing additional constraints on the generating
functions $\ ^{1}\tilde{\Phi},\ ^{1}n,^{1}\check{A},$ we can construct
warped/ trapped configurations as in brane gravity models and
generalizations, see similar examples in \cite{vex3,vsingl1,vgrg,vbranef}.

\subsubsection{8--d deformations and Finsler like configurations}

Next, we generate a 8--d metric with nontrivial induced torsion describing
nonholonomic deformations $(\mathbf{\mathring{g}},\mathbf{\mathring{N},\ }%
^{v}\mathring{\Upsilon}=0,\mathring{\Upsilon}=0)\rightarrow (\widetilde{%
\mathbf{g}},\widetilde{\mathbf{N}}\mathbf{,\ }^{v}\widetilde{\Upsilon }%
=\Lambda ,\widetilde{\Upsilon }=\Lambda ,\mathbf{\ }^{v_{1}}\widetilde{%
\Upsilon }=\Lambda ,\mathbf{\ }^{v_{2}}\widetilde{\Upsilon }=\Lambda )$. A
similar 4--d example is given by (\ref{ofindtmg}) but here we use a
different source (in this subsection, we take the source as a cosmological
constant $\Lambda $ in 8--d). This class of solutions is parameterized in
the form {\small
\begin{eqnarray}
&& ds^{2} =e^{\psi (x^{k^{\prime }})}[(dx^{1^{\prime }})^{2}+(dx^{2^{\prime
}})^{2}] -\frac{\Phi ^{2}}{4\Lambda }\overline{A}[dy^{3^{\prime }}+\left( \
_{1}n_{k^{\prime }}(x^{i^{\prime }})+\ _{2}n_{k^{\prime }}(x^{i^{\prime }})%
\frac{4\mu _{g}(\Phi ^{\ast })^{2}}{\Phi ^{5}}-\partial _{k^{\prime }}(%
\widehat{y}^{3^{\prime }}+\varphi \overline{B}/\overline{A})\right)
dx^{k^{\prime }}]^{2}  \notag \\
&&+\frac{(\partial _{\varphi }\Phi )^{2}}{\ \Lambda \Phi ^{2}}(\overline{C}-%
\overline{B}^{2}/\overline{A})[d\varphi +\frac{\partial _{i^{\prime }}\Phi }{%
\partial _{\varphi }\Phi }dx^{i^{\prime }}]^{2} +\frac{\ ^{1}\tilde{\Phi}^{2}%
}{4\ \Lambda }\left[ dy^{5}+(\partial _{\tau }\ ^{1}n)du^{\tau }\right]
^{2}+\ \frac{(\partial _{6}\ ^{1}\tilde{\Phi})^{2}}{\ \Lambda \ ^{1}\tilde{%
\Phi}^{2}}\left[ dy^{6}+(\partial _{\tau }\ ^{1}\check{A})du^{\tau }\right]
^{2}  \notag \\
&&+\frac{\ ^{2}\tilde{\Phi}^{2}}{4\ \Lambda }\left[ dy^{7}+(\partial _{\tau
_{1}}\ ^{2}n)du^{\tau _{1}}\right] ^{2}+\ \frac{(\partial _{8}\ ^{2}\tilde{%
\Phi})^{2}}{\ \Lambda \ ^{2}\tilde{\Phi}^{2}}\left[ dy^{8}+(\partial _{\tau
_{1}}\ ^{2}\check{A})du^{\tau _{1}}\right] ^{2},  \label{8dfd}
\end{eqnarray}%
} where the generating functions are chosen
\begin{eqnarray}
\Phi &=&\Phi (x^{k^{\prime }},\varphi ),\ ^{1}\tilde{\Phi}(u^{\beta
},y^{6})=\ ^{1}\tilde{\Phi}(x^{k^{\prime }},t,\varphi ,y^{6}),\ \ ^{2}\tilde{%
\Phi}(u^{\beta _{1}},y^{8})=\ ^{2}\tilde{\Phi}(x^{k^{\prime }},t,\varphi
,y^{5},y^{6},y^{8});  \label{genf8fd} \\
\ ^{1}n &=&\ ^{1}n(u^{\beta },y^{6}),\ ^{2}n=\ ^{2}n(u^{\beta _{1}},y^{8}),
\notag \\
\ ^{\eta }\widetilde{A} &=&\ ^{\eta }\widetilde{A}(x^{k^{\prime }},\varphi
),\ ^{1}\check{A}=\ ^{1}\check{A}(x^{k^{\prime }},t,\varphi ,y^{6}),\ ^{2}%
\check{A}=\ ^{2}\check{A}(x^{k^{\prime }},t,\varphi ,y^{5},y^{6},y^{8}).
\notag
\end{eqnarray}

The generating functions for the class of solutions (\ref{8dfd}) are chosen
in a form when the nonholonomically induced torsion (\ref{dtors}) is
effectively modeled on a 4--d pseudo--Riemannian spacetime but on the higher
shells $s=1$ and $s=2$ the torsion fields are zero. We can generate extra
dimensional torsion N--adapted coefficients if nontrivial integration
functions of the type $\ _{2}n_{k^{\prime }}(x^{i^{\prime }})$ are considered
for the higher dimensions.

Metrics of type (\ref{8dfd}) can be re--parameterized to define exact
solutions in the so--called Einstein--Finsler gravity and fractional
derivative modifications constructed on tangent bundles to Lorentz
manifolds, see details in Refs. \cite{vacarfinslcosm,vgrg,vbranef,vfracrf}
and following different Finsler, or fractional \ models, \cite%
{stavr,mavr,castro,calcagni,gratia}. For Finsler like theories, we have to
consider $y^{5},y^{6},y^{7},y^{8}$ as fiber coordinates for a tangent bundle
with local coordinates $x^{i^{\prime }},y^{3^{\prime }},\varphi $ when the $%
\ ^{1}v+\ ^{2}v$ coefficients of the metric and other geometric/physical
objects can be transformed into standard ones in Finsler geometry via frame
and coordinate transformations. In some sense, Finsler-like theories with small
corrections to GR are extra--dimensional ones with "velocity/momentum"
coordinates and with low "speed/energy" nonlinear corrections.

Finally we note that the class of metrics (\ref{8dfd}) contains a subclass
of the 6d$\rightarrow $8d generalization of (\ref{6dks}) to those configurations with
zero torsion if we choose $\Phi =e^{2\varpi }$ and impose on the N--coefficients
respective constraints which are necessary for selecting LC--configurations.

\subsubsection{Kerr massive deformations and vacuum extra dimensions}

In this subsection, we momentarily return to the vacuum ellipsoid solutions (%
\ref{vacum4del}) and extend the metric to extra dimensions when the source
is of type $\mathbf{\Upsilon =\widetilde{\lambda }+\ }\varepsilon (%
\widetilde{\Lambda }+\Lambda )=0,$ $\ ^{\mu }\tilde{\Lambda}\mathbf{=}\mu
_{g}^{2}|\ \lambda |\mathbf{,}$ and result in ellipsoidal off--diagonal
configurations in GR, where $\varepsilon =-\ ^{\mu }\tilde{\Lambda}\mathbf{/(%
}\widetilde{\Lambda }+\Lambda )\ll 1$ can be considered as an eccentricity
parameter. We can construct models of off--diagonal extra dimensional
interactions when the $f$--modifications $\widetilde{\Lambda }$ compensate an
extra dimensional contribution via the effective constant $\widetilde{\Lambda }$
and which are related to the configurations of massive gravity deformations of a Kerr
solution. We select a subclass of solutions for $\varepsilon $--deformations
of the vacuum solutions and described by the ansatz for the target metrics {\small
\begin{eqnarray}
ds^{2} &=&e^{\psi (x^{k^{\prime }})}(1+\varepsilon \chi (x^{k^{\prime
}}))[(dx^{1^{\prime }})^{2}+(dx^{2^{\prime }})^{2}]-\frac{e^{2\varpi }}{4\
^{\mu }\tilde{\Lambda}}\overline{A}[1+\varepsilon \chi _{3^{\prime
}}][dy^{3^{\prime }}+\left( \partial _{k^{\prime }}\ ^{\eta }n(x^{i^{\prime
}})-\partial _{k^{\prime }}(\widehat{y}^{3^{\prime }}+\varphi \overline{B}/%
\overline{A})\right) dx^{k^{\prime }}]^{2} +  \notag \\
&&\frac{(\partial _{4}\varpi )^{2}\eta _{4^{\prime }}}{\ ^{\mu }\tilde{%
\Lambda}}(\overline{C}-\overline{B}^{2}/\overline{A})[1+\varepsilon \chi
_{4^{\prime }}][d\varphi +(\partial _{i^{\prime }}\ \widetilde{A}%
+\varepsilon \partial _{i^{\prime }}\ \ ^{1}\check{A})dx^{i^{\prime }}]^{2}+%
\frac{\ ^{1}\tilde{\Phi}^{2}}{4(\ \widetilde{\Lambda }+\Lambda )}\left[
dy^{5}+(\partial _{\tau }\ ^{1}n)du^{\tau }\right] ^{2} +  \label{kmasedvac}
\\
&&\frac{(\partial _{6}\ ^{1}\tilde{\Phi})^{2}}{(\ \widetilde{\Lambda }%
+\Lambda )\ ^{1}\tilde{\Phi}^{2}}\left[ dy^{6}+(\partial _{\tau }\ ^{1}%
\check{A})du^{\tau }\right] ^{2}+\frac{\ ^{2}\tilde{\Phi}^{2}}{4\ (%
\widetilde{\Lambda }+\Lambda )}\left[ dy^{7}+(\partial _{\tau _{1}}\
^{2}n)du^{\tau _{1}}\right] ^{2}+\ \frac{(\partial _{8}\ ^{2}\tilde{\Phi}%
)^{2}}{\ (\ \widetilde{\Lambda }+\Lambda )\ ^{2}\tilde{\Phi}^{2}}\left[
dy^{8}+(\partial _{\tau _{1}}\ ^{2}\check{A})du^{\tau _{1}}\right] ^{2}.
\notag
\end{eqnarray}%
} The extra-dimensions components of this metric are generated by the functions $%
\ ^{1}\tilde{\Phi},$ $^{2}\tilde{\Phi}$ and the N--coefficients similarly to (%
\ref{8dfd}) but with modified effective sources in the extra dimensions, $\Lambda
\rightarrow \ \widetilde{\Lambda }+\Lambda .$ This result shows that extra
dimensions can mimic the $\varepsilon $--deformations in order to compensate
contributions from the $f$--modifications and even effective vacuum
configurations of the 4--d horizontal part.  In general, vacuum metrics (\ref%
{kmasedvac}) encode extra-dimensions modifications/ polarizations of the physical
constants and coefficients of the metrics under nonlinear polarizations of an
effective 8-d vacuum distinguishing the 4--d nonholonomic configurations and
massive gravity contributions. Extra-dimensions and $f$--modified
contributions are described by terms proportional to the eccentricity $%
\varepsilon .$

\subsubsection{Extra dimension massive ellipsoid Kerr -- de Sitter
configurations}

Combining the solutions (\ref{elkdscon}) \ and (\ref{8dfd}), we construct a
class of non--vacuum 8--d solutions with rotoid configurations if we
constrain $\chi _{3}$ in the $\varepsilon $--deformations (for 4--d, see a
similar formula (\ref{edefcel})) to be of the form
\begin{equation*}
\chi _{3}=2\ ^{1}\tilde{\Phi}/\tilde{\Phi}-\mathbf{\ }(\ \widetilde{\Lambda }%
+\Lambda )/\ ^{\mu }\tilde{\Lambda}=2\underline{\zeta }\sin (\omega
_{0}\varphi +\varphi _{0}).
\end{equation*}%
We reexpress $\ ^{1}\tilde{\Phi}=e^{\varpi }[\mathbf{\ }(\ \widetilde{%
\Lambda }+\Lambda )/2\ ^{\mu }\tilde{\Lambda}+\underline{\zeta }\sin (\omega
_{0}\varphi +\varphi _{0})],$ for $\tilde{\Phi}=e^{\varpi }$ and (\ref%
{genf8fd}), and generate a class of generic off--diagonal exra dimensional
metrics for ellipsoid Kerr -- de Sitter configurations {\small
\begin{eqnarray*}
&&ds^{2}=e^{\psi (x^{k^{\prime }})}(1+\varepsilon \chi (x^{k^{\prime
}}))[(dx^{1^{\prime }})^{2}+(dx^{2^{\prime }})^{2}]- \\
&&\frac{e^{2\varpi }}{4|\ ^{\mu }\tilde{\Lambda}|}\overline{A}%
[1+2\varepsilon \underline{\zeta }\sin (\omega _{0}\varphi +\varphi
_{0})][dy^{3^{\prime }}+\left( \partial _{k^{\prime }}\ ^{\eta
}n(x^{i^{\prime }})-\partial _{k^{\prime }}(\widehat{y}^{3^{\prime
}}+\varphi \overline{B}/\overline{A})\right) dx^{k^{\prime }}]^{2}+ \\
&&\frac{(\varpi ^{\ast })^{2}}{\ \ ^{\mu }\tilde{\Lambda}}(\overline{C}-%
\overline{B}^{2}/\overline{A})[1+\varepsilon (\partial _{4}\varpi \frac{\
\widetilde{\Lambda }+\Lambda }{\ ^{\mu }\tilde{\Lambda}}+2\partial
_{4}\varpi \underline{\zeta }\sin (\omega _{0}\varphi +\varphi _{0})+2\omega
_{0}\mathbf{\ }\underline{\zeta }\cos (\omega _{0}\varphi +\varphi
_{0}))][d\varphi +(\partial _{i^{\prime }}\ \widetilde{A}+\varepsilon
\partial _{i^{\prime }}\ \ ^{1}\check{A})dx^{i^{\prime }}]^{2} \\
&&+\frac{\ ^{1}\tilde{\Phi}^{2}}{4\ (\ \widetilde{\Lambda }+\Lambda )}\left[
dy^{5}+(\partial _{\tau }\ ^{1}n)du^{\tau }\right] ^{2}+\ \frac{(\partial
_{6}\ ^{1}\tilde{\Phi})^{2}}{\ (\ \widetilde{\Lambda }+\Lambda )\ ^{1}\tilde{%
\Phi}^{2}}\left[ dy^{6}+(\partial _{\tau }\ ^{1}\check{A})du^{\tau }\right]
^{2} \\
&&+\frac{\ ^{2}\tilde{\Phi}^{2}}{4\ (\widetilde{\Lambda }+\Lambda )}\left[
dy^{7}+(\partial _{\tau _{1}}\ ^{2}n)du^{\tau _{1}}\right] ^{2}+\ \frac{%
(\partial _{8}\ ^{2}\tilde{\Phi})^{2}}{\ (\ \widetilde{\Lambda }+\Lambda )\
^{2}\tilde{\Phi}^{2}}\left[ dy^{8}+(\partial _{\tau _{1}}\ ^{2}\check{A}%
)du^{\tau _{1}}\right] ^{2}.
\end{eqnarray*}%
}

Such non--vacuum solutions can be also modelled for Einstein--Finsler spaces
if the extra-dimension coordinates are treated as velocity/momentum ones. The
metrics possess a respective Killing symmetry in $\partial /\partial y^{7}.$
They define $\varepsilon $--deformations of Kerr -- de Sitter black holes
into ellipsoid configurations with effective cosmological constants
determined by the constants in massive gravity, $f$--modifications and extra
dimensional contributions.

\section{ Concluding Remarks}

\label{s5}

In this work, we have elaborated the anholonomic frame deformation
method, AFDM, in constructing exact solutions in gravity theories, which we
formulated and developed in \cite{vpars,vex1,vex2,vex3,veym}, see also
references therein. The method is based on a general decoupling property of
the gravitational field equations which is possible for certain classes of
noholonomic $2+2+...$ splitting of the spacetime dimensions. Such solutions are
generic off--diagonal, with zero or non--zero torsion structure, and may
depend on all (higher dimensions, or 4--d) spacetime coordinates. In the
simplest form, the constructions can be performed by using an "auxiliary"
metric-compatible connection which is constructed along  with the
"standard" Levi--Civita connection and from the same metric structure. Both
connections are related via a distortion tensor which is completely determined by
the coefficients of the metric and the frame splitting. After a class of
off--diagonal solutions is constructed in general, we can impose certain
conditions on the structure of the nonholonomic frames,  when the coefficients of both the
auxiliary and standard connections are the same, and we can extract solutions
with zero torsion, for instance, in general relativity theory.

In general form, the off--diagonal metrics and nonlinear and linear
connections constructed following the AFDM method depend on various classes
of generating and integration functions, certain symmetry parameters and on
possible nonzero sources and/or (polarized) cosmological constants. This is
possible because in our approach the (generalized/modified) Einstein
equations are transformed (after choosing the corresponding ansatz for the metrics) into systems
of nonlinear partial differential equations which can be integrated in a very
general form and depending on certain classes of generating/integration
functions. This is different from the case of a diagonal ansatz, for instance,
for the Schwarzschild metric when the gravitational field equations
transform into a system of nonlinear ordinary differential equations
depending on certain integration constants. We can construct chains of nonholonomic
frame deformations in order to transform a given primary metric (it may be
an exact solution, or not, in a gravity theory) into other classes of target
metrics and which can be fixed to be exact solutions in a "metric compatible"
gravity theory. From a formal point of view, the chains' metrics can correspond to spaces with
nontrivial topology,  have a singular/stochastic/evolution etc behaviour and various
types of horizons, symmetries and boundary conditions. In general, it is not
possible to formulate some uniqueness property or limiting/ asymptotic conditions.
Certain geometric data and physical information of "intermediary" metrics
is encoded, step by step, into the target metrics. We can impose certain
nonholonomic constraints on such integral varieties in order to relate a new
class of target metric solutions to some well--defined primary metrics. However, it
is not clear what physical importance these "very general" classes of target metric
exact solutions may have.

In a series of works \cite{vp,vt,vsingl1,vgrg} (see details and references
in \cite{veym}) we studied various examples. When using the AFDM we can
construct locally anisotropic black ellipsoid/hole, spinning and/or
solitonic spaces etc. Certain configurations seem to be stable \cite{vex1}
and mantain, for instance, the main properties of the Schwarzschild metric but
for small rotoid deformations. \vskip5pt

The goal of this article was fourfold:

\begin{enumerate}
\item to elaborate the AFDM in a form which allows us to construct generic
off--diagonal solutions with Killing symmetries and the generalizations to
non--Killing configurations using extensions to higher dimensions and
so--called "vertical" conformal factors;

\item to study off--diagonal modifications of the Kerr metric under massive
gravity and $f$--modified nonlinear interactions, via higher dimensions, and
state the conditions when such configurations can be modelled as effective
ones in general gravity, or via nonholonomically induced torsion fields etc;

\item to show how the well--known and physically important exact solution
for the Kerr black hole can be constructed, for some special class-types of
integration functions, following the AFDM; and

\item to provide certain examples when the solutions in point 2 can be
generalized to various vacuum and non--vacuum configurations with
ellipsoidal symmetries.
\end{enumerate}

In some cases of rotoidal deformations with small eccentricity parameter, we
have been able to prove that the physical properties of the primeval metrics
are preserved but with certain effective polarizations of the physical constants
and deformation to ellipsoidal configurations. It is possible to construct
exact solutions for very general off--diagonal deformations (not depending
of small parameters) but the physical properties are not clear if, for
instance, additional smooth, symmetry, Cauchy and/or boundary conditions are
not imposed. A very important property is that off--diagonal nonlinear
gravitational interactions can mimic effective modified gravity theories,
with anisotropies and re--scalings,  which can find applications in  modern cosmology
and/or elaborate new models of quantum gravity \cite{vgrg,odints1,vepl}.

\vskip5pt

\textbf{Acknowledgments:\ } The work of TG and SV is partially supported by
the Program IDEI, PN-II-ID-PCE-2011-3-0256. SV wrote a part of this
article during a recent visit at TH-CERN. He is grateful for important
discussions, support and collaboration to S. Basilakos, S. Capozziello, E.
Elizalde, N. Mavromatos, D. Singleton and P. C. Stavrinos,

\appendix

\setcounter{equation}{0} \renewcommand{\theequation}
{A.\arabic{equation}} \setcounter{subsection}{0}
\renewcommand{\thesubsection}
{A.\arabic{subsection}}

\section{ The Conditions for Zero Torsion}

\label{zt} We can consider frame transformations to the N--coefficients and the ansatz (%
\ref{ansk}) when all coefficients of a nonholonomically induced torsion (\ref%
{dtors}) are zero and $\ _{\shortmid }\Gamma _{\ \alpha _{s}\beta
_{s}}^{\gamma _{s}}=\widehat{\mathbf{\Gamma }}_{\ \alpha _{s}\beta
_{s}}^{\gamma _{s}}.$ For simplicity, we analyze such conditions for the
shell $s=0,$ i.e. for 4--d spacetimes.

In N--adapted frames, the coefficients of d--torsion (\ref{dtors}) are $%
\widehat{T}_{\ jk}^{i}=\widehat{L}_{jk}^{i}-\widehat{L}_{kj}^{i}=0,~\widehat{%
T}_{\ ja}^{i}=\widehat{C}_{jb}^{i}=0,~\widehat{T}_{\ bc}^{a}=\ \widehat{C}%
_{bc}^{a}-\ \widehat{C}_{cb}^{a}=0$ for any ansatz (\ref{ansk}). The
nontrivial coefficients are $\widehat{T}_{aj}^{c}=\widehat{L}%
_{aj}^{c}-e_{a}(N_{j}^{c})$ and $\widehat{T}_{\ ji}^{a}=-\Omega _{\ ji}^{a}.$
The values
\begin{eqnarray*}
\widehat{L}_{bi}^{a} &=&\partial _{b}N_{i}^{a}+\frac{1}{2}h^{ac}(\partial
_{i}h_{bc}-N_{i}^{e}\partial _{e}h_{bc}-h_{dc}\partial
_{b}N_{i}^{d}-h_{db}\partial _{c}N_{i}^{d}), \\
\widehat{T}_{aj}^{c} &=&\frac{1}{2}h^{ac}(\partial
_{i}h_{bc}-N_{i}^{e}\partial _{e}h_{bc}-h_{dc}\partial
_{b}N_{i}^{d}-h_{db}\partial _{c}N_{i}^{d}).
\end{eqnarray*}%
are computed for $N_{i}^{3}=n_{i}(x^{k},y^{4}),N_{i}^{4}=w_{i}(x^{k},y^{4});$
$h_{bc}=diag[h_{3}(x^{k},y^{4}),h_{4}(x^{k},y^{4})];$ $\
h^{ac}=diag[(h_{3})^{-1},(h_{4})^{-1}]$. We have
\begin{equation*}
\widehat{T}_{bi}^{3}=\frac{1}{2}h^{3c}(\partial _{i}h_{bc}-N_{i}^{e}\partial
_{e}h_{bc}-h_{dc}\partial _{b}N_{i}^{d}-h_{db}\partial _{c}N_{i}^{d})=\frac{1%
}{2h_{3}}(\partial _{i}h_{b3}-w_{i}\partial _{4}h_{b3}-h_{3}\partial
_{b}n_{i}),
\end{equation*}%
i.e. $\widehat{T}_{3i}^{3}=\frac{1}{2h_{3}}(\partial _{i}h_{3}-w_{i}\partial
_{4}h_{3}),\ \widehat{T}_{4i}^{3}=\frac{1}{2}\partial _{4}n_{i}.$

Similarly, we get
\begin{eqnarray*}
\widehat{T}_{bi}^{4} &=&\frac{1}{2}h^{4c}(\partial
_{i}h_{bc}-N_{i}^{e}\partial _{e}h_{bc}-h_{dc}\partial
_{b}N_{i}^{d}-h_{db}\partial _{c}N_{i}^{d}) \\
&=&\frac{1}{2h_{4}}(\partial _{i}h_{b4}-w_{i}\partial
_{4}h_{b4}-h_{4}\partial _{b}w_{i}-h_{3b}\partial _{4}n_{i}-h_{4b}\partial
_{4}w_{i})
\end{eqnarray*}%
i.e. $\widehat{T}_{3i}^{4}=-\frac{h_{3}}{2h_{4}}\partial _{4}n_{i},\
\widehat{T}_{4i}^{4}=\frac{1}{2h_{4}}(\partial _{i}h_{4}-w_{i}\partial
_{4}h_{4})-\partial _{4}w_{i}.$

The coefficients $\Omega _{ij}^{a}=\mathbf{e}_{j}\left( N_{i}^{a}\right) -%
\mathbf{e}_{i}(N_{j}^{a})$ are computed
\begin{eqnarray*}
\Omega _{ij}^{a} &=&\mathbf{\partial }_{j}\left( N_{i}^{a}\right) -\partial
_{i}(N_{j}^{a})-N_{j}^{b}\partial _{b}N_{i}^{a}+N_{i}^{b}\partial
_{b}N_{j}^{a} \\
&=&\mathbf{\partial }_{j}\left( N_{i}^{a}\right) -\partial
_{i}(N_{j}^{a})-w_{j}\partial _{4}N_{i}^{a}+w_{i}\partial _{4}N_{j}^{a}.
\end{eqnarray*}%
We obtain such nontrivial values
\begin{eqnarray}
\Omega _{12}^{3} &=&-\Omega _{21}^{3}=\mathbf{\partial }_{2}n_{1}-\partial
_{1}n_{2}-w_{2}\partial _{4}n_{1}+w_{1}\partial _{4}n_{2}{},  \notag \\
\Omega _{12}^{4} &=&-\Omega _{21}^{4}=\mathbf{\partial }_{2}w_{1}-\partial
_{1}w_{2}-w_{2}\partial _{4}w_{1}+w_{1}\partial _{4}w_{2}.  \label{omeg}
\end{eqnarray}

Summarizing the above formulas for $\partial _{4}n_{i}=0$ and $\mathbf{\partial }%
_{2}n_{1}-\partial _{1}n_{2}=0,$ we get the condition of zero torsion for
the ansatz in (\ref{ansk}) with $n_{k}=\partial _{k}n(x^{i}),$%
\begin{eqnarray}
\frac{1}{2h_{3}}(\partial _{i}h_{3}-w_{i}\partial _{4}h_{3}) &=&0,
\label{qa1} \\
\frac{1}{2h_{4}}(\partial _{i}h_{4}-w_{i}\partial _{4}h_{4}) &=&\partial
_{4}w_{i},  \label{qa2} \\
\mathbf{\partial }_{2}w_{1}-\partial _{1}w_{2}-w_{2}\partial
_{4}w_{1}+w_{1}\partial _{4}w_{2} &=&0.  \label{qa3}
\end{eqnarray}
In this form we can define a LC--configuration. The final step is to impose the
condition that the coefficients $n_{k}$ do not depend on $y^{4}.$ This can
be fixed in the form $_{1}n_{k}(x^{i})=\partial _{k}n(x^{i})$ and $_{2}n_{k}=0,$
i.e. $n_{k}=\partial _{k}n(x^{i}).$

Finally, we note that the LC-conditions can be formulated recurrently, in
similar forms, for higher order shells both for zero and non-zero sources.

\end{document}